\RequirePackage[normalem]{ulem} 
\RequirePackage{color}\definecolor{RED}{rgb}{1,0,0}\definecolor{BLUE}{rgb}{0,0,1} 

\documentclass[useAMS,usenatbib]{mnras}
\usepackage{url,ulem,amsmath,epsfig,epstopdf}
\usepackage{graphicx}
\usepackage{epsf}
\usepackage{color}
\usepackage{rotating}
\usepackage[T1]{fontenc}
\usepackage{array,multirow}

\usepackage{aas_macros}
\usepackage{amsmath}
\usepackage{amssymb}
\usepackage{amsfonts}
\usepackage{txfonts}
\def\lsim{\mathrel{\lower0.6ex\hbox{$\buildrel {\textstyle <}
 \over {\scriptstyle \sim}$}}}
\def\gsim{\mathrel{\lower0.6ex\hbox{$\buildrel {\textstyle >}
 \over {\scriptstyle \sim}$}}}

\def\hmpc{~h^{-1}\rm Mpc}

\newcommand{\mc}[1]{\textcolor{red}{\bf #1}}

\newcommand{\bisous}{Bisous}
\newcommand{\classic}{\textsc{CLASSIC}}
\newcommand{\disperse}{\textsc{DisPerSE}}
\newcommand{\fine}{\textsc{FINE}}
\newcommand{\origami}{\textsc{ORIGAMI}}
\newcommand{\mst}{\textsc{MST}}
\newcommand{\dtfe}{\textsc{DTFE}}
\newcommand{\Nexus}{\textsc{NEXUS}}
\newcommand{\nexus}{\textsc{NEXUS+}}
\newcommand{\tweb}{T-web}
\newcommand{\vweb}{V-web}
\newcommand{\mswa}{\textsc{MSWA}}
\newcommand{\mmft}{\textsc{MMF-2}}
\newcommand{\spine}{Spineweb}
\newcommand{\logFilter}{Log-Gaussian}
\newcommand{\Vector}[1]{\mathbf{#1}}

\begin{document}
\title[Tracing the Cosmic Web]{{Tracing the Cosmic Web}\mc{}\thanks{This paper is the outcome of the ``Tracing the Cosmic Web'' Lorentz Center workshop, held in Leiden, 17-21 February of 2014.}}
\author[Libeskind et al.]
{\parbox{\textwidth}{Noam I Libeskind$^{1}$\thanks{email: nlibeskind@aip.de}, Rien van de Weygaert$^{2}$, Marius Cautun$^{3}$, Bridget Falck$^{4}$, Elmo Tempel$^{1,5}$, Tom Abel$^{6,7}$, Mehmet Alpaslan$^{8}$, Miguel~A. Arag\'on-Calvo$^{9}$, Jaime E. Forero-Romero$^{10}$, Roberto Gonzalez$^{11,12}$, Stefan Gottl\"{o}ber$^{1}$, Oliver Hahn$^{13}$, Wojciech A. Hellwing$^{14,15}$, Yehuda Hoffman$^{16}$, Bernard J. T. Jones$^{2}$, Francisco Kitaura$^{17,18}$, Alexander Knebe$^{19,20}$, Serena Manti$^{21}$, Mark Neyrinck$^{3}$, Sebasti\'an~E.~Nuza$^{22,23,1}$ 
, Nelson Padilla$^{11,12}$, Erwin Platen$^{2}$,  Nesar Ramachandra$^{24}$,  Aaron Robotham$^{25}$, Enn Saar$^{5}$, Sergei Shandarin$^{24}$, Matthias Steinmetz$^{1}$, Radu S. Stoica$^{26,27}$, Thierry Sousbie$^{28}$, Gustavo Yepes$^{18}$\newline
\emph{\normalsize Affiliations are listed at the end of the paper}}}

%

\maketitle

 \begin{abstract}
The cosmic web is one of the most striking features of the distribution of galaxies and dark matter on the largest scales in the Universe. It is composed of dense regions packed full of galaxies, long filamentary bridges, flattened sheets and vast low density voids. The study of the cosmic web has focused primarily on the identification of such features, and on understanding the environmental effects on galaxy formation and halo assembly. As such, a variety of different methods have been devised to classify the cosmic web -- depending on the data at hand, be it numerical simulations, large sky surveys or other. In this paper we bring twelve of these methods together and apply them to the same data set in order to understand how they compare. In general these cosmic web classifiers have been designed with different cosmological goals in mind, and to study different questions. Therefore one would not {\it a priori} expect agreement between different techniques however, many of these methods do converge on the identification of specific features. In this paper we study the agreements and disparities of the different methods. For example, each method finds that knots inhabit higher density regions than filaments, etc. and that voids have the lowest densities. For a given web environment, we find substantial overlap in the density range assigned by each web classification scheme. We also compare classifications on a halo-by-halo basis; for example, we find that 9 of 12 methods classify around a third of group-mass haloes (i.e. $M_{\rm halo}\sim10^{13.5}h^{-1}M_{\odot}$) as being in filaments. Lastly, so that any future cosmic web classification scheme can be compared to the 12 methods used here, we have made all the data used in this paper public.
\end{abstract}

\begin{keywords}
{Cosmology: theory -- dark matter -- large-scale structure of the Universe. Methods: data analysis}
\end{keywords}

\section{Introduction}
\label{section:intro}
On Megaparsec scales the matter and galaxy distribution is not uniform, but defines an intricate multi-scale inter-connected network 
which is known as the \emph{cosmic web} \citep{bondweb1996}. It represents the fundamental spatial organization of 
matter on scales of a few up to a hundred Megaparsec. Galaxies, intergalactic gas and dark matter arrange themselves in a salient wispy 
pattern of dense compact clusters, long elongated filaments, and sheetlike tenuous walls surrounding near-empty void regions. Ubiquitous throughout  
the entire observable Universe, such patterns  exist at nearly all epochs, albeit at smaller scales. It defines a complex spatial 
pattern of intricately connected structures, displaying a rich geometry with multiple morphologies and shapes. This complexity is 
considerably enhanced by its intrinsic multiscale nature, including objects over a considerable range of spatial scales and densities. 
For, a recent up-to-date report on a wide range of relevant aspects of the cosmic web, we refer to the volume by \cite{iau308}.

The presence of the weblike pattern can be easily seen in the spatial distribution of galaxies. Its existence was  suggested 
by early attempts to map the nearby cosmos in galaxy redshift surveys \citep{gregthomp1978,joeveer1978b,lapparent1986,geller1989,shectman1996} 
Particularly iconic was the publication of the \emph{slice of the Universe} by \cite{lapparent1986}. Since then, the impression 
of a weblike arrangement of galaxies has been confirmed many times by large galaxy redshift surveys such as 2dFGRS 
\citep{colless2003,weyschaap2009}, the Sloan Digital Sky Survey SDSS \citep{tegmark2004} and the 2MASS redshift survey \citep{huchra2012}, 
as well as by recently produced maps of the galaxy distribution at larger cosmic depths such as VIPERS \citep{vipers2014}. 
From cosmological N-body simulations \citep[e.g.][]{springmillen2005,illustris2014,eagle2015} and recent Bayesian reconstructions of the 
underlying dark matter distribution in the Local Universe \citep{hess2013,kitaura2013,2014MNRAS.445..988N,leclercq2015,sorce2016}, we have 
come to realize that the weblike pattern is even more pronounced and intricate in the distribution of dark matter.

\subsection{The Components of the Cosmic Web}
The most prominent and defining features of the cosmic web are the filaments. The most outstanding specimen in the local Universe 
is the Pisces-Perseus chain \citep{giovanelli1985}. A recent systematic inventory of filaments in the SDSS galaxy redshift 
distribution has been catalogued by \cite{tempel2014} \cite[also see][]{jones2010,sousbie2011b}. Filaments appear to be the highways of the Universe, 
the transport channels along which mass and galaxies get channelled into the higher density cluster regions \citep{haarlemwey1993,2004ApJ...603....7K} and 
which define the connecting structures between higher density complexes \citep{bondweb1996,colberg2005,weybond2008,aragon2010}. On the largest 
scales, filaments on scales of 10 up to 100~Mpc, are found to connect complexes of superclusters - such as the great attractor 
\citep{1988ApJ...326...19L}, the Shapley concentration 
\citep{1930BHarO.874....9S,2006A&A...447..133P} or more recently the Vela supercluster \citep{2017MNRAS.466L..29K} - as was, for example, indicated by the work of \cite{2004ApJ...606...25B}, \cite{romanodiaz2007} and \cite{2015MNRAS.452.1052L}. 

By contrast, the tenuous sheetlike membranes are considerably more difficult to find in the spatial mass distribution traced by 
galaxies. Their low surface density renders them far less conspicuous than the surrounding filaments, while they are populated 
by galaxies with a considerably lower luminosity \citep[see e.g.][]{cautun2014}. When looking at the spatial structure outlined by 
clusters, we do recognise more prominent flattened supercluster configurations, often identified as \emph{Great Walls}, which 
is a reflection of their dynamical youth. Particularly outstanding specimens are the 
CfA Great Wall \citep{geller1989}, the Sloan Great Wall \citep{2005ApJ...624..463G}, and most recently the BOSS 
Great Wall \citep{2016A&A...588L...4L} and the well established supergalactic plane \citep{1953AJ.....58...30D,2000MNRAS.312..166L}. 

Along with filaments, the large void regions represent the most prominent aspect of the Megaparsec scale Universe. These are enormous regions 
with sizes in the range of $20-50h^{-1}$~Mpc that are practically devoid of any galaxy, usually roundish in shape and occupying the major 
share of space in the Universe \citep[see][for a recent review]{weyiau308}.  Forming an essential and prominent aspect of the 
{\it cosmic web} \citep{bondweb1996}, voids are instrumental in the spatial organisation of the cosmic web \citep{icke1984,sahni1994,
shethwey2004,einasto2011b,aragon2013}. The first indications for their existence was found in early galaxy redshift samples 
\citep{chincarini1975,gregthomp1978,zeldovich1982}, while the discovery of the 50 Mpc size Bo\"otes void by \cite{kirshner1981}, \cite{kirshner1987} 
and the CfA study by \cite{lapparent1986} established them as key aspects of the large scale galaxy distribution. Recent studies have 
been mapping and cataloguing the void population in the Local Universe \citep{fairall1998,pan2012,sutter2012}, and even that in the implied dark 
matter distribution \citep{leclercq2015b}. In the immediate vicinity of our Milky Way, one of the most interesting features 
is in fact the Local Void whose diameter is around 30~Mpc \citep{tully_atlas1987}. Its effectively repulsive dynamical influence 
has been demonstrated in studies of cosmic flows in the local volume \citep{tully2008}, while a recent study even indicated the 
dominant impact of a major depression at a distance of more than 100 Mpc \citep[the so-called ``dipole repeller'',][]{hoffman2017}. 

\subsection{Physics and Dynamics of the Cosmic Web}
The cosmic web is a direct result of two physical drivers, which are at the heart of the current paradigm of structure formation. 
The first is that the initial density field is a Gaussian random field, described by a power spectrum of density fluctuations \citep{adler1981,bbks}. The second is that these perturbations evolve entirely due to gravity \citep{peebles1980}. Gravitational instability is responsible 
for increasing the contrast in the universe, as rich over-dense regions grow in mass and density while shrinking in physical size, and as empty 
voids expand and come to dominate the volume of the universe. Once the gravitational clustering process begins to go beyond the 
linear growth phase, we see the emergence of complex patterns and structures in the density field.

Within the gravitationally driven emergence and evolution of cosmic structure the weblike patterns in the overall cosmic matter 
distribution do represent a universal but possibly transient phase. As borne out by a large array of N-body computer experiments 
of cosmic structure formation \citep[e.g.][]{springmillen2005,illustris2014,eagle2015}, web-like patterns 
defined by prominent anisotropic filamentary and planar features --- and with characteristic large underdense void regions --- are the 
natural outcome of the gravitational cosmic structure formation process. They are the manifestation of the anisotropic nature of gravitational 
collapse, and mark the transition from the primordial (Gaussian) random field to highly nonlinear structures that have fully collapsed into halos 
and galaxies. 
Within this context, the formation and evolution of anisotropic structures are the product of anisotropic deformations accurately described by the Zeld'd'dovich formalism in the mildly nonlinear stage, driven by gravitational tidal forces induced by the inhomogeneous mass distribution.
%
 In other words, it is the anisotropy of the force field and the resulting deformation of 
the matter distribution which are at the heart of the emergence of the weblike structure of the mildly nonlinear mass distribution \citep[also see][]{bondweb1996,hahn2007a,weybond2008,forero2009}.

This idea was first pointed out by \citet[][also see \citealt{icke1973}]{zeldovich1970} who described, in the now seminal \emph{"Zel'dovich approximation"} framework, how gravitational collapse amplifies any initial anisotropies and gives rise to 
highly anisotropic structures. Accordingly, the final morphology of a structure depends on 
the eigenvalues of the deformation tensor.  Sheets, filaments and clusters correspond to domains with one, two and three positive 
eigenvalues, while voids correspond to regions with all negative eigenvalues. Based on this realization, \cite{doroshkevich1970} derived 
a range of analytical predictions for structure emerging from an initial field of Gaussian perturbations. In the emerging picture of 
structure formation, also known as Zel'dovich's pancake picture, anisotropic collapse has a well defined sequence, with regions first 
contracting along one axis to form sheets, then along the second axis to produce filaments and only at the end to fully collapse along 
each direction \citep{shandzeld1989,shandsuny2009}. 

Following up on this, the early evolution of the cosmic web can be understood in detail in terms of the singularities and caustics that 
are arising in the matter distribution as a result of the structure of the corresponding flow field \citep[see][]{shandzeld1989,hidding2014}. 
Indeed, one of the most interesting recent developments in our understanding of the dynamical evolution of the cosmic web has been the 
uncovering of the intimate link between the emerging anisotropic structures and the multistream migration flows involved in the buildup of 
cosmic structure \citep[][]{shandarin2011,shandarin2012,falck2012,neyrinck2012,abel2012}. 

Also recent observational advances have enabled new 
profound insights into the dynamical processes that are shaping the cosmic web in our Local Universe. In particular the Cosmicflows-2 and 
Cosmicflows-3 surveys of galaxy peculiar velocities in our Local Universe have produced tantalizing results \citep{courtois2013,tully2014}, opening up a window on the flows of mass along and towards structures in the local cosmic web. Amongst others, these studies show 
the sizeable impact of low-density void regions on the dynamics in the vicinity of the Milky Way and have allowed 
the velocity shear based V-web identification of weblike components in the local Universe \citep{2015MNRAS.452.1052L,2015ApJ...812...17P,hoffman2017}. 

The extension of the Zel'dovich approximation, the \emph{adhesion approximation}, allows further insights into the hierarchical buildup 
of the cosmic web \citep{gurbatov1989,kofman1990,kofman1992,hidding2012}. By introducing an artificial viscosity term, the adhesion 
approximation mitigates some of the late-time limitations of the Zel'dovich approximation. It also leads to a profound understanding 
of the link between the evolving phase-space structure of the cosmic matter distribution and the tendency to continuously morph the 
emerging spatial structure into one marked by ever larger structures \citep[see also][for a review of analytical extensions to the Zel'dovich approximation]{1995PhR...262....1S}. 

Interestingly, for a considerable amount of time the emphasis on anisotropic collapse as agent for forming and shaping structure in 
the Zel'dovich pancake picture was seen as the rival view to the purely hierarchical clustering picture. In fact, the successful 
synthesis of both elements culminated in the \emph{cosmic web} theory \citep{bondweb1996}, which stresses the dominance of filamentary 
shaped features and appears to provide a successful description of large scale structure formation in the $\Lambda$CDM cosmology. This theoretical 
framework pointed out the dynamical relationship between the filamentary patterns and the compact dense clusters that stand out 
as the nodes within the cosmic matter distribution: filaments as cluster-cluster bridges \citep[also see][]{bondweb1996,weyedb1996,colberg2005,
weybond2008}. In the overall cosmic mass distribution, clusters --- and the density peaks in the primordial density field that are their 
precursors --- stand out as the dominant features for determining and outlining the anisotropic force field that generates the cosmic web. 
The cosmic web theory embeds the anisotropic evolution of structures in the cosmic web within the context of the hierarchically evolving 
mass distribution \citep{bondmyers1996}. Meanwhile, complementary analytical descriptions of a hierarchically evolving cosmic web within the 
context of excursion set theory form the basis for a statistical evaluation of its properties \citep{shethwey2004,shen2006}.

\subsection{Significance and Impact of the Cosmic Web}
Understanding the nature of the cosmic web is important for a variety of reasons. Quantitative measures of the cosmic web 
may provide information about the dynamics of gravitational structure formation, the background cosmological model, the 
nature of dark matter and ultimately the formation and evolution of galaxies. Since the cosmic web defines the fundamental spatial organization of matter and galaxies on scales of one to tens of Megaparsecs, its structure probes a wide variety of scales, form the linear to the nonlinear regime. This suggests that quantification of the cosmic web at these scales should provide a significant amount of 
information regarding the structure formation process. As yet, we are only at the beginning of systematically exploring the various 
structural aspects of the cosmic web and its components towards gaining deeper insights into the emergence of spatial 
complexity in the Universe \citep[see e.g.][]{cautun2014}. 

The cosmic web is also a rich source of information regarding the underlying cosmological model. The evolution, structure and dynamics of the 
cosmic web are to a large extent dependent on the nature of dark matter and dark energy. As the evolution of the cosmic web 
is directly dependent on the rules of gravity, each of the relevant cosmological variables will leave its imprint on the 
structure, geometry and topology of the cosmic web and the relative importance of the structural elements of the 
web, i.e. of filaments, walls, cluster nodes and voids. A telling illustration of this is the fact that  void regions of the cosmic web offer one of the cleanest probes and measures of dark energy as well as tests of gravity and General Relativity. Their structure and shape, as well as mutual alignment, are direct 
reflections of dark energy \citep{parklee2007,platen2008,leepark2009,lavaux2010,lavaux2012,bos2012,2015MNRAS.446L...1S,pisani2015}. Given that the measurement of cosmological parameters depends on the observer's web environment \citep[e.g.][]{2014MNRAS.438.1805W}, one of our 
main objectives is to develop means of exploiting our measures of filament structure and dynamics, and the connectivity 
characteristics of the weblike network, towards extracting such cosmological information. 

\begin{table*}
 \caption{An overview of the methods compared in this study.}
 \begin{tabular}{@{}lcccl}
  \hline\hline
  Method & Web types & Input & Type & Main References \\
  \hline
\ \\
  Adapted Minimal Spanning Tree (MST) & filaments & haloes & Graph \& Percolation & \citet{2014MNRAS.438..177A} \\
\ \\
  Bisous  & filaments & haloes & Stochastic & \citet{tempel2014,2016AC....16...17T} \\
  FINE  & filaments & haloes & Stochastic & \citet{gonzalez2010} \\
\ \\
  Tidal Shear Tensor (T-web) & all & particles & Hessian & \citet{forero2009} \\
  Velocity Shear Tensor (V-web) & all & particles & Hessian & \citet{2012MNRAS.425.2049H} \\
  CLASSIC & all & particles & Hessian & \citet{2012MNRAS.425.2443K} \\
\ \\
  NEXUS+ & all & particles & Scale-Space, Hessian & \citet{cautun2013} \\
  Multiscale Morphology Filter-2 (MMF-2) & all except knots & particles & Scale-Space, Hessian& \citet{aragon2007} \\
  &&&&\citet{aragon2014}\\
\ \\
  Spineweb	 & all except knots & particles & Topology & \citet{aragon2010a} \\
  DisPerSE & all except knots & particles & Topology & \citet{sousbie2011} \\
\ \\
  ORIGAMI & all & particles & Phase-Space & \citet{falck2012,falck2015} \\
  MultiStream Web Analysis (MSWA)& all & particles & Phase-Space & \citet{Ramachandara_Shandarin:15} \\
\ \\
  \hline
 \end{tabular}
 \label{table:methods}
\end{table*}

Perhaps the most prominent interest in developing more objective and quantitative measures of large-scale cosmic web 
environments concerns the environmental influence on the formation and evolution of galaxies, and the dark matter 
halos in which they form \citep[see e.g.][]{hahn2007b,hahnphd2009,cautun2014}. The canonical example of such 
an influence is that of the origin of the rotation of galaxies: the same tidal forces responsible for the torquing of collapsing 
protogalactic halos \citep{hoyle1951,peebles1969,doroshkevich1970} are also directing the anisotropic contraction of matter in 
the surroundings. We may therefore expect to find an alignment between galaxy orientations and large scale filamentary 
structure, which indeed currently is an active subject of investigation \citep[e.g.][]{aragon2007a,leepen2000,jones2010,codis2012,tempel2012,2013MNRAS.428.2489L,
tempel2013,trowland2013,trowlandphd2013,aragon2014,2016MNRAS.457..695P,hirv2017,2017MNRAS.464.4666G}. Some studies even claim this implies an instrumental 
role of filamentary and other weblike environments in determining 
the morphology of galaxies \citep[see e.g.][for a short review]{pichon2016}. Indeed, the direct impact of the structure and 
connectivity of filamentary web on the star formation activity of forming galaxies has been convincingly demonstrated 
by \citet[][see also \citealt{2009ApJ...703..785D,2015MNRAS.449.2087D,2015MNRAS.454..637G,aragon2016}]{2009Natur.457..451D}. Such studies point out the instrumental importance of the filaments as transport conduits of cold 
gas on to the forming galaxies, and hence the implications of the topology of the network in determining the evolution and 
final nature. Such claims are supported by a range of observational findings, of which the morphology-density relation 
\citep{dressler1980} is best known as relating intrinsic galaxy properties with the cosmic environment in which the 
galaxies are embedded \citep[see e.g.][]{kuutma2017}. A final example of a possible influence of the cosmic web on the nature of 
galaxies concerns a more recent finding that has lead to a vigorous activity in seeking to understand it. The satellite galaxy 
systems around the Galaxy and M31 have been found to be flattened. It might be that their orientation points at a direct influence 
of the surrounding large scale structures \citep[see][]{ibata2013,cautun2015,2015MNRAS.452.1052L,2015ApJ...800...34G,2015ApJ...799...45F,2016ApJ...829...58G}, for example a reflection of local filament or local sheet. 

\subsection{Detection and Classification of Cosmic Web Structure}
To enable further advances in the astronomical issues addressed above, we need to establish a more objective description 
and quantification of the structure seen in the cosmic web. However, extracting such topological and morphological information 
from a discrete set of points, provided by either an $N$-body simulation or a galaxy survey, is very difficult. As such, many different methods have been developed to tackle this problem (reviewed in depth in Section~4). Some of the problems faced by observational surveys include sampling errors, projection effects, observational errors, incomplete sky coverage, magnitude limits, as well as various biases (e.g. Malmquist bias, selection bias). On the other hand, $N$-body simulations return the full 6-dimensional phase space and density field of the simulated universe at any desired epoch. 
 A method that takes full advantage of this often unobservable information cannot be directly applied to observations, but can be applied to simulations constrained to match observations \citep[e.g.][]{2016arXiv160100093L}. For this reason, methods that are developed specifically for the analysis of numerical simulations, may be completely inapplicable to current observational data sets and vice versa. Yet the numerous articles in the literature which attempt to study the cosmic web often refer to the same structural hierarchy: knots, filaments, sheets and voids. Here, we use a numerical simulation to compare classifiers, that, regardless of their position on the theoretical to observational spectrum, speak the same language of knots, filaments, sheets, and voids.

In the spirit of previous structure finder comparison projects~\citep[][etc.]{Colberg2008,Knebe2011}, we present a comparison of cosmic web identification codes and philosophies. However, our comparison differs significantly from e.g. the seminal Santa Barbara comparison project \citep{1999ApJ...525..554F} or other tests of codes which purport to model the same physical process \citep[e.g.][]{2012MNRAS.423.1726S,2013MNRAS.435.1618K}. Instead, the methods compared here were developed for very different purposes, to be applied to different kinds of data and with different goals in mind. Some of the methods are based on treating galaxies (haloes) as points; while others were developed to be applied to density or velocity fields. Furthermore, unlike halo finders seeking collapsed or bound objects, there is no robust analytical theory \cite[such as the spherical top hat collapse model of][]{1999MNRAS.308..119S} which we may use as a guide for how we expect different cosmic web finders to behave. Therefore, we enter into this comparison fully expecting large disagreements between the methods examined.

\subsection{Outline}
This paper is laid out as follows: in Section~2 we group the different methods into ``families'' that follow broadly similar approaches. In Section~3, we present the test dataset that has been used as the basis for our comparison. In Section~4, we review each method that has taken part in the comparison. In Section~5, we describe the results of the comparison. In Section 6 we summarise our results and draw conclusions.

\section{Web Identification Methods: \ \ \ Classification}
\label{section:methods}

It is a major challenge to characterise the structure, geometry and connectivity of the cosmic web. The complex spatial pattern -- marked  
by a rich geometry with multiple morphologies and shapes, an intricate connectivity, a lack of  symmetries, an intrinsic multiscale nature and 
a wide range of densities -- eludes a sufficiently relevant and descriptive analysis by conventional statistics to quantify the arrangement of 
mass and galaxies. 

Many attempts to analyze the clustering of mass and galaxies at Megaparsec scales have been rather limited in their ability to describe and quantify, let 
alone identify, the features and components of the cosmic web. Measures like the two-point correlation function, which has been the mainstay of many 
cosmological studies over the past nearly forty years \citep{peebles1980}, are not sensitive to the spatial complexity of patterns in the mass and galaxy distribution.  
The present paper seeks to compare the diverse range of more sophisticated techniques that have been developed over the past few years to address the 
spatially complex Megaparsec scale patterns delineated by mass and galaxies in the Universe. 

In the present study we compare the results and web evaluations and identifications of 12 different formalisms. 
They are diverse, involving different definitions for the physical identity of the structural features, as well 
as employing different means of turning these definitions into practical identification tools. The various 
different methods that have been developed can largely be grouped into five main classes:

\begin{itemize}
\item[1.] {\bf Graph and Percolation techniques.} The connectedness of elongated supercluster structures in the cosmic matter distribution 
was first probed by means of percolation analysis, introduced and emphasized by Zel'dovich and coworkers \citep{zeldovich1982,shandzeld1989,
shandarin2004}. A related graph-theoretical construct, the minimum spanning tree (MST) of the galaxy distribution, was extensively 
analyzed by Bhavsar and collaborators \citep{barrow1985,graham1995,colberg2007} in an attempt to develop an objective measure of filamentarity. 
\citet{colberg2007} set out to identify filaments and their adjoining clusters, using an elaborate set of criteria for the identification of 
features based on the branching of MSTs. In our study, we involve the MST based algorithm developed by \cite{2014MNRAS.440L.106A} for identification 
of filaments and void regions in the GAMA survey \citep{2014MNRAS.438..177A}. 

\item[2.] {\bf Stochastic methods.} This class of methods involves the statistical evaluation of stochastic geometric concepts. Examples are 
 filament detection algorithms based on the Bayesian sampling of well-defined and parameterized stochastic spatial (marked) point 
processes that model particular geometric configurations. \citet{stoica2005,2007JRSSC..56....1S,2010A&A...510A..38S} and \citet{2016AC....16...17T} use the Bisous model as an object point process of connected and aligned cylinders to locate and catalogue filaments in galaxy surveys. One of the advantages of this approach is that it can be applied directly on the original galaxy point field, given by the positions of the galaxies centres, without requiring the computation of a continuous density field. These 
methods are computationally very demanding. A thorough mathematical nonparametric formalism involving the medial axis of a point cloud, as yet 
for 2-D point distributions, was proposed by \cite{genovese2010}. It is based on a geometric representation of filaments as the medial axis of 
the data distribution. Also solidly rooted within a geometric and mathematical context is the more generic geometric inference formalism 
developed by \cite{chazal2009}. It allows the recovery of geometric and topological features of the supposedly underlying density field from a 
sampled point cloud on the basis of distance functions. In addition, we also see the proliferation of tessellation-based algorithms. Following 
specific physical criteria, \cite{gonzalez2010} put forward a promising combination of a tessellation-based density estimator 
and a dynamical binding energy criterion \citep[also see][]{weyschaap2009}. We may also include another recent development in this broad 
class of methods. \cite{leclercq2015,leclercq2015c} describe a highly interesting framework for the classification of 
geometric segments using information theory. \cite{leclercq2016} have  previously compared a few cosmic-web classifiers to each other, judging them on the basis of their information content.
\end{itemize}

\begin{figure*}
 \mbox{\hskip -1.27truecm\includegraphics[width=1.1\textwidth]{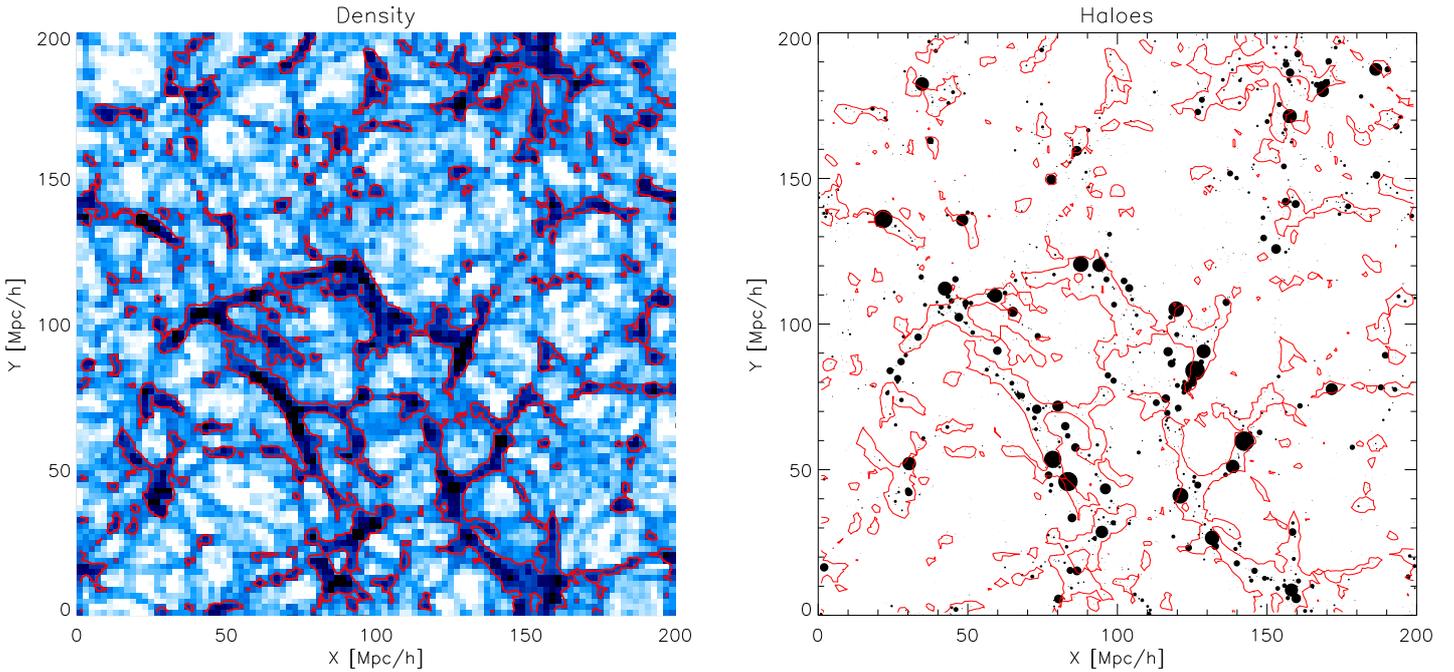}}
 \vspace{-0.6cm}
 \caption{ A thin slice through the cosmological simulation used for comparing the web identification methods. The left panel shows the density field in a $2\hmpc$ slice with darker colours corresponding to higher density regions. The red lines show the $\delta=0$ contours (dividing over and under dense regions, with respect to the mean) and are reproduced in the right panel (and in Fig.~\ref{fig:env_slice_a} as black lines). The right panel shows the positions of haloes in a $10\hmpc$ slice, where symbol sizes are scaled by halo mass. This same slice will be used to showcase the web identification methods in Figs.~\ref{fig:env_slice_a} and ~\ref{fig:env_slice_b} as well as the level of agreement across web finders in Fig.~\ref{fig:halo_agree}. }
 \label{fig:den_slice}
\end{figure*}

\begin{itemize}
\item[3.] {\bf Geometric, Hessian-based methods.} A large class of approaches exploits the morphological and (local) geometric  information included in the Hessian of the density, tidal or velocity shear fields
\citep[e.g.][]{aragon2007,hahn2007a,forero2009,bond2010a,2012MNRAS.421L.137L,cautun2013}. Based on the realization that the formation and 
dynamical evolution of the cosmic web is tied to the tidal force field \citep[see][]{bondweb1996}, \cite{hahn2007a} 
developed an elaborate classification scheme based on the signature of the tidal tensor \citep[also see][]{hahn2007b}. 
A further extension and elaboration of this tidal field based scheme was developed by \cite{forero2009}, while also the 
multiscale Nexus formalism incorporates versions that classify weblike features on the tidal tensor signature \citep[][see below]{cautun2013}

Following a similar rationale and focusing on the link between emerging weblike structures and the nature of the 
velocity flow in and around these features, in a sense following up on the classic realization of such a connection by \cite{zeldovich1970}, 
Libeskind, Hoffman and collaborators forwarded the V-web technique \citep{2012MNRAS.425.2049H,2012MNRAS.421L.137L,2013MNRAS.428.2489L,2013ApJ...766L..15L,2014MNRAS.441.1974L,2014MNRAS.443.1274L,2015MNRAS.452.1052L,2015MNRAS.453L.108L,2015MNRAS.446.1458M,2016MNRAS.460..297M,2016MNRAS.458..900C,2016MNRAS.457..695P}. Its classification is explicitly based on the 
signature of the velocity shear field. 

Instead of using the tidal or velocity sheer field configuration, one may also try to link directly to the morphology of the density field itself 
\citep{aragon2007,bond2010a,cautun2013}. Though this allows a more detailed view of the  multiscale matter distribution, it 
is usually more sensitive to noise and less directly coupled to the underlying dynamics of structure formation than the tidal field morphology. 
A single scale dissection of the density field into its various morphological components has been defined by \cite{bond2010a}, and applied to N-body 
simulations and galaxy redshift samples \citep[also see][]{bond2010a,bond2010b,choi2010}. 

\item[3b.] {\bf Scale-space Multiscale Hessian-based methods.} While most of the Hessian-based formalisms are defined 
on one particular (smoothing) scale for the field involved, explicit multiscale versions have also been developed. 
The MMF/Nexus Multiscale Morphology Filter formalism of \cite{aragon2007} and \cite{cautun2013}  look at structure from a {\it Scale Space} point of view, where the  (usually Gaussian) smoothing scale of the field, defines an extra dimension.
This formalism takes into account the multiscale character of the 
cosmic mass distribution by assessing at each spatial location the prominence of structural signatures, set by the signature of the Hessian of 
the field involved \citep{aragon2007,cautun2013}. A somewhat similar multiscale approach was followed by the Metric Space Technique described 
by \cite{wu2009}, who applied it to a morphological analysis of SDSS-DR5.  While the original MMF method \citep{aragon2007} only 
involved the density field, the Nexus formalism extended this to a versatile algorithm that classifies the cosmic web on the basis 
of a multiscale filter bank applied to  either the density, tidal, velocity divergence or velocity shear fields. Applying the technique to the logarithm of the density increases its sensitivity and dynamical range and allows the approach to attain its optimal form, the so called \nexus{} method, revealing both major filamentary arteries as 
well as tiny branching tendrils \citep{cautun2013}. 

\item[4.] {\bf Topological methods.} While the Hessian-based methods concentrate on criteria of the local geometric structure of density, 
velocity or tidal field, another family of techniques seeks to assess the cosmic web by studying the connectivity and topological 
properties of the field involved. A typical example involves the delineation of underdense void basins in the large scale mass 
distribution by means of the {\it Watershed Transform}, in the form of the Watershed Void Finder \citep{platen2007} and ZOBOV 
\citep{neyrinck2008}. The Spineweb procedure \citep{aragon2010} extends this to an elaborate scheme for tracing the various 
weblike features -- filaments, sheets and voids -- on purely topological grounds. Spineweb achieves this by identifying the 
central axis of filaments and the core plane of walls with the boundaries between the watershed basins of the density field. 
While the basic Spineweb procedure involves one single scale, the full multiscale spineweb procedure allows a multiscale topological 
characterization of the cosmic web \citep{aragon2010b,aragon2013}. 

In essence, the Spineweb procedure is a practical implementation of the mathematics of Morse theory \citep{morse1934}. Morse theory 
describes the spatial connectivity of the density field on the basis of its singularity structure, i.e. on the location and identity 
of the singularities - maxima, minima and saddle points - and their spatial connectivity by means of the characteristic lines 
defined by the gradient field. \cite{colombi2000} first described the role of Morse theory in a cosmological context, which 
subsequently formed the basis of the {\it skeleton analysis} by \cite{novikov2006} (2-D) and \cite{sousbie2008} (3-D). This 
defined an elegant and mathematically rigorous tool for filament identification. In a considerably more versatile elaboration 
of this, invoking the power of topological persistence to identify topologically significant features, \cite{sousbie2011}  
has formulated the sophisticated \disperse~formalism that facilitates the detection of the full array of structural features 
in the cosmic mass distribution \citep[also see][]{sousbie2011b}. Nonetheless, most of its applications are directed towards outlining the filaments. A further 
development along these lines, invoking the information provided by persistence measures, is that advocated 
by \cite{shivashankar2016}.   

\item[5.] {\bf Phase-space methods.} Most closely connected to the dynamics of the cosmic web formation process  
are several recently proposed formalisms that look at the phase-space structure of the evolving mass distribution  
\citep[][]{abel2012,falck2012,shandarin2012}. They are based on the realization that -- in cosmologies in which 
the intrinsic velocity dispersion of particles in the primordial universe is small -- the evolving 
spatial mass distribution has the appearance of a 3D sheet folding itself in 6D phase space, 
a {\it phase space sheet}. By assessing its structure in full phase space, these formalisms 
trace the mass streams in the flow field reflecting the emergence of nonlinear structures. Noting that the 
emergence of nonlinear structures occurs at locations where different streams of the corresponding 
flow field cross each other, these phase-space methods provide a dynamically based morphological 
identification of the emerging structures. 


This class of methods contains the \origami\ formalism \citep{falck2012,falck2015}, the phase-space sheet methods of \cite{shandarin2011} 
\citep[also see][]{Ramachandara_Shandarin:15} and \cite{abel2012}, and the Claxon formalism \citep{hidding2017}. The Claxon approach incorporates the modelling of 
the nonlinear evolution of the cosmic mass distribution by means of the adhesion 
formalism \citep{gurbatov1989,hidding2012}, in order to identify and classify the singularities -- shocks -- emerging 
in the evolving structure. Claxon states that these singularities trace the skeleton of the cosmic web. 
\end{itemize}
\section{Test data:  Simulation and Data set}
\label{section:simulation}

Each of the participants applied their web identification methods to the same Gadget-2~\citep{2005MNRAS.364.1105S} dark matter only $N$-body simulation, with a box size of 200 $\hmpc$ and $512^3$ particles. The $\Lambda$CDM cosmological parameters are taken from Planck \citep{2014A&A...571A..16P}: $h=0.68$, $\Omega_M = 0.31$, $\Omega_\Lambda = 0.69$, $n_s = 0.96$, and $\sigma_8 = 0.82$. Haloes in the simulation are identified using a standard FOF algorithm \citep{1985ApJ...292..371D}, with a linking length of $b=0.2$ and a minimum of 20 particles per halo. Fig. \ref{fig:den_slice} shows a thin slice through the density field and the halo population of this simulation.

The main output of the methods is the classification of the dark matter density field into one of four web components: knot, filament, wall and void. This classification is performed for either volume elements (e.g. the Hessian methods), dark matter mass elements (e.g. the phase-space methods), or for the haloes (e.g. the point process methods). The exact choice was left to the discretion of the authors to better reflect the procedure used in the studies employing those methods. 

Though the output format of the web identification methods may vary, each participant was asked to provide two datasets: the web identification tag defined on a regular grid with a $2\hmpc$ cell size ($100^3$ cells) and the web classification of each FOF halo. Most methods returned both datasets except for some of the point-process methods (MST, FINE), for which assigning a environment tag to each grid cell would not make sense. These return information regarding the filamentary environment of just the FOF haloes.

The simulation is made publicly available\footnote{http://data.aip.de/tracingthecosmicweb/ \\  doi:10.17876/data/2017\_1} for exploitation by interested parties. We have included the $z=0$ Gadget snapshots, the FOF halo catalogue as well as the output of each cosmic web method included in this work. Where available, each method's classification is returned on a regular grid. Included in the data set is also the FOF catalogue appended with the classification of each halo for each method. We encourage other methods not included in this paper, to use this data set as a bench mark of the community's current status.


\section{Web Identification Methods:\ \ \  \\ \ \ \ \ \ \ Description \& Details}
The following section describes each method as well as the practical details in the analysis of this data set. See Table~\ref{table:methods} for a brief summary.

\subsection{Adapted minimal spanning tree\\ \hskip 0.75cm(Alpaslan \& Robotham)}
\label{section:Alpaslan}

The adapted minimal spanning tree algorithm \citep[][see also \citealt{barrow1985,2004A&A...418....7D,colberg2007}]{2014MNRAS.438..177A}, uses a multiple pass approach to detecting large scale structure, similar to \cite{Murphy2011}. 

Designed to be run on galaxy survey data, the adapted MST algorithm begins by identifying filamentary networks by using galaxy group centroids as nodes for an initial MST; in doing so, redshift-space distortion effects typically present in such data are successfully removed. The maximal allowable distance $b$ between two group (or halo) centres is selected such that at least 90\% of groups or haloes with $M_{\rm halo} \geq 10^{11} M_{\odot}$ are considered to be in filaments. A large $b$ will cause galaxies in voids to be associated with filaments, and a small $b$ will only identify close pairings of groups to be in filaments and ignore the expansive structures visible in the data.

Following the identification of filaments from group centres, galaxies that are within an orthogonal distance $r$ of filaments are associated with those filaments. Additionally, the topological structure of the MST that forms each filament is analysed, with the principal axis of each filament (the so-called `backbone') identified as the longest contiguous path of groups that spans the entirety of the filament, along with tributary `branches' that link to it. The size and shapes of these pathways are used to successfully compare observational results to simulated universes in \cite{2014MNRAS.438..177A}. Galaxies associated with each filament are further associated with the branch of the filament they are closest to, allowing for a detailed analysis of galaxy properties as a function of filament morphology \citep{2015MNRAS.451.3249A,2016MNRAS.457.2287A}.

Galaxies that are too distant from filaments are reprocessed under a second MST which identifies smaller-scale interstitial structures dubbed `tendrils' \citep{2014MNRAS.440L.106A}. Tendrils typically contain a few tens of galaxies, and typically exist within voids, or bridge the gap between two filaments within underdense regions. The properties of galaxies in these structures are often similar to those in more dense filaments \citep{2015MNRAS.451.3249A}.

Finally, galaxies that are beyond a distance $q$ from tendrils are identified as isolated void galaxies. The distances $r$ and $q$ are selected such that the integral over the two-point correlation, $\int R^2 \xi(R) \,\mathrm{d} R$, of void galaxies is minimized. This definition of a void galaxy ensures that the algorithm identifies a population of very isolated galaxies; this differs from searching for void galaxies in low density regions, which does allow for clustering. 

\subsection{Bisous\\ \hskip 0.75cm (Tempel, Stoica \& Saar)}

The detection of cosmic web filaments is performed by applying an object (marked) point process with interactions \citep[the Bisous process;][]{stoica2005} to the spatial distribution of galaxies or haloes. This algorithm provides a quantitative classification that complies with the visual impression of the cosmic web and is based on a robust and well-defined mathematical scheme. More detailed descriptions of the Bisous model can be found in \citet{2007JRSSC..56....1S,2010A&A...510A..38S} and \citet{tempel2014, 2016AC....16...17T}. A brief and intuitive summary is provided below.

The model approximates the filamentary web by a random configuration of small segments (cylinders). It is assumed that locally, galaxy conglomerations can be probed with relatively small cylinders, which can be combined to trace a filament if the neighboring cylinders are oriented similarly. An advantage of the approach is that it relies directly on the positions of galaxies and does not require any additional smoothing for creating a continuous density field.

The solution provided by the model is stochastic. Therefore, it is found some variation in the detected patterns for different Markov chain Monte Carlo (MCMC) runs of the model. On the other hand, thanks to the stochastic nature of the method simultaneously a morphological and a statistical characterization of the 
filamentary pattern is gained.

In practice, after fixing an approximate scale of the filaments, the algorithm returns the filament detection probability field together with the filament orientation field. Based on these data, filament spines are extracted and a filament catalogue is built in which every filament is represented by its spine as a set of points that defines the axis of the filament.

The spine detection follows two ideas. First, filament spines are located at the highest density regions outlined by the filament probability maps. Second, in these regions of high probability for the filamentary network, the spines are oriented along the orientation field of the filamentary network. See \citet{tempel2014, 2016AC....16...17T} for more details of the procedure.

The \bisous{} model uses only the coordinates of all haloes. These were analyzed using a uniform prior for filament radius between $0.4-1.0\hmpc$, which determines the scale of the detected structures. This scale has a measurable effect on properties of galaxies \citep{2015ApJ...800..112G, 2015A&A...576L...5T,2015MNRAS.450.2727T}. Using the halo distribution, the \bisous{} model generates two fields -- the filament detection and the filament orientation fields. These two fields are continuous and have a well defined value at each point. To generate the datasets required by the comparison project, each grid cell on the target $100^3$ mesh and each FOF halo was tagged as either part of a filament or not. For the visitmap\footnote{In mathematics the visitmap is also called a ``level set'', and refers to a probabilistic filament detection map, see \cite{Heinrich:12}.} a threshold value 0.05 was used, which selects regions that are reasonably covered by the detected filamentary network. To exclude regions where the filament orientation is not well defined (e.g. regions at intersection of filaments), it is required that orientation strength parameter is higher than 0.7. The same values were used in previous studies \citep[e.g.][]{2015A&A...583A.142N}.

\subsection{FINE \\ \hskip 0.75cm(Gonzalez \& Padilla)}
\label{section:gonzalez}

The filamentary structure in the cosmic web can be found by following the highest density paths between density peaks.
The Filament Identification using NodEs (\fine) method described in \citet{gonzalez2010} looks for filaments in halo or galaxy distributions.

The method requires halo/galaxy positions and masses (luminosities for galaxies), and we define as Nodes, the haloes/galaxies above a given mass/luminosity.
The mass of the nodes will define the scale of the filaments in the search. The smaller the node masses, the smaller the filaments that will be found between them.

The density field is computed using Voronoi Tessellations similar to \citet{schaapwey2000}.
The method looks first for a filament skeleton between any node pair by following the highest density path and a minimum separation; those two parameters characterize the filament quality.
Filament members are selected by binding energy in the plane perpendicular to the filament; this condition is associated to characteristic orbital times. However, if one assumes a fixed orbital timescale for all filaments, the resulting filament properties show only marginal changes, indicating that the use of dynamical information is not critical for this criterion.
Filaments detected using this method are in good agreement with \citet{colberg2005} who use  by-eye criteria.

In this comparison we define nodes as the haloes with masses above $5\times10^{13}M_{\odot}$, and the minimum density threshold for the skeleton search is $5$ times the mean Voronoi density.

\subsection{V-web: Velocity Shear Tensor\\ \hskip 0.75cm (Libeskind, Hoffman, Knebe \& Gottl\"ober)}
\label{section:libeskind}
The cosmic web may be quantified directly using the cosmic velocity field, as suggested by \cite{2012MNRAS.425.2049H}. This method is ideally suited to numerical simulations but may be applied to any cosmic velocity field, for example reconstructed ones from redshift or velocity data.

The method is similar to that suggested by \cite{hahn2007a} but uses the shear field instead of the Hessian of the potential. In the linear regime these two methods give similar results. First, a grid is superimposed on the particle distribution. A ``clouds in cells'' (CIC) technique is used to obtain a smoothed density and velocity distribution at each point on the grid. The CIC of the velocity field is then Fast Fourier Transformed into $k$-space and smoothed with a Gaussian kernel. The size of the kernel determines the scale of the computation and must be at least equal to one grid cell (i.e. in this case we use a $256^3$ grid and so $r_{\rm smooth}\ge L_{\rm box}/256$) in order to wash out artificial effects introduced by the preferential axes of the Cartesian grid. Using the Fourier Transform of the velocity field the normalized shear tensor is calculated as:
\begin{equation}
\Sigma_{\alpha\beta} = -\frac{1}{2H_{0}}\left(\frac{\partial v_{\alpha}}{\partial r_{\beta}}+\frac{\partial v_{\beta}}{\partial r_{\alpha}}\right)
\end{equation}
where $\alpha, \beta$ are the $x,y,z$ components of the positions $r$ and velocity $v$ and $H_{0}$ is the Hubble constant. Note that the shear tensor is simply the symmetric part of the velocity deformation tensor \citep[the anti-symmetric part being the curl or vorticity, see][]{2013ApJ...766L..15L,2014MNRAS.441.1974L}. The shear tensor is then diagonalised and the eigenvalues are sorted, according to convention ($\lambda_{\rm1}>\lambda_{\rm2}>\lambda_{\rm3}$). The eigenvalues and corresponding eigenvectors (${\bf e}_{\rm 1}$, ${\bf e}_{\rm 2}$, ${\bf e}_{\rm 3}$) of the shear field are obtained at each grid cell. Note that the eigenvectors (${\bf e}_{i}$'s) define non-directional lines and as such the +/- orientation is arbitrary and degenerate.

A web classification scheme based on how many eigenvalues are above an arbitrary threshold may be carried out at each grid cell. If none, one, two or three eigenvalues are above this threshold, the grid cell may be classified as belonging to a void, sheet, filament or knot. The threshold may be taken to be zero \citep[as in][]{hahn2007a} or may be fixed to another value to, e.g., reproduce the visual impression of the matter distribution; for the purposes of $\Lambda CDM$ simulations, such as this one, the threshold is chosen to be 0.44 \citep{forero2009,2012MNRAS.421L.137L,2013MNRAS.428.2489L,2014MNRAS.443.1274L}.

\subsection{T-web: Tidal Shear Tensor\\ \hskip 0.75cm (Forero-Romero, Hoffman \& Gottl\"ober)}
\label{section:forero-romero}

This method \citep[T-web,][]{forero2009} works on density field grids
obtained either from numerical simulations or reconstructions from
redshift surveys. 

The method builds on the work by \cite{hahn2007a}. It
also uses the Hessian of the gravitational potential 
\begin{equation}
T_{\alpha\beta} = \frac{\partial^2\phi}{\partial x_\alpha\partial x_\beta},
\end{equation}
where the physical gravitational potential has been normalized by
$4\pi G\bar{\rho}$ so that $\phi$ satisfies the Poisson
equation
\begin{equation}
\nabla^2\phi=\delta,
\end{equation}
with $\delta$ the dimensionless matter overdensity, $G$ the
gravitational constant and $\bar{\rho}$ the average density of the
Universe.

This tidal tensor can be represented by a real symmetric $3\times 3$
matrix with eigenvalues $\lambda_1>\lambda_2>\lambda_3$ and
eigenvectors ${\bf e}_1$, ${\bf e}_2$ and ${\bf e}_3$. The eigenvalues
are indicators of orbital stability along the directions defined by
the eigenvectors. 

This method introduces a threshold $\lambda_{\rm th}$ to gauge the
strength of the eigenvalues of the tidal shear tensor. The number of
eigenvalues larger than the threshold is used to classify the cosmic
web into four kinds of environments: voids (3 eigenvalues smaller than
$\lambda_{\rm th}$), sheets (2), filaments (1) and knots (0).

In practice the density is interpolated over a grid using the particle
data and a Cloud-In-Cell scheme. The Poisson equation is solved in
Fourier space to obtain the potential over a grid. At each grid cell
the shear tensor is computed to obtain and store the corresponding
eigenvalues and eigenvectors. The grid cell has a size of $\sim
1h^{-1}$Mpc and the threshold is fixed to be $\lambda_{\th}=0.2$ as
suggested by previous studies that aim at capturing the visual
impression of the cosmic web \citep{forero2009}.




\subsection{MMF/Nexus: the Multiscale Morphology Filter \\ \hskip 0.75cm(Arag\'on-Calvo, Cautun, van de Weygaert \& Jones)} 
The MMF/Nexus Multiscale Morphology Filter technique \citep[][]{aragon2007,aragon2010b,cautun2013,cautun2014,aragon2014} performs the morphological 
identification of the cosmic web using a \textit{Scale-Space formalism} that ensures the detection of structures present at all scales. The formalism 
consists of a fully adaptive framework for classifying the matter distribution on the basis of local variations in the density field, velocity 
field or gravity field encoded in the Hessian matrix in these scales. Subsequently, a set of morphological filters is used to classify the spatial 
matter distribution into three basic components, the clusters, filaments and walls that constitute the cosmic web. The final product of the 
procedure is a complete and unbiased characterization of the cosmic web components, from the prominent features present in overdense regions 
to the tenuous networks pervading the cosmic voids.

Instrumental for this class of MMF cosmic web identification methods is that it simultaneously pays heed to two principal 
aspects characterizing the weblike cosmic mass distribution. The first aspect invokes the Hessian of the corresponding 
fields to probe the existence and identity of the mostly anisotropic structural components of the cosmic web. 
The second, equally important, aspect uses a scale-space analysis to probe the multiscale character of the cosmic mass distribution, 
the product of the hierarchical evolution and buildup of structure in the Universe. 

The Scale Space representation of a data set consists of a sequence of copies of the data having different resolutions 
\citep{florack1992,lindeberg1998}. A feature searching algorithm is applied to all of these copies, and the features are extracted in a 
scale independent manner by suitably combining the information from all copies. A prominent application of Scale Space analysis involves 
the detection of the web of blood vessels in a medical image \citep{sato1998,li2003}. The similarity to the structural patterns seen on 
Megaparsec scales is suggestive. The Multiscale Morphology Filter has translated, extended and optimized the Scale Space technology to 
identifying the principal characteristic structural elements in the cosmic mass and galaxy distribution.  
The final outcome of the MMF/Nexus procedure is a field which at each location $\Vector{x}$ 
specifies what the local morphological signature is, cluster node, filaments, wall or void. 
The MMF/Nexus algorithms perform the environment detection by applying the above steps first to knots, then to filaments and 
finally to walls. Each volume element is assigned a single environment characteristic by requiring that filament regions 
cannot be knots and that wall regions cannot be either knots or filaments. The remaining regions are classified as voids.

Following the basic version of the MMF technique introduced by \cite{aragon2007}, it was applied to the analysis of the cosmic web 
in simulations of cosmic structure formation \citep{aragon2010b} and for finding filaments and galaxy-filament alignments in the SDSS galaxy 
distribution \citep{jones2010}. The principal technique, and corresponding philosophy, has subsequently branched into several 
further elaborations and developments. In this survey, we describe the Nexus formalism developed by \cite{cautun2013} and the 
MMF2 method developed by \cite{aragon2014}. Nexus has extended the MMF formalism to a substantially wider range of 
physical agents involved in the formation of the cosmic web, along with a substantially firmer foundation for the 
criteria used in identifying the various weblike structures. MMF-2 not only focusses on the multiscale nature 
of the cosmic web itself, but also addresses the nesting relations of the hierarchy. 

\subsubsection{\nexus{} \\ \hskip 0.9cm (Cautun, van de Weygaert \& Jones)}
\label{section:nexus}
The \nexus{} version of the MMF/Nexus formalism \citep{cautun2013,cautun2014} builds upon 
the original Multiscale Morphology Filter \citep{aragon2007,aragon2010b} algorithm and was developed with the goal of obtaining 
a more physically motivated and robust method. 

\nexus{} is the principal representative of the full \Nexus{} suite of cosmic web identifiers \citep[see][]{cautun2013}. 
These include the options for corresponding multiscale morphology identifiers on the basis of the raw density, the 
logarithmic density, the velocity divergence, the velocity shear and tidal force field. \Nexus{} has incorporated these 
options in a versatile code for the analysis of cosmic web structure and dynamics following the realization that they 
are significant physical influences in shaping the cosmic mass distribution into the complexity of the cosmic web. 

\nexus{} takes as input a regularly sampled density field. In a first step, the input field is Gaussian smoothed 
over using a \logFilter{} filter \citep[see][]{cautun2013} that is applied over a set of scales $[R_0,R_1,...,R_N]$, 
with $R_n=2^{n/2}R_0$. 
\nexus{} then computes an environmental signature for 
each volume element. 

The \Nexus{} suite of MMF identifiers pays particular attention to the key aspect of setting the detection 
thresholds for the environmental signature. Physical criteria are used to determine a detection threshold. All points 
with signature values above the threshold are valid structures. For knots, the threshold is given by the requirement 
that most knot-regions should be virialized. For filaments and walls, the threshold is determined on the basis of the 
change in filament and wall mass as a function of signature. The peak of the mass variation with signature delineates 
the most prominent filamentary and wall features of the cosmic web.

For the \nexus{} implementation, the Delaunay Tessellation Field Estimator \dtfe{} method \citep{schaapwey2000,weyschaap2009} is  
used to interpolate the dark matter particle distribution to a continuous density field defined on a regular grid of size 
$600^3$ (grid spacing of $0.33\hmpc$). \nexus{} was applied to the resulting density field using a set of 7 smoothing scales 
from $0.5$ to $4\hmpc$ (in increments of $\sqrt{2}$ factors). This resulted in an environment tag for each grid cell that, in a second step, 
was down sampled to the target $100^3$ grid using a mass weighted selection scheme. For each cell of the coarser grid, we computed the mass 
fraction in each environment using all the fine level cells ($6^3$ in total) that overlap the coarser one. Then, the coarser cell was assigned 
the environment corresponding to the largest mass fraction. Each FOF halo was assigned the web tag corresponding to the fine grid cell in 
which the halo centre was located.

\subsubsection{MMF-2: Multiscale Morphology Filter-2\\ \hskip 0.9cm (Arag\'on-Calvo)}
\label{section:MMF}
The MMF-2 implementation of the MMF formalism differs from the \Nexus{} formalism in that it focusses on the multiscale character 
of the initial density field, instead of that of the evolved mass distribution. 
In order to account for hierarchical nature of the cosmic web, MMF-2 introduces the concept of \textit{hierarchical space} 
\citep{aragon2010, aragon2013}. While the conventional scale-space approach emphasizes the scale of the structures, it does 
not addresss their nesting relations. To accomplish this, MMF-2 resorts to the alternative of \textit{hierarchical space} 
\citep{aragon2010, aragon2013, aragon2014}. 

\textit{Hierarchical space} is created in the 
first step in the MMF-2 procedure \citep{aragon2010, aragon2013}. It is obtained by Gaussian-smoothing the initial conditions, 
and in principal concerns a continuum covering the full range of scales in the density field. For practical purposes however, a 
small set of linear-regime smoothed initial conditions is generated. Subsequently, by means of an N-body code these conditions 
are gravitationally evolved to the present time. 

By applying to linear-regime smoothing, \textit{hierarchical space} involves density field Fourier modes that are independent. This 
allows the user to target specific scales in the density field before Fourier mode-mixing occurs. The subsequent gravitational 
evolution of these smooth initial conditions results in a mass distribution that contains all the anisotropic features of the Cosmic Web, 
while it lacks the small-scale structures below the smoothing scale. Dense haloes corresponding to these small scales are absent. This reduces the 
dynamic range in the density field and greatly limits the contamination produced by dense haloes in the identification of filaments 
and walls. 

In line with the MMF procedure, for each realisation in the hierarchical space a set of 
morphology filters is applied, defined by ratios between the eigenvalues of the Hessian matrix 
\citep[$\lambda_1 < \lambda_2 < \lambda_3$, see][]{aragon2007}.  It also involves the applications of a threshold to the response from 
each morphology filter. This leads to a  final product consisting of a set of binary masks sampled on a regular grid indicating 
which voxels belong to a given morphology at a given hierarchical level. 

\bigskip
\subsection{CLASSIC\\ \hskip 0.75cm (Manti, Nuza \& Kitaura)}
\label{section:CLASSIC}

The CLASSIC approach is based on performing a prior linearization to the cosmological density field and later a cosmic web classification of the resulting matter distribution. The method is implemented in two steps: first, a linearization is made to better fulfil the mathematical conditions of the original idea of cosmic web classification, which is based on a linear Taylor expansion of the gravitational field \citep[see][]{1974FizSz..24..304Z,hahn2007a}, and then, cosmological structures are divided into voids, sheets, filaments and knots. The linearization is done using higher order Lagrangian perturbation theory as proposed by \cite{2012MNRAS.425.2443K}. In this framework, a given density field can be expressed as the sum of a linear and a non-linear component which are tightly coupled to each other by the tidal field tensor. The cosmic web classification is performed on a grid cell in a similar way as suggested by \cite{hahn2007a}, i.e. counting the number of eigenvalues of the Hessian of the gravitational potential above a given threshold \citep[see also][]{forero2009}. In particular, the threshold adopted was chosen to obtain a volume filling fraction (VFF) of voids of about $70\%$ as done by \cite{2014MNRAS.445..988N} for their reconstruction on the local universe based on peculiar velocity fields. As a result, the corresponding VFFs of sheets, filaments and knots are uniquely determined by this choice. 

\subsection{Spine Web \\ \hskip 0.75cm (Arag\'on-Calvo, Platen \& van de Weygaert)}
\label{section:Spine}

The Spine method \citep{aragon2010a} produces a characterization of space based on the topology of the density field, catalogs of individual voids, walls and filaments and their connectivity. Its hierarchical implementation \citep{aragon2010b,aragon2013} allows us to describe the nesting properties of the elements of the cosmic web in a quantitative way. The Spine can be applied to both simulations and galaxy catalogs with minimal assumptions. Given its topological nature it is highly robust against geometrical deformations (e.g. fingers of God or polar grid sampling) as long as the topology of the field remains unchanged.

The Spine method extends the idea introduced in the watershed void finder \citep{platen2007} to identify voids as the contiguous regions sharing the same local minima. Walls are then identified as the two-dimensional regions where two voids meet and filaments correspond to the one-dimensional intersection of two or more walls. Nodes correspond to the intersection of two or more filaments but due to the finite size of voxels in practice they are difficult to recover and therefore we merge them with the filaments into the filament-node class. 

The Spine method can be extended to a fully hierarchical analysis as explained in  \citet{aragon2010b} and \citet{aragon2013}. In this approach voids regions are identified at several hierarchical levels (see MMF-2 method), then voids identified at large scales (high in the hierarchical space) are reconstructed in terms of the voids they contain at smaller scales in order to recover their original boundaries lost by the smoothing procedure used to create the hierarchical space. From the reconstructed voids we compute the watershed transform and identify walls and filament-nodes as described above. 

We use the fact that walls are the intersection of two voids to identify voxels belonging to a unique wall (voxels at the boundary between the same pair of walls). The same can be done for filaments in order to obtain a catalog of voids, walls and filaments. The same connectivity relations can be used to reconstruct the full graph describing the elements of the cosmic web.

\subsection{DisPerSE \\ \hskip 0.75cm(Sousbie)}
\label{section:Disperse}
\disperse{} is a formalism designed for analyzing the cosmic web, and in particular its filamentary network, 
on the basis of the topological structure of the cosmic mass distribution \citep{sousbie2011,sousbie2011b}. 
The elaborate framework of \disperse{} is based on three mathematical and computational pillars. These are 
Morse theory \citep{morse1934,milnor1963}, Discrete Morse theory \citep{forman1998,gyulassy2008} and the 
Delaunay Tessellation Field Estimator \dtfe{} \citep{schaapwey2000,weyschaap2009,cautun2011}. 
The formalism uses  three concepts in computational 
topology: Persistent Homology and Topological Simplification \citep{edelsbrunner2002,gyulassy2008,edelsbrunner2010}. 
These are used  for removal of noise and the selection of the significant morphological features from a discretely 
sampled cosmic mass distribution. 

\disperse{} analyzes and characterizes the cosmic web in terms of a spatial segmentation of space defined by the  
singularity structure of the cosmic mass distribution, the \emph{Morse complex}. 
The morphological components of the cosmic web are identified with the various $k$-dimensional manifolds that 
outline this uniquely defined segmentation. Filaments are identified with the ascending 
$1$-manifold. Voids, walls and clusters are identified with the ascending $3$-, $2$- and $0$-manifolds 
\citep[also see][]{aragon2010a,shivashankar2016}.  \disperse{} follows earlier 
applications of Morse theory to structural classification in astrophysical datasets \citep[][]{colombi2000}. 
The immediate precursor to \disperse{} is the skeleton formalism \citep{sousbie2008,sousbie2008b,sousbie2009}. 

Morse theory, which forms the basis for \disperse{}, looks at the singularity structure of the density field $f(x)$. It leads to 
the translation of the spatial distribution and connectivity of maxima, minima and saddle points in the density fields into a 
geometric segmentation of space that is known as the \emph{Morse complex}. This is a uniquely defined segmentation of space in a 
spatial tessellation of cells, faces, edges and nodes defined by the singularities and their connections by integral lines. 
The \emph{index} of a non-degenerate critical point is the number of negative eigenvalues of its Hessian: a \emph{minimum} of a 
field $f$ has index $0$, a \emph{maximum} has index $3$, while there are two types of \emph{saddles}, with index $1$  and $2$.  
Morse theory captures the connectivity of a field $f$ via the flowlines defined by the field gradient $\nabla f$, the \emph{integral lines}.
The field minima and maxima are the divergence and convergence points of these lines. It leads to a natural segmentation of space 
into distinct regions of space called ascending and descending manifolds. The \emph{ascending} $k$-manifold of a critical point $P$ defines 
the $k$-dimensional region of space defined by the set of points along integral lines that emanate from origin $P$. Conversely, the 
\emph{descending} $m$-manifold of a critical point $P$ is the $m$-dimensional region of space defined by the set of points along 
integral lines for which $P$ is the destination. 

Since astronomical datasets (N-body simulations, galaxy catalogs, etc.) are discrete and intrinsically noisy tracers of the density field,  \disperse{} utilizes \emph{Discrete Morse theory}  \citep[see e.g.][]{forman1998}. It consists of a combinatorial formulation of Morse theory 
in terms of intrinsically discrete functions defined over a simplicial complex\footnote{In essence, a simplicial complex 
is a geometric assembly of cells, faces, edges and vertices marking a discrete map of the volume. Cells, faces, edges and 
nodes/vertices are $3-$, $2-$, $1$- and $0$-dimensional simplices.}. A well-known example 
is the Delaunay tessellation $\cal D$ \citep{weygaert1994,okabe2000}. \disperse{} uses the Delaunay tessellation, and 
specifically its role as functional basis in the \dtfe{} formalism \citep{schaapwey2000,weyschaap2009,cautun2011}. 
\disperse{} uses the \dtfe{} density estimate at each sample point, while 
it assigns a density value $f(\sigma_k)$ to each simplex $\sigma_k$ of the Delaunay simplicial complex $\cal D$. On the basis of 
these values, and the mutual connections between the various simplices, one may identify discrete simplicial analogues to the 
singularity points, gradient vector field, integral lines and Morse complex \citep[see e.g.][for a detailed 
treatment]{gyulassy2008,sousbie2011}. 

The finite sampling of the density field introduces noise into the detection of structural features. Instead of resorting to a simplistic feature-independent filtering 
operation, which tends to suppress or even annihilate real structural features, \disperse{} makes use of persistent homology \citep{edelsbrunner2002,edelsbrunner2010}. Topological persistence 
is the language that allows the identification of features according to their significance \citep{edelsbrunner2002}. 
Persistence theory defines a topological criterion for the birth and death of features, and the persistence of a feature, i.e. its significance, is
quantified according to the interval between its appearance and demise. For removal of insignificant 
features \disperse{} augments the persistence measurement with the \emph{topological simplification} of 
the discrete Morse complex \citep{edelsbrunner2002,gyulassy2008}, consisting of an ordered elimination of simplicial singularities and  their connections.

The final product of \disperse{} is a simplicial complex with appropriately adapted gradient lines and corresponding 
ascending manifolds. It provides a map of the morphological structures that make up the weblike 
arrangement of galaxies and mass elements on Megaparsec scales, identified in terms of the ascending 
manifolds of a discrete and topologically filtered Morse complex. Most outstanding is the filamentary network 
corresponding to the index $1$ ascending manifolds.

\subsection{ORIGAMI \\ \hskip 0.85cm (Falck \& Neyrinck)}
\label{section:origami}

The gravitational collapse of dark matter can be thought of as the six-dimensional distortion, in phase-space, of an initially flat three-dimensional manifold. Folds in this manifold occur at caustics and mark out regions of shell-crossing within which the velocity field is multi-valued. \origami\ uses the association between shell-crossing and nonlinear structure formation to identify the different components of the cosmic web, which are fundamentally distinguished by the dimensionality of their collapse: haloes are collapsing along three orthogonal axes, filaments along two, walls along one, and voids are instead expanding~\citep{falck2012}. 

\origami\ determines whether, and in how many dimensions, shell-crossing has occurred for each dark matter particle in the simulation by checking whether particles are out of order with respect to their initial orientation on the Lagrangian grid. For computational efficiency \origami\ currently requires grid instead of `glass' initial conditions; we check for crossings along the Cartesian grid of the simulation box and additionally along three sets of rotated axes~\citep[for details see][]{falck2012}. For each particle we test for crossings with respect to all particles along a given Lagrangian axis, extending up to $1/4$th of the box size in each direction. The number of orthogonal axes along which shell-crossing is detected is counted for each set of axes, and the maximum among all sets of axes is the particle's morphology index: 0, 1, 2, and 3 crossings indicate void, wall, filament, and halo particles respectively.

\origami\ thus provides a cosmic web classification for each dark matter particle in the simulation without using a density parameter or smoothing scale. Since we apply no smoothing or otherwise impose a scale, the cosmic web classification thus depends on the simulation resolution: there is more small-scale structure present in higher resolution simulations, so ORIGAMI identifies a higher fraction of halo particles and lower fraction of void particles as resolution increases~\citep{falck2012,falck2015}.

For the purpose of this comparison project, the cosmic web identification for each particle is converted to classification on a regular grid as follows: for each grid cell, select the morphology having the maximum number of particles as the morphology of that cell. If there are no particles in a cell, that cell may be designated as a void. If there is a tie, with two morphologies having the maximum number of particles, assign the lower morphology to the cell (i.e. void $<$ wall $<$ filament $<$ halo). Ties can be quite common between void and wall particles, especially for low resolution simulations and fine grids.

Because of the particle-based web definition, most halo particles identified by \origami\ correspond to haloes identified by FOF. In previous work~\citep{2014JCAP...07..058F}, the web environment of ORIGAMI haloes is defined according to the morphology of particles that neighbor the haloes, but this does not work for FOF haloes since the neighbor particles of FOF haloes are most often included as part of ORIGAMI haloes. For this comparison project, then, we classify each FOF halo according to the web identification of the grid cell it is in.



\subsection{Multi-Stream Web Analysis (MSWA)\\ \hskip 0.85cm (Shandarin \& Ramachandra)}
\label{section:shandarin}

The growth of CDM density perturbations results in the emergence of regions where the velocity has distinct multiple values.
These regions are also of high densities. The DM web can be viewed as a multi stream field. For example, voids are the regions 
where the velocity field has a single value. This is because no gravitationally bound DM object can form prior to origin of shell crossing, which 
corresponds to the formation of regions with at least three streams. Three-stream flows associated with Zel'dovich's pancakes are
gravitationally bound only in one direction, roughly perpendicular to the pancake. Filaments are bound in two directions orthogonal to the
filament, and haloes are of course fully bound.
The generic geometrical structures of the web, i.e. walls aka pancakes, filaments and haloes, 
cannot be uniquely defined by any particular threshold of the multi stream field, which is also true for the density field.
Generally all three-stream regions belong to the walls, but the transition to the filaments and to the haloes may occur at different levels.

The multi stream field can be easily computed from the final and initial coordinates of the particles in cosmological $N$-body simulations
\citep{ abel2012,shandarin2012}. 
A cold collisionless matter represents an extremely thin three-dimensional sheet called a Lagrangian submanifold in six-dimensional space 
made by  three  initial and three final coordinates at a chosen state of the simulation. Similarly to the three-dimensional sheet in six-dimensional phase space, the
Lagrangian submanifold contains full dynamic information about the system. It needs to be tetrahedralized only once by using initial positions of the particles
on a regular grid as vertices of the tetrahedra. During the following evolution the tessellation remains intact. It always remains continuous, and
its projection in 3D coordinate space fully tiles it at least by one layer in voids and many times in the web regions.
The tetrahedra during the evolution are deformed, but the deformation has no effect on the connectivity assignments  between the particles.
The number of streams can be computed on an arbitrary set  of spatial points by simply counting how many tetrahedra contain a given point. 
The first study of the multi-stream environment of DM haloes has been recently described in \citet{Ramachandara_Shandarin:15}.

Delineating the web components (walls, filaments and haloes) in multi-stream fields is not straightforward, and could be done using various approaches. By studying the scaling of multi-stream variation around dark-matter haloes, \cite{Ramachandara_Shandarin:15} showed that the geometries of structures change from sheets to filaments at a multi-stream value of $n_{\rm str} \gtrsim 17$. The next transition from filaments to knots is seen at around $n_{\rm str} \gtrsim 90$ -- which also roughly corresponds to the virial mass density $\Delta_{\rm vir} = 200$. These thresholds are heuristic -- the analysis may have to be repeated for different simulations for the calibration of thresholds. On the other hand, local Hessian-based geometric methods were recently used to identify multi-stream structures by \cite{Ramachandra2017}. This approach hints towards a portrait of structures in multi-stream fields that are free of ad hoc thresholds. For instance, the haloes could be identified simply as convex surfaces enclosing a local maxima of the multi-stream field (Ramachandra \& Shandarin, in preparation).

A summary of the classification scheme used for this comparison project is as follows: Voids are simply the regions with $n_{str} = 1$. The web components are delineated by utilizing the first approach of calibrating thresholds, i.e, sheets: $3 \leq n_{\rm str} < 17$, filaments: $ 17 \leq n_{\rm str} < 90$ and knots: $n_{\rm str} \geq 90$.

\begin{figure*}
 \includegraphics[width=\textwidth]{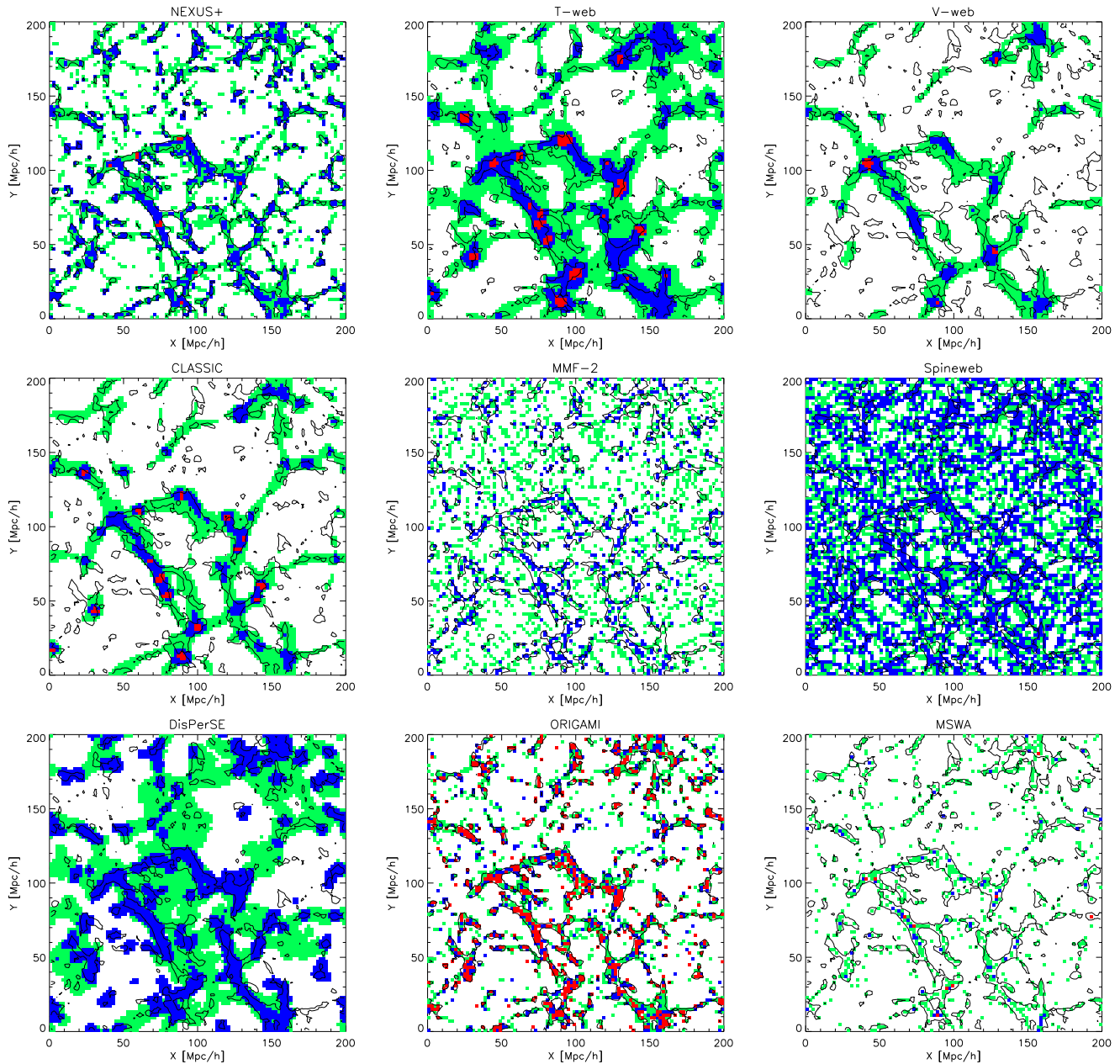}
 \vspace{-.4cm}
 \caption{Visual comparison of environments as detected by the different cosmic web finders. All panels show a thin, $2\hmpc$ thick slice, where the various colours indicate: knots (red), filaments (blue), walls (green) and voids (white). Each panel has a set of solid lines which indicate the $\delta=0$ contours (see the density distribution in Fig.~ \ref{fig:den_slice}). The simulation is purposefully coarse grained with cells of size 2$h^{-1}$Mpc, as it is on this scale that the methods returned a classification.}
 \label{fig:env_slice_a}
\end{figure*}

\begin{figure*}
 \includegraphics[width=\textwidth]{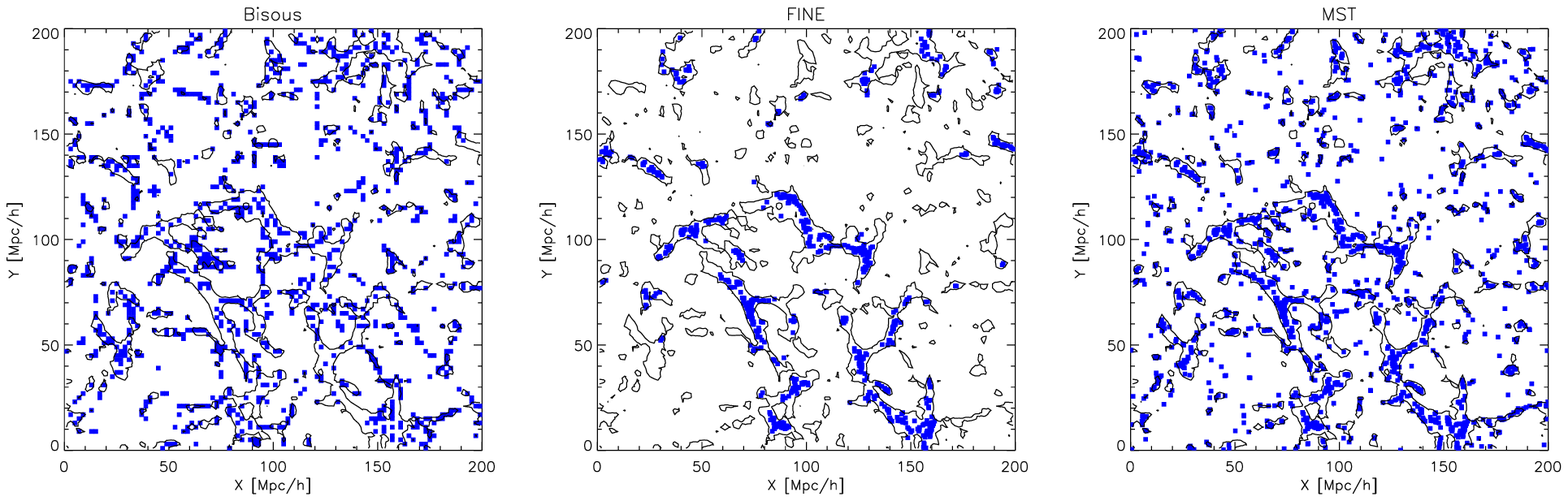}
 \vspace{-.4cm}
 \caption{Same as Fig.~\ref{fig:env_slice_a} but for the three methods that  identify only filaments.}
 \label{fig:env_slice_b}
\vskip -0.5truecm
 \includegraphics[width=\textwidth]{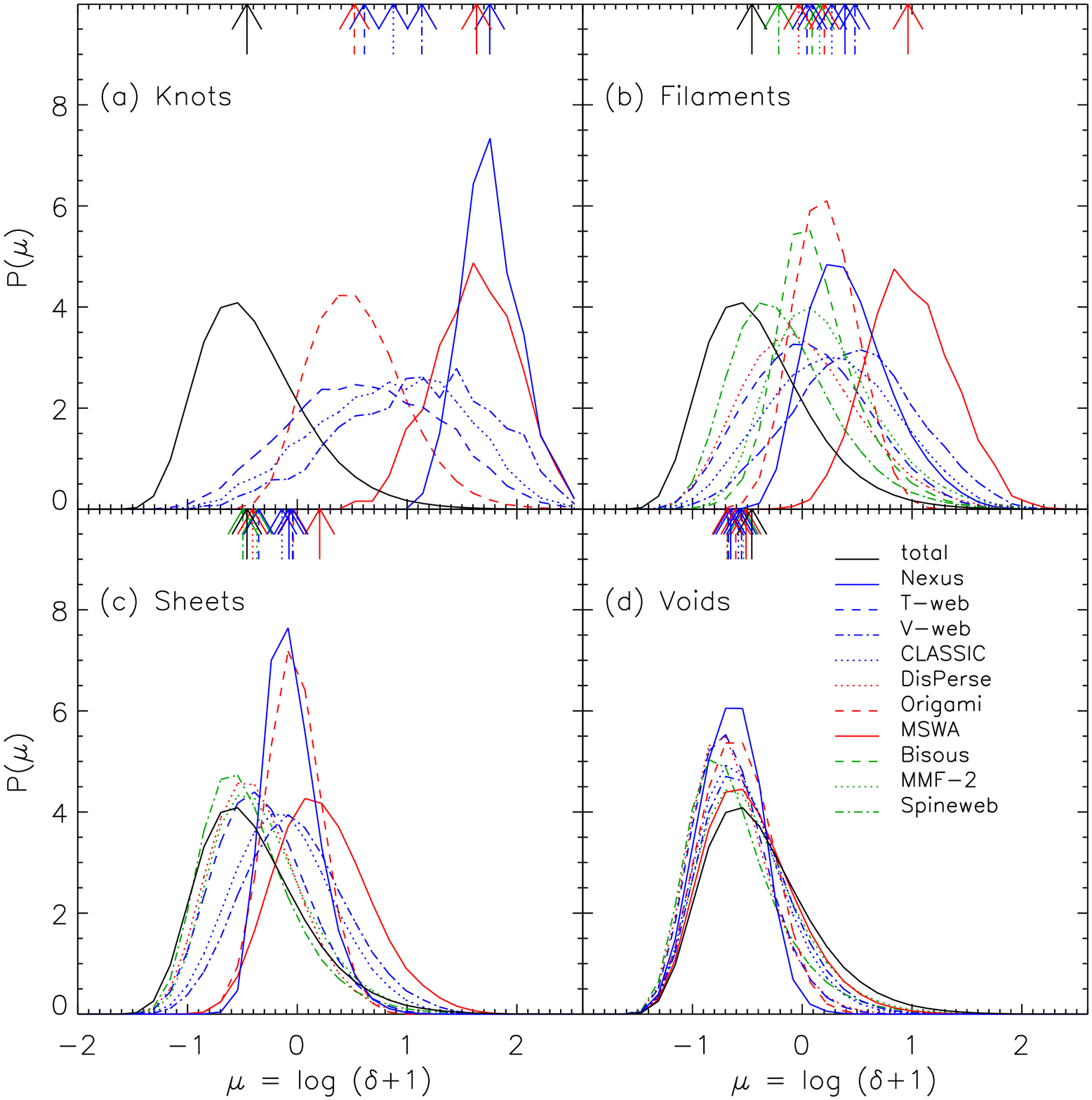}
 \vspace{-1.cm}
 \caption{Comparison of the density contrast, $1+\delta$, PDF as a function of environment for the different cosmic web finders. The panels show the density PDF for: knots (top-left), filaments (top-right), sheets (bottom-left) and voids (bottom-right). The vertical arrows indicate the median of each distribution. Each PDF is normalized to unity and thus does not correspond to the volume filling fraction.}
 \label{fig:den_PDF}
\end{figure*}

\begin{center}
\begin{table*}
\begin{flushright}
\begin{tabular}{|c|cccc|cccc|cccc|}
\hline
&\multicolumn{4}{|c|} {} &\multicolumn{4}{|c|} {}&\multicolumn{4}{|c|} {}\\
&\multicolumn{4}{|c|} {Volume fraction (cells)} &\multicolumn{4}{|c|} {Mass fraction (cells)}&\multicolumn{4}{|c|} {Mass fraction (haloes)}\\
&\multicolumn{4}{|c|} {} &\multicolumn{4}{|c|} {}&\multicolumn{4}{|c|} {}\\
\hline

\hline
Method&Knots&Filaments&Sheets&Voids&Knots&Filaments&Sheets&Voids&Knots&Filaments&Sheets&Voids\\
\hline
 & &&&&&&&&&&&\\
  MST  & -- &-- & -- & -- & --& -- & -- & -- & -- &0.941& -- &  0.023\\
 & &&&&&&&&&&&\\
  Bisous  & -- &0.051 & -- & -- & -- &0.286&-- & -- &--&0.377& -- & --\\
 FINE  & -- &-- & -- & -- & --& -- & -- & -- & -- &0.411& -- &  --\\
 & &&&&&&&&&&&\\
  V-web  &0.003&0.034&0.204&0.755&0.097&0.235&0.331&0.337&0.231&0.317& 0.293 & 0.159\\
  T-web  &0.013&0.149&0.413&0.425&0.166&0.380&0.319&0.135&0.328&0.415& 0.211 &  0.045\\
  CLASSIC &0.006&0.053&0.238&0.703&0.121&0.239&0.324&0.315&0.271&0.276& 0.290& 0.163\\
 & &&&&&&&&&&&\\
  \nexus{}  &0.001&0.113&0.228&0.657&0.084&0.488&0.250&0.178&0.245&0.658& 0.088 &  0.006\\
  MMF-2  & -- &0.078 & 0.190 & 0.732 & --& 0.295 &  0.197 & 0.508 & -- &0.909& 0.072 & 0.019\\
 & &&&&&&&&&&&\\
  Spineweb  & -- &0.361 & 0.307 & 0.332 & -- & 0.600 & 0.235 & 0.165 & -- & 0.971 & 0.027 & 0.001\\
DisPerSE  &--&0.239&0.373&0.388&--&0.621&0.254&0.125&--&0.797& 0.158 &  0.044\\
 & &&&&&&&&&&&\\
 ORIGAMI	  &0.074&0.064&0.123&0.738&0.489&0.131&0.137&0.243&0.898&0.067& 0.024 & 0.010\\
  MSWA  &0.001&0.007&0.088&0.903&0.070&0.106&0.264&0.560&0.641&0.219& 0.130 & 0.009\\
 & &&&&&&&&&&&\\
\hline
\end{tabular}
\vskip 0.5cm
\caption{The fraction of the volume, total mass and mass in haloes (with $M_{\rm halo} > 10^{11}h^{-1}M_{\odot}$) in each web environment for each method. Note that two methods (MST and FINE) identify filaments in the halo (not particle) distribution and do not provide an environment characterization of individual volume elements. MST assigns all haloes not ascribed to a filament as being in voids.}
  \label{tab:mff}
\end{flushright}
\end{table*}
\end{center}

\section{Comparison and Results}
\label{sec:results}

Here we present a visual and quantitative comparison of the different methods. We focus on comparing general features of the cosmic web: mass and volume filling fractions, density distributions and halo mass functions in each environment. As already mentioned, all methods were applied to the same simulation and they all used, depending on the method, either the dark matter particle distribution or the FOF halo catalogue.

\subsection{Visual comparison}
\label{sec:cube}
We begin our analysis by performing a visual comparison of the various web finders. Fig.~\ref{fig:env_slice_a} and Fig.~\ref{fig:env_slice_b} show the environments returned by the web identification methods that took part in the comparison. Each panel shows the same $2\hmpc$ thick slice through the simulation box. Broadly speaking, there are two types of methods: the ones that return multiple cosmic web environments (i.e. voids, sheets, filaments and possibly knots; these are shown in the Fig.~\ref{fig:env_slice_a}) and the ones that identify only filaments (shown in Fig.~\ref{fig:env_slice_b}). Among the first type, \disperse{}, MMF and Spineweb do not identify knots. For the second type of methods, we show either the grid cells identified as filaments (the \bisous{} method) or the positions of the haloes associated to filaments for the methods that did not return a web classification for each volume element (the \fine{} and \mst{} methods). A number of general points are immediately visible from inspection of Figs.~\ref{fig:env_slice_a} and ~\ref{fig:env_slice_b} (in no particular order): 
\begin{itemize}
\item \disperse{} provides no knots, and its filaments are relatively thick compared to the other methods.
\item \mmft{} and \spine{} fill much of the simulation's volume with sheets and filaments.
\item \origami{} ascribes much of the over-dense volume as knots -- owing primarily to the fact that these regions contains haloes which have undergone shell crossing along three orthogonal axes.
\item The Hessian methods (\nexus{}, \tweb{}, \vweb{} and \classic{}) have a mix of knots, filaments and sheets, with voids dominating the under dense volume.
\item The \bisous{} model and \mst{} seem to more or less agree with each other, whereas the \fine{} method ascribes far fewer haloes to filaments.
\end{itemize} 

\begin{figure*}
 \includegraphics[width=\textwidth]{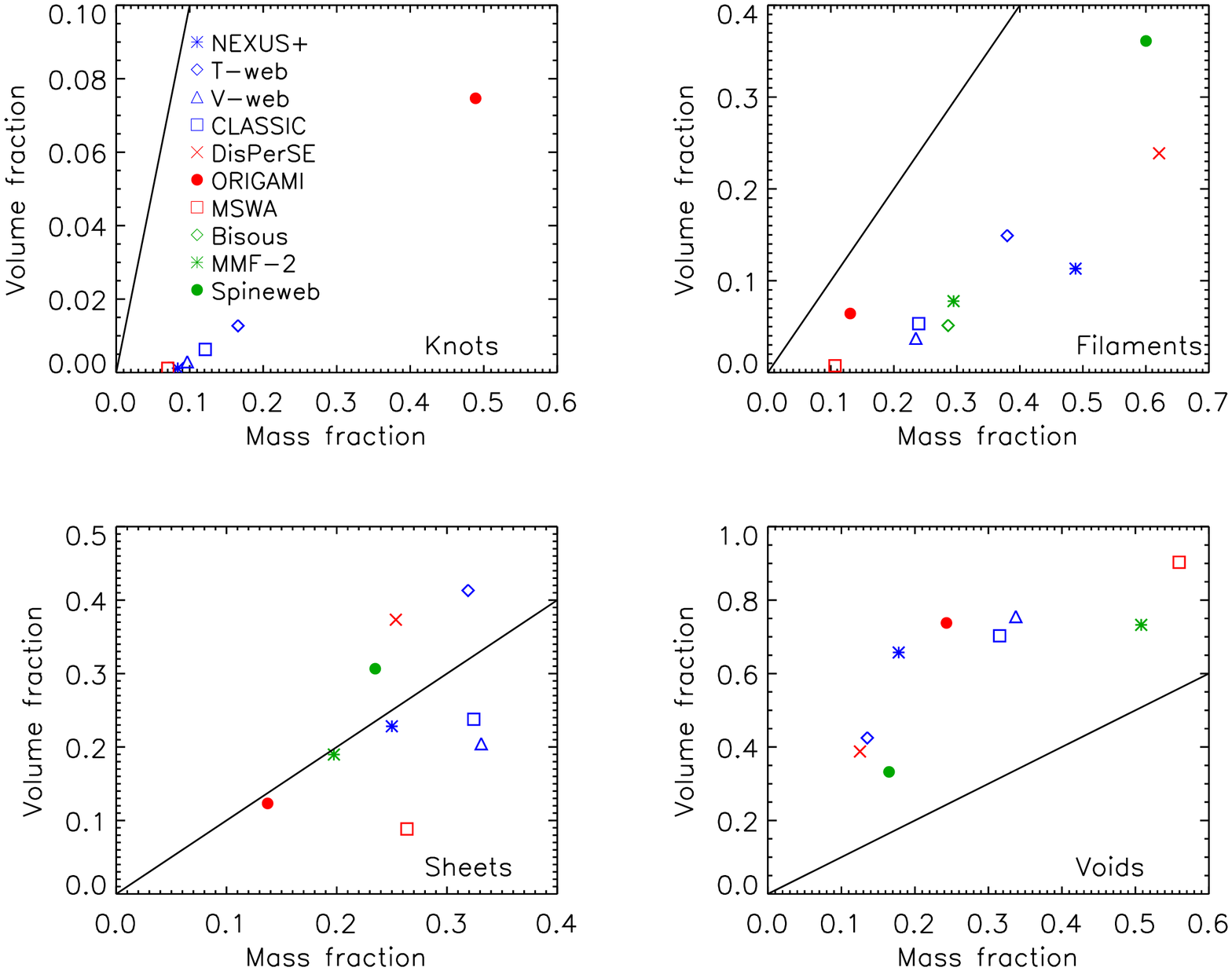}
 \vspace{-.7cm}
 \caption{ The mass and volume filling fraction of knots (top-left), filaments (top-right), sheets (bottom-left) and voids (bottom-right) as identified by the various cosmic web finders. These quantities were computed using a regular grid with a cell spacing of $2\hmpc$. The solid line shows the mean filling fraction, i.e. a slope of unity, where the volume filling fraction equals the mass filling fraction. Namely, points above this line lie in under-densities, points below it in over-densities. }
 \label{fig:vff_mff}
\end{figure*}

It is important to note that some of these methods (specifically \nexus{}, \mmft{}, \tweb{} and \vweb{}) have been designed, to various degrees, to reproduce the visual impression of the cosmic web. Furthermore, given that voids are by definition under-dense regions, it is ideologically unlikely that a given method would be designed to identify clusters deep inside voids\footnote{The measure of a density depends on scale: large enough volumes that include relatively small over-densities can, on average, be well below the mean density and thus considered voids.}.

\subsection{Density PDF}
The relationship between the cosmic web and the density field can be quantified by studying the probability distribution function (PDF) of the density field for each volume element (grid cell) as a function of web environment. This is shown in Fig.~\ref{fig:den_PDF}, where the total density PDF for this simulation (computed on a regular grid with cell spacing of 2$\hmpc$) is shown in black and is the same in all panels; we quantify the density by normalising to the mean density of the universe, $\delta=\rho/\bar{\rho}$. Note that only those methods that assign web classification to volume elements are included here -- the \fine{} and \mst{} methods assign a cosmic web environment only to haloes and are therefore excluded. The median of each PDF is denoted by the corresponding arrow.

\subsubsection{Knots} 
In Fig.~\ref{fig:den_PDF}(a) we show that knots are characterised by a wide variety of environmental densities. Although the \tweb{}, \vweb{} and \classic{} roughly agree, they differ substantially from the fourth Hessian method, \nexus{}, which has a much narrower and higher distribution of densities. Indeed \nexus{} is in closer agreement with \mswa{}. \origami{} peaks at roughly the same density as \vweb{}, although is a little narrower.

Perhaps the strongest conclusion we can draw from Fig.~\ref{fig:den_PDF}(a) is that the local density by itself is a poor proxy for being considered a knot by any given method. Or, conversely, where knots are found, their density may differ by an order of magnitude or more.

\subsubsection{Filaments} 
In Fig.~\ref{fig:den_PDF}(b) we show the PDF of densities for cells identified as filaments. Qualitatively, the picture is similar to that for knots, but pushed to slightly lower densities. There also appears to be a weak convergence of the median density among methods. Namely, although the widths of the PDF are similar, their medians are more strongly in agreement (with the exception of \mswa{}), and span less than an order of magnitude. \mswa{} stands out here in labelling higher density cells as filamentary; the \bisous{} model (the only filament only model that can participate in this test) closely resembles \origami{}, while the PDFs of three of the Hessian methods (\tweb{}, \vweb{} and \classic{}) have similar shapes but are offset with respect to each other. The PDF of \spine{} peaks at the lowest density.

\subsubsection{Sheets}
The density PDF for cells labelled as sheets, shown in Fig.~\ref{fig:den_PDF}(c), displays more coherence  than those of knots or filaments. Despite the PDFs still varying widely among the web finders, the median densities of the PDFs are roughly similar and take values around $\delta=0$.
The median of the set of density PDFs moves to lower values, although, like the PDFs for knots and filaments, there is still a wide variety of permissible environments. Three pairs of methods produce nearly identical PDFs: \nexus{} and \origami{}, \disperse{}, \tweb{}, and \mmft{}, and \vweb{} and \classic{}. Again, the PDF of \spine{} peaks at the lowest density.

\subsubsection{Voids}
The best agreement between methods is found in regions denoted as voids, as shown in Fig.~\ref{fig:den_PDF}(d). The void density PDFs show less diversity and generally have the same shape. The spread in medians is small: less than 0.2 dex. As voids purport to be the most under-dense regions in the universe, they also make up the greatest fraction of the simulation's volume (as can be inferred by the overlap between the void density PDF and the total density PDF). It can thus be said that the methods studied here all agree that the majority of the simulation volume is indeed categorised as void.

\subsubsection{Trends in the density PDFs}
The cosmic web classification is layered: knots are embedded in filaments, which, in turn, reside in sheets, which, in turn constitute the boundaries between different void basins. As our analysis of the cosmic web moves from knots to voids, the median of the density distribution of each method and for each web type moves to lower values {\it in tandem.} Although for a given web type there may be a wide variety of permissible density environments across the analysed methods, each method follows a similar trend. The peak of the density PDF moves to lower and lower densities, with most methods converging in the lowest density and most abundant environment in the simulation: voids.

\subsection{Mass and volume fraction}
We continue the cosmic web finder's comparison with a study of the volume and mass filling fractions that are ascribed to a specific cosmic web type. These quantities are shown in Fig.~\ref{fig:vff_mff} for knots, filaments, sheets and voids. The mass fraction is found by summing up the particles in all the cells with the same cosmic web type and dividing by the total number of particles in the simulation. The volume fraction is found by counting all the cells with the same cosmic web type and dividing by the total number of cells. Note that for these tests we have a 100$^3$ grid with (2$\hmpc)^3$ cells.
 
\begin{figure*}
 \includegraphics[width=\textwidth]{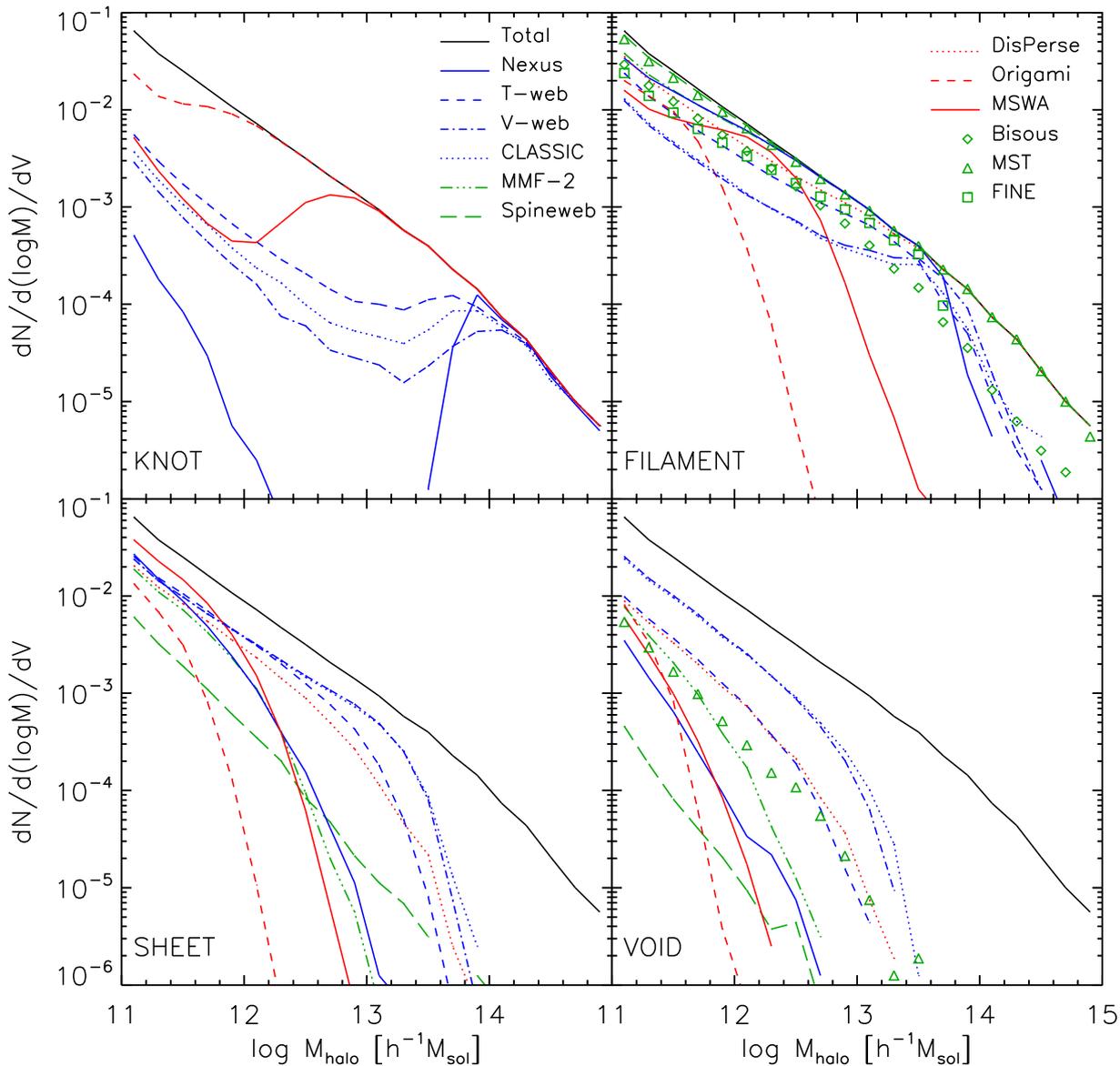}
 \vspace{-1.cm}
 \caption{ Comparison of the halo mass function as a function of environment for the various cosmic web finders. The panels show the mass function for: knots (top-left), filaments (top-right), sheets (bottom-left) and voids (bottom-right). The black solid line shows the total halo mass function. }
 \label{fig:HMF}
\end{figure*}

\begin{figure*}
 \includegraphics[width=\textwidth]{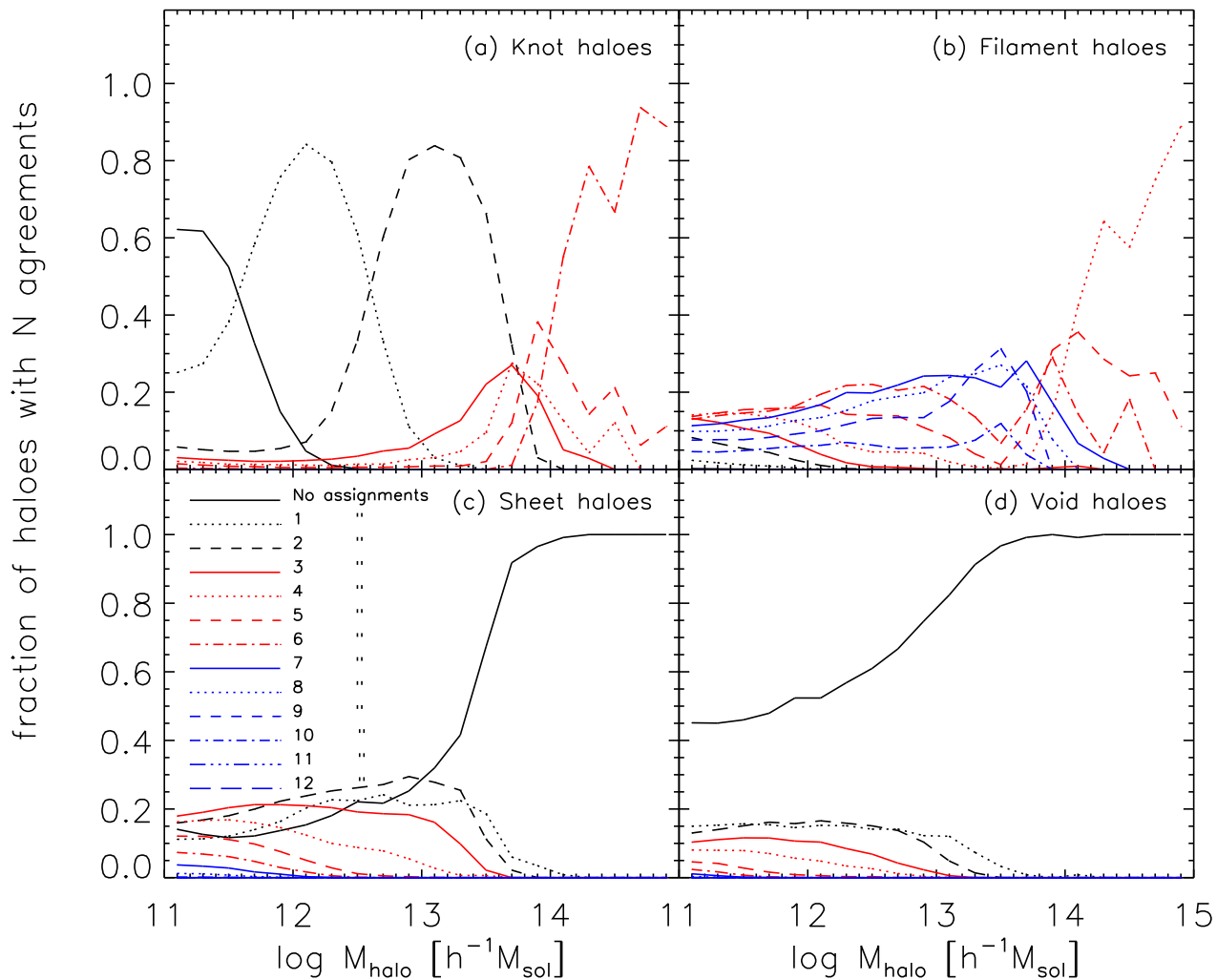}
 \vspace{-.4cm}
 \caption{The level of agreement, on a halo by halo basis, in assigning a web classification to a given halo. For each halo in a given mass bin, we ask how many methods have assigned it the same web type. We plot the fraction of these haloes as a function of halo mass for the four web environments: knots (top-left), filaments (top-right), sheets (bottom-left) and voids (bottom-right). Each line shows the fraction of haloes at fixed halo mass that were assigned by N methods to that environment, with N from 0 (no assignments) to 10 (all methods agree). Note that not all methods identify all web types, so that the maximum number of agreements varies with web type: 6 for knots, 12 for filaments, 9 for sheets and 10 voids.}
  \label{fig:halo_agree}
\end{figure*}

\begin{figure*}
 \includegraphics[width=\textwidth]{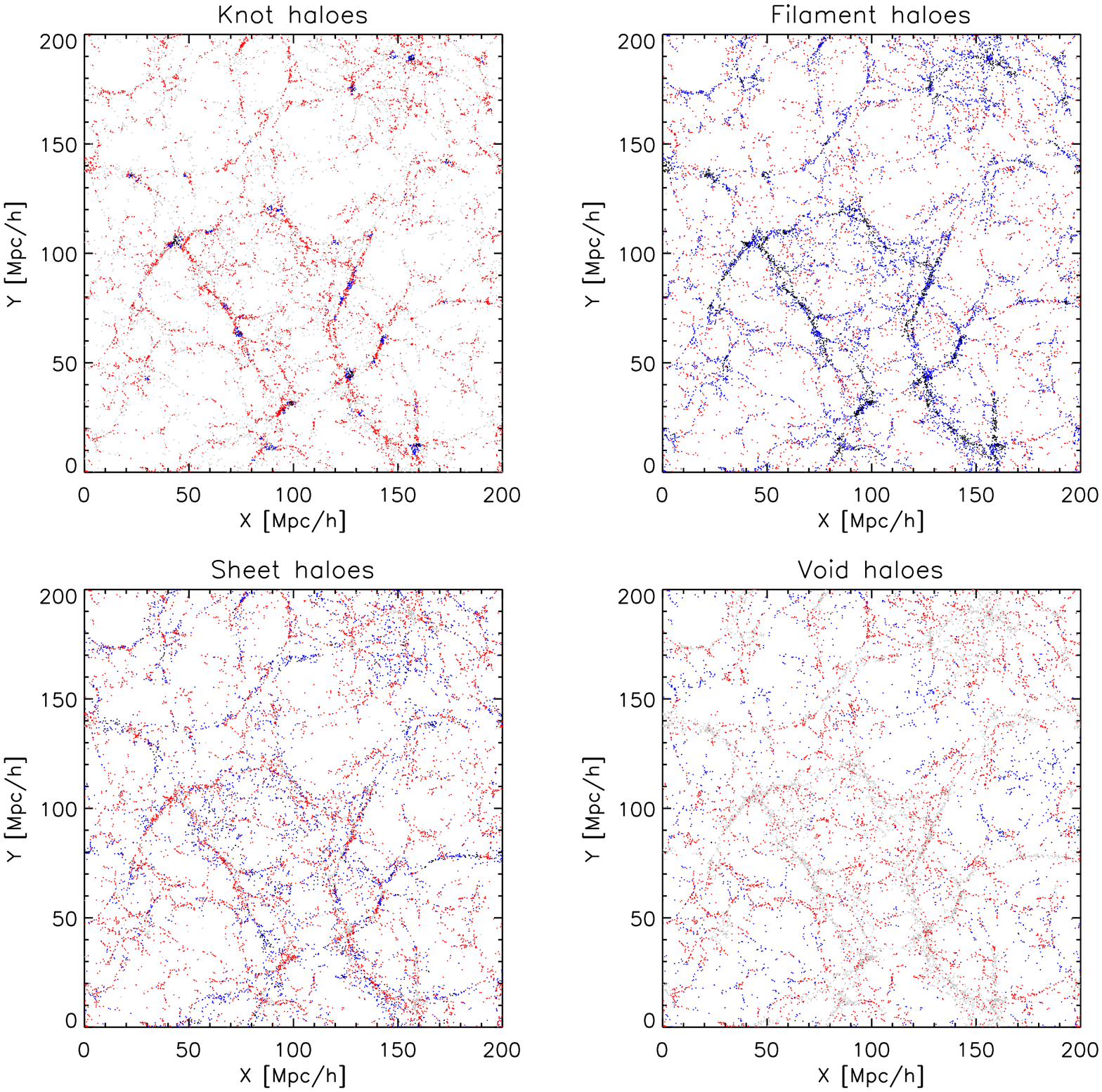}
 \vspace{-.4cm}
 \caption{A visualisation of the agreement across methods regarding a specific halo's classification shown in the same 10$h^{-1}$Mpc thick slice as in Figs.~\ref{fig:den_slice}--~\ref{fig:env_slice_b}. Not all web types are identified by each method, so for each panel the various colours indicate a different number of agreements. In the ``Knot haloes'' panel, a halo is plotted in black if 5--6 methods agree, blue if 3--4 methods agree and red if 1--2 methods agree that a halo is in a knot. In the ``Filament haloes'' panel colours represent: black if 9--12 methods agree, blue if 4--8 methods agree and red if 1--3 methods agree that a halo is in a filament. For ``Sheet haloes'' the colours represent: black if 7--9 methods agree, blue if 4--6 methods agree and red if 1--3 methods agree that a halo is in a sheet. For the ``Void haloes'' the colours represent: black if 7--10 methods agree, blue if 4--6 methods agree and red if 1--3 methods agree that a halo is in a void. In all panels, haloes not assigned that web classification are shown in grey. }
  \label{fig:halo_agree_dot}
\end{figure*}

\begin{itemize} 
\item {\bf Knots:} as \origami{} makes no distinction between knots and haloes, it is perhaps unsurprising that this method finds that nearly half the simulation's mass is confined in $\sim7\%$ of the volume. 
Most other methods tag far fewer cells as knots, claiming they constitute below $\sim1\%$ of the volume with between 10\%-20\% of the mass. Interestingly the mass-volume fraction relation for knots follows a fairly tight linear proportionality -- the more mass found in knots, the more volume, regardless of method used. 
\item {\bf Filaments:} A similar, but slightly weaker proportionality between mass and volume fraction is found for filament regions. Here, \spine{} and \disperse{} place roughly 60\% of the simulations mass in filaments which occupy some $\sim35\%$ and $\sim25\%$ of the simulation volume, respectively. Unlike knots, there is considerably more spread in the relationship between mass and volume fractions amongst the methods, although a linear relationship is still discernible to the eye. Similar to knots, \mswa{} continues to place virtually none of the volume and roughly $\sim10\%$ of the simulation's mass in filaments. The \bisous{} model -- the only one of the filament-only models that can participate in this comparison -- finds very similar filament volume and mass filling fractions as \classic{},  \vweb{} and \mmft{}, with some $\sim5\%$ of the simulations volume and $\sim30\%$ of the simulations mass labelled as filaments. To summarize, the filament volume filling fraction spans from virtually nothing (\mswa{}) to more than a third of the volume (\spine{}); while the filament mass filling fraction spans roughly double that range, from $\sim10\%$ (\mswa{}) to $\sim60\%$ (\disperse{}).
\item {\bf Sheets:} The spread of the sheet mass filling fraction is quite tight, with most methods assigning $\sim30\%\pm5\%$ of the total mass to sheets (with the exception of \origami{} and \mmft{}, which find lower values). However, the sheet volume filling fractions vary substantially between methods, ranging from less than $\sim10\%$ for MWSA to more than 40\% for T-web. As in knots and filaments, MWSA continues to assign only a small volume fraction to sheets.
\item {\bf Voids:} The volume fraction associated to voids shows three distinct groups: three methods with $\sim 40\%$ (\disperse{}, \spine{} and \tweb{}), five with $\sim 70\%$ (\nexus{}, \origami{}, \vweb{}, \mmft{} and \classic{}), and one with $\sim90\%$ (\mswa{}).  For the first group of finders (\disperse{}, \spine{} and \tweb{}), the mass fraction is more or less the same at around $10-15\%$. For the second group (\nexus{}, \origami{}, \vweb{}, \mmft{} and \classic{}), the mass fraction in voids spans a large range from $\sim 15\%$ for \nexus{} to $\gsim50\%$ for \mmft{}. In general it is apparent that the mass fraction assigned to void regions spans a large range. It is interesting to note how the void mass filling fractions of these methods have flipped compared to their estimate for the filament mass fraction.
\end{itemize}

In summary, the various methods predict fairly large ranges for the volume and mass fractions assigned to a given web type. Given the substantial differences in how these methods identify the web components, it is not very surprising that there are large discrepancies in these fractions. That said, in each plot of Fig.~\ref{fig:vff_mff}, clusters of methods can be identified which have similar values of either the volume fraction, mass fraction or both. The values for the mass and volume fractions are shown in the first two columns of Table~\ref{tab:mff}.

\subsection{Halo assignment and mass functions}
We now compare how the web environment assigned to haloes varies across cosmic web finders.  For most methods, each halo is assigned the cosmic web environment of the cell in which its centre is located in. For the filament only methods (\bisous{}, \fine{}, and \mst{}), the methods themselves directly identify which haloes are part of a filamentary structure. 

In Fig.~\ref{fig:HMF} we show the halo mass function for the entire halo sample and for each web type. We find a mixed picture, with substantial variations in the halo mass function of web types. Despite this, there are also agreements. For example, all the methods place the most massive haloes (i.e. $M \gsim 10^{14}M_{\odot}$) into knots. Similar trends are visible in how the filament halo mass function behaves -- the mass functions are similarly valued at low masses and show a ``knee'' that precipitates a quick decline in the mass function. The agreement of mass functions in filaments is strongest (except the phase-space methods, \origami{} and \mswa{}), and the shape of the halo mass function in filaments is the closest one to the total halo mass function. Indeed, \mmft{} and \spine{} place nearly all haloes in filaments: the green dashed and dot-dashed curves are only visibly separate from the black line below $\sim10^{12.5}h^{-1}M_{\odot}$.

%
%

The last column of Table~\ref{tab:mff} shows how much mass is locked up in haloes of a given web type for each web finder. The four Hessian methods (\nexus{}, \tweb{}, \vweb{} and \classic{}) agree that around 20-30\% of mass in haloes is found in knot haloes. This is dramatically different from \disperse{}, \mmft{} and \spine{} which do not identify haloes as belonging to knots, and the phase-space methods (\origami{} and \mswa{}), which place the bulk of halo mass in knots. It is interesting to note that the methods that do not identify knots, but do identify filaments, sheets and voids (\spine{}, \mmft{}, and \disperse{}) place the overwhelming bulk of halo mass in filaments (with $\gsim 80\%$ of all halo mass in filaments). All methods also agree that haloes in voids have the least amount of total halo mass, although they disagree on exactly how much this is, with methods predicting either $\sim 15\%$  (\vweb{}, \classic{}), $\sim 5\%$ (\tweb{}, \disperse{}) or $\lsim 1\%$ (\nexus{}, \origami{}, \mswa{}, \spine, \mmft) of halo mass in voids.

It is important to compare the environment tag associated to haloes on a halo per halo basis too, not only globally as is the case when comparing halo mass functions. To accomplish this, we ask the following question: for haloes in a given mass range, how many methods agree that some fraction of these have the same cosmic web environment? The answer to this question is shown in Fig.~\ref{fig:halo_agree}.

To better understand our analysis, lets consider the panel of Fig.~\ref{fig:halo_agree}(a), which gives the agreement across methods for individual knot haloes. For high halo masses, $M\gsim10^{14}h^{-1}M_{\odot}$, the panel shows that most such haloes ($\sim90\%$) are assigned to knots by all the six methods (namely: \nexus{}, \tweb{}, \vweb{}, \classic{}, \origami{}, and \mswa{}) that identify knot environments (dot-dashed red curve). Conversely, 60\% of the smallest haloes (with $M\sim10^{11}h^{-1}M_{\odot}$) are not assigned by any method to the knot environment (solid black curve). In between the two extreme masses, we find two bell-like curves where haloes with $M\sim10^{12}h^{-1}M_{\odot}$ are assigned to knots by only one method (black dotted curve), and haloes with $M\sim10^{13}h^{-1}M_{\odot}$ (black dashed curve) are assigned to knots by the two phase-space methods, \mswa{} and \origami{}.

In Fig.~\ref{fig:halo_agree}(b), we show the agreement among filament haloes. Note the two peaks in the blue dashed and blue dot-dashed lines at $M_{\rm halo}\approx 10^{13.5}$: nine methods agree that $\sim30\%$ of haloes of this mass are in filaments while 10 methods agree that at least 10\% of halos of this mass are in filaments. Here, four methods (\disperse{}, \spine{}, \mmft{} and \mst{}) place the most massive haloes in filaments. Fig.~\ref{fig:halo_agree}(c) and Fig.~\ref{fig:halo_agree}(d) indicate that no method puts the most massive haloes in sheets or voids. Specifically this means that no haloes with $M\gsim10^{14}h^{-1}M_{\odot}$ are found in sheets, and no haloes with $M\gsim10^{13.5}h^{-1}M_{\odot}$ are found in voids, by any method.

The degree of agreement of web classifiers on a halo per halo basis varies accordingly to the spatial distribution of haloes, as we illustrate in Fig.~\ref{fig:halo_agree_dot}. Here, each halo is coloured by how many methods agree on its given classification. Because the number of methods capable of assigning haloes to a given web type changes (e.g filament only finders can't identify knot haloes, etc) the colour scheme is not identical in each panel (see caption for exact colour explanation). In general if many of the capable methods agree on a specific halo's classification the halo is shown in black; if around half of the capable methods agree, the halo is plotted in blue. If a small number of capable methods agree, the halo is plotted in red. If no method assigns a halo a given classification, the halo is plotted in grey.

Fig.~\ref{fig:halo_agree_dot} ``Knot'' and ``Filament'' halo panels shows quite clearly that the haloes where the most methods agree belong to a biased set and are not simply random. Knot haloes find the most agreement in the densest areas of the simulation -- a reassuring result. Similarly, those haloes which by eye appear to define the filamentary network too have the most agreements. Accordingly none of the haloes in either the densest parts of the simulation or in the filaments are assigned as void haloes (appearing as grey points). Sheets appear, as often is the case, as tenuous structures. Fig.~\ref{fig:halo_agree_dot} indicates that most or many methods are likely to agree on a specific halo's classification based on its location.

\section{Summary and Conclusions}
Large galaxy redshift surveys (e.g. 2dFGRS, SDSS, 2MRS) reveal that at Megaparsec scales
the Universe has a salient weblike structure. On these scales, the galaxies and  the matter distribution in the
universe has arranged itself into a complex weblike network of dense, interconnected knots,
elongated filaments, two-dimensional sheets, and  large nearly-empty voids. These cosmic environments characterise the universe on the largest scales. 
One important aspect of the cosmic web is its multi-scale character, manifesting itself in the existence of weblike
structures over a sizeable range of scale. High-resolution simulations have revealed
that such structures can be found down to very small scales, even down to the virial radius
of haloes, and that they play a prominent role in the accretion of cold gas onto young and
assembling protogalaxies in the early Universe \citep{2012MNRAS.422.1732D}. It ties in with the results of a range of recent studies that have analysed the role of environment on the formation and the evolution of galaxies \citep[e.g.][]{2013ApJ...776...71C,2015MNRAS.448.3665E,2015ApJ...800..112G,2015ApJ...800L...4C,2016MNRAS.455..127M,2017A&A...597A..86P}. Furthermore, theoretical studies have suggested that around half of the warm gas in the Universe is hiding in a ``warm-hot- intergalactic medium'', presumably in the filaments of the cosmic web \citep[e.g.][]{2015Natur.528..105E}. It has therefore become of key importance to gain more insight into the structure and dynamics of the weblike universe, and into the interaction of the cosmic web with galaxy scale processes.

The cosmic web is one of the most intriguing and striking patterns found in nature, rendering
its analysis and characterization far from trivial. This is evidenced by the many elaborate
descriptions that have been developed. The absence of an objective and quantitative procedure for identifying and isolating knots, filaments, sheets and voids in the cosmic matter distribution has been a major obstacle in investigating the structure and dynamics of the cosmic web. The overwhelming complexity of the individual structures and their connectivity, the huge range of densities and the intrinsic multi-scale nature prevent the use of simple tools. Over the past years, we have seen the introduction and proliferation of many new approaches and techniques. These methods are very varied in how they identify the cosmic web environments; being designed with different cosmological data in mind and to answer different questions. These issues are compounded since the techniques available to theorists and simulators differ substantially from those employed by observers. This makes it even more important to understand how the various web identification methods compare with each other.

The main driver of this paper is to quantify in a systematic way both the similarities and differences between cosmic web finders. There is no well motivated common framework to objectively define the constituents of the cosmic web, so there is no way of judging which methods are successful or which ones are  - in some objective way - ``better''. As such, the goal is to compare the output of the various methods to better relate studies that make use of different web identification methods. We proceeded by comparing several basic properties of the cosmic web: the mass and volume filling fraction of each component, the density distribution and the halo mass function in each environment, and a halo by halo comparison of their environment tag. For this, we asked the authors of each method to apply their technique to the same data, the output of an N-body simulation, and to return the resulting web classification in a common format.


We find a substantial diversity in the properties of the cosmic web across the various methods. This is to be expected given the challenges inherent in identifying the cosmic web and the multitude of approaches undertaken in doing so. In spite of this, we also find many similarities across the methods. Some of the most important agreements are:
\begin{itemize}
	\item Voids correspond to the most underdense regions and are consistently identified as such by all the methods. The voids occupy the largest volume fraction, with the majority of methods finding a ${\sim}70\%$ volume filling fraction.
	\item Most methods, except \origami{} and \tweb{}, find that knots contain ${\sim}10\%$ of the total mass in less than $1\%$ of the volume of the universe.
	\item All the methods find that the density PDF systematically shifts towards lower densities as we go from knots to filaments, than to sheets and voids. Despite this trend, there is still a substantial overlap between the density PDF of different environments, which suggests that a simple density is inadequate for cosmic web identification. 
	\item Most massive haloes, $M\gsim10^{14}h^{-1}M_{\odot}$, are classified as residing in knot environments by all the methods that identify knots.
	\item The voids are only sparsely populated with haloes and they lack completely massive haloes with $M\gsim10^{13.5}h^{-1}M_{\odot}$.
\end{itemize}

We have a very incomplete knowledge of what is the effect of environment on galaxy formation and evolution or of what is the cosmological information encoded in the cosmic web pattern. The lack of knowledge is a result of the limitations of analytical approaches in modelling these non-linear processes. Each web finder captures different aspects of this very complex pattern, i.e. the cosmic web, so     it is a worthwhile pursuit to analyze the connection between the environments identified by each method and the effect on galaxies and cosmological constraints.

\section*{Acknowledgments}

The authors thank the Lorentz Center in Leiden for the organizational and financial support of this meeting, which ultimately led to this 
paper. In addition, we are grateful to NOVA and NWO for the financial support to facilitate the workshop. The authors would like to thank Adi Nusser, Christoph Pichon and Dmitri Pogosyan for useful discussions. NIL acknowledges and thanks Jenny Sorce for useful conversations. RvdW and SS, and (with a separate grant) MN and MAC, acknowledge support from the New Frontiers of Astronomy and Cosmology program at the Sir John Templeton Foundation. MC is supported by ERC Advanced Investigator grant COSMIWAY (grant number GA 267291) and the UK STFC grant ST/L00075X/1. MN was supported by the UK Science and Technology Facilities Council (ST/L00075X/1). BF acknowledges financial support from the Research 
Council of Norway (Programme for Space Research). ET acknowledges the support by the ETAg grants IUT26-2, IUT40-2, and by the European 
Regional Development Fund (TK133). FK and SEN thank Arman Khalatyan for assistance. 
AK is supported by the {\it Ministerio de Econom\'ia y Competitividad} and the {\it Fondo Europeo de Desarrollo Regional} (MINECO/FEDER, UE) in Spain through grant AYA2015-63810-P as well as the Consolider-Ingenio 2010 Programme of the {\it Spanish Ministerio de Ciencia e Innovaci\'on} (MICINN) under grant MultiDark CSD2009-00064. He also acknowledges support from the {\it Australian Research Council} (ARC) grant DP140100198. He further thanks William Fitzsimmons for fortune. SEN acknowledges support by the Deutsche Forschungsgemeinschaft under grant NU 332/2-1. GY acknowledge  financial support from MINECO/FEDER under research grant AYA2015-63810-P.

\bibliography{Allrefs}

\begin{thebibliography}{}
\makeatletter
\relax
\def\mn@urlcharsother{\let\do\@makeother \do\$\do\&\do\#\do\^\do\_\do\%\do\~}
\def\mn@doi{\begingroup\mn@urlcharsother \@ifnextchar [ {\mn@doi@}
  {\mn@doi@[]}}
\def\mn@doi@[#1]#2{\def\@tempa{#1}\ifx\@tempa\@empty \href
  {http://dx.doi.org/#2} {doi:#2}\else \href {http://dx.doi.org/#2} {#1}\fi
  \endgroup}
\def\mn@eprint#1#2{\mn@eprint@#1:#2::\@nil}
\def\mn@eprint@arXiv#1{\href {http://arxiv.org/abs/#1} {{\tt arXiv:#1}}}
\def\mn@eprint@dblp#1{\href {http://dblp.uni-trier.de/rec/bibtex/#1.xml}
  {dblp:#1}}
\def\mn@eprint@#1:#2:#3:#4\@nil{\def\@tempa {#1}\def\@tempb {#2}\def\@tempc
  {#3}\ifx \@tempc \@empty \let \@tempc \@tempb \let \@tempb \@tempa \fi \ifx
  \@tempb \@empty \def\@tempb {arXiv}\fi \@ifundefined
  {mn@eprint@\@tempb}{\@tempb:\@tempc}{\expandafter \expandafter \csname
  mn@eprint@\@tempb\endcsname \expandafter{\@tempc}}}

\bibitem[\protect\citeauthoryear{{Abel}, {Hahn}  \& {Kaehler}}{{Abel}
  et~al.}{2012}]{abel2012}
{Abel} T.,  {Hahn} O.,   {Kaehler} R.,  2012, \mn@doi [MNRAS]
  {10.1111/j.1365-2966.2012.21754.x}, \href
  {http://adsabs.harvard.edu/abs/2012MNRAS.427...61A} {427, 61}

\bibitem[\protect\citeauthoryear{{Adler}}{{Adler}}{1981}]{adler1981}
{Adler} R.~J.,  1981, {The Geometry of Random Fields}

\bibitem[\protect\citeauthoryear{{Alpaslan} et~al.,}{{Alpaslan}
  et~al.}{2014a}]{2014MNRAS.438..177A}
{Alpaslan} M.,  et~al., 2014a, \mn@doi [\mnras] {10.1093/mnras/stt2136}, \href
  {http://adsabs.harvard.edu/abs/2014MNRAS.438..177A} {438, 177}

\bibitem[\protect\citeauthoryear{{Alpaslan} et~al.,}{{Alpaslan}
  et~al.}{2014b}]{2014MNRAS.440L.106A}
{Alpaslan} M.,  et~al., 2014b, \mn@doi [\mnras] {10.1093/mnrasl/slu019}, \href
  {http://adsabs.harvard.edu/abs/2014MNRAS.440L.106A} {440, L106}

\bibitem[\protect\citeauthoryear{{Alpaslan} et~al.,}{{Alpaslan}
  et~al.}{2015}]{2015MNRAS.451.3249A}
{Alpaslan} M.,  et~al., 2015, \mn@doi [\mnras] {10.1093/mnras/stv1176}, \href
  {http://adsabs.harvard.edu/abs/2015MNRAS.451.3249A} {451, 3249}

\bibitem[\protect\citeauthoryear{{Alpaslan} et~al.,}{{Alpaslan}
  et~al.}{2016}]{2016MNRAS.457.2287A}
{Alpaslan} M.,  et~al., 2016, \mn@doi [\mnras] {10.1093/mnras/stw134}, \href
  {http://adsabs.harvard.edu/abs/2016MNRAS.457.2287A} {457, 2287}

\bibitem[\protect\citeauthoryear{{Aragon-Calvo} \& {Szalay}}{{Aragon-Calvo} \&
  {Szalay}}{2013}]{aragon2013}
{Aragon-Calvo} M.~A.,  {Szalay} A.~S.,  2013, \mn@doi [\mnras]
  {10.1093/mnras/sts281}, \href
  {http://adsabs.harvard.edu/abs/2013MNRAS.428.3409A} {428, 3409}

\bibitem[\protect\citeauthoryear{{Arag{\'o}n-Calvo} \&
  {Yang}}{{Arag{\'o}n-Calvo} \& {Yang}}{2014}]{aragon2014}
{Arag{\'o}n-Calvo} M.~A.,  {Yang} L.~F.,  2014, \mn@doi [\mnras]
  {10.1093/mnrasl/slu009}, \href
  {http://adsabs.harvard.edu/abs/2014MNRAS.440L..46A} {440, L46}

\bibitem[\protect\citeauthoryear{{Arag{\'o}n-Calvo}, {Jones}, {van de Weygaert}
   \& {van der Hulst}}{{Arag{\'o}n-Calvo} et~al.}{2007a}]{aragon2007}
{Arag{\'o}n-Calvo} M.~A.,  {Jones} B.~J.~T.,  {van de Weygaert} R.,   {van der
  Hulst} J.~M.,  2007a, \mn@doi [\aap] {10.1051/0004-6361:20077880}, \href
  {http://adsabs.harvard.edu/abs/2007A%26A...474..315A} {474, 315}

\bibitem[\protect\citeauthoryear{{Arag{\'o}n-Calvo}, {van de Weygaert}, {Jones}
   \& {van der Hulst}}{{Arag{\'o}n-Calvo} et~al.}{2007b}]{aragon2007a}
{Arag{\'o}n-Calvo} M.~A.,  {van de Weygaert} R.,  {Jones} B.~J.~T.,   {van der
  Hulst} J.~M.,  2007b, \mn@doi [\apjl] {10.1086/511633}, \href
  {http://adsabs.harvard.edu/abs/2007ApJ...655L...5A} {655, L5}

\bibitem[\protect\citeauthoryear{{Arag{\'o}n-Calvo}, {van de Weygaert},
  {Araya-Melo}, {Platen}  \& {Szalay}}{{Arag{\'o}n-Calvo}
  et~al.}{2010a}]{aragon2010b}
{Arag{\'o}n-Calvo} M.~A.,  {van de Weygaert} R.,  {Araya-Melo} P.~A.,  {Platen}
  E.,   {Szalay} A.~S.,  2010a, \mn@doi [\mnras]
  {10.1111/j.1745-3933.2010.00841.x}, \href
  {http://adsabs.harvard.edu/abs/2010MNRAS.404L..89A} {404, L89}

\bibitem[\protect\citeauthoryear{{Arag{\'o}n-Calvo}, {van de Weygaert}  \&
  {Jones}}{{Arag{\'o}n-Calvo} et~al.}{2010b}]{aragon2010}
{Arag{\'o}n-Calvo} M.~A.,  {van de Weygaert} R.,   {Jones} B.~J.~T.,  2010b,
  \mn@doi [\mnras] {10.1111/j.1365-2966.2010.17263.x}, \href
  {http://adsabs.harvard.edu/abs/2010MNRAS.408.2163A} {408, 2163}

\bibitem[\protect\citeauthoryear{{Arag{\'o}n-Calvo}, {Platen}, {van de
  Weygaert}  \& {Szalay}}{{Arag{\'o}n-Calvo} et~al.}{2010c}]{aragon2010a}
{Arag{\'o}n-Calvo} M.~A.,  {Platen} E.,  {van de Weygaert} R.,   {Szalay}
  A.~S.,  2010c, \mn@doi [\apj] {10.1088/0004-637X/723/1/364}, \href
  {http://adsabs.harvard.edu/abs/2010ApJ...723..364A} {723, 364}

\bibitem[\protect\citeauthoryear{{Arag{\'o}n-Calvo}, {Neyrinck}  \&
  {Silk}}{{Arag{\'o}n-Calvo} et~al.}{2016}]{aragon2016}
{Arag{\'o}n-Calvo} M.~A.,  {Neyrinck} M.~C.,   {Silk} J.,  2016, preprint,
  \href {http://adsabs.harvard.edu/abs/2016arXiv160707881A} {} (\mn@eprint
  {arXiv} {1607.07881})

\bibitem[\protect\citeauthoryear{Bardeen, Bond, Kaiser  \& Szalay}{Bardeen
  et~al.}{1986}]{bbks}
Bardeen J.~M.,  Bond J.~R.,  Kaiser N.,   Szalay A.~S.,  1986, \apj, 304, 15

\bibitem[\protect\citeauthoryear{{Barrow}, {Bhavsar}  \& {Sonoda}}{{Barrow}
  et~al.}{1985}]{barrow1985}
{Barrow} J.,  {Bhavsar} S.,   {Sonoda} D.,  1985, \mnras, 216, 17

\bibitem[\protect\citeauthoryear{{Bharadwaj}, {Bhavsar}  \&
  {Sheth}}{{Bharadwaj} et~al.}{2004}]{2004ApJ...606...25B}
{Bharadwaj} S.,  {Bhavsar} S.~P.,   {Sheth} J.~V.,  2004, \mn@doi [\apj]
  {10.1086/382140}, \href {http://adsabs.harvard.edu/abs/2004ApJ...606...25B}
  {606, 25}

\bibitem[\protect\citeauthoryear{{Bond} \& {Myers}}{{Bond} \&
  {Myers}}{1996}]{bondmyers1996}
{Bond} J.,  {Myers} S.,  1996, \apjs, 103, 1

\bibitem[\protect\citeauthoryear{{Bond}, {Kofman}  \& {Pogosyan}}{{Bond}
  et~al.}{1996}]{bondweb1996}
{Bond} J.~R.,  {Kofman} L.,   {Pogosyan} D.,  1996, \mn@doi [\nat]
  {10.1038/380603a0}, \href {http://adsabs.harvard.edu/abs/1996Natur.380..603B}
  {380, 603}

\bibitem[\protect\citeauthoryear{{Bond}, {Strauss}  \& {Cen}}{{Bond}
  et~al.}{2010a}]{bond2010a}
{Bond} N.~A.,  {Strauss} M.~A.,   {Cen} R.,  2010a, \mn@doi [\mnras]
  {10.1111/j.1365-2966.2010.16823.x}, \href
  {http://adsabs.harvard.edu/abs/2010MNRAS.406.1609B} {406, 1609}

\bibitem[\protect\citeauthoryear{{Bond}, {Strauss}  \& {Cen}}{{Bond}
  et~al.}{2010b}]{bond2010b}
{Bond} N.~A.,  {Strauss} M.~A.,   {Cen} R.,  2010b, \mn@doi [\mnras]
  {10.1111/j.1365-2966.2010.17307.x}, \href
  {http://adsabs.harvard.edu/abs/2010MNRAS.409..156B} {409, 156}

\bibitem[\protect\citeauthoryear{{Bos}, {van de Weygaert}, {Dolag}  \&
  {Pettorino}}{{Bos} et~al.}{2012}]{bos2012}
{Bos} E.~G.~P.,  {van de Weygaert} R.,  {Dolag} K.,   {Pettorino} V.,  2012,
  \mn@doi [MNRAS] {10.1111/j.1365-2966.2012.21478.x}, \href
  {http://adsabs.harvard.edu/abs/2012MNRAS.426..440B} {426, 440}

\bibitem[\protect\citeauthoryear{{Carlesi} et~al.,}{{Carlesi}
  et~al.}{2016}]{2016MNRAS.458..900C}
{Carlesi} E.,  et~al., 2016, \mn@doi [\mnras] {10.1093/mnras/stw357}, \href
  {http://adsabs.harvard.edu/abs/2016MNRAS.458..900C} {458, 900}

\bibitem[\protect\citeauthoryear{{Carollo} et~al.,}{{Carollo}
  et~al.}{2013}]{2013ApJ...776...71C}
{Carollo} C.~M.,  et~al., 2013, \mn@doi [\apj] {10.1088/0004-637X/776/2/71},
  \href {http://adsabs.harvard.edu/abs/2013ApJ...776...71C} {776, 71}

\bibitem[\protect\citeauthoryear{{Cautun} \& {van de Weygaert}}{{Cautun} \&
  {van de Weygaert}}{2011}]{cautun2011}
{Cautun} M.~C.,  {van de Weygaert} R.,  2011, {The DTFE public software: The
  Delaunay Tessellation Field Estimator code}, Astrophysics Source Code Library
  (\mn@eprint {arXiv} {1105.0370})

\bibitem[\protect\citeauthoryear{{Cautun}, {van de Weygaert}  \&
  {Jones}}{{Cautun} et~al.}{2013}]{cautun2013}
{Cautun} M.,  {van de Weygaert} R.,   {Jones} B.~J.~T.,  2013, \mn@doi [\mnras]
  {10.1093/mnras/sts416}, \href
  {http://adsabs.harvard.edu/abs/2013MNRAS.429.1286C} {429, 1286}

\bibitem[\protect\citeauthoryear{{Cautun}, {van de Weygaert}, {Jones}  \&
  {Frenk}}{{Cautun} et~al.}{2014}]{cautun2014}
{Cautun} M.,  {van de Weygaert} R.,  {Jones} B.~J.~T.,   {Frenk} C.~S.,  2014,
  \mn@doi [\mnras] {10.1093/mnras/stu768}, \href
  {http://adsabs.harvard.edu/abs/2014MNRAS.441.2923C} {441, 2923}

\bibitem[\protect\citeauthoryear{{Cautun}, {Bose}, {Frenk}, {Guo}, {Han},
  {Hellwing}, {Sawala}  \& {Wang}}{{Cautun} et~al.}{2015}]{cautun2015}
{Cautun} M.,  {Bose} S.,  {Frenk} C.~S.,  {Guo} Q.,  {Han} J.,  {Hellwing}
  W.~A.,  {Sawala} T.,   {Wang} W.,  2015, \mn@doi [\mnras]
  {10.1093/mnras/stv1557}, \href
  {http://adsabs.harvard.edu/abs/2015MNRAS.452.3838C} {452, 3838}

\bibitem[\protect\citeauthoryear{{Chazal}, {Cohen-Steiner}  \&
  {M\'erigot}}{{Chazal} et~al.}{2009}]{chazal2009}
{Chazal} F.,  {Cohen-Steiner} D.,   {M\'erigot} Q.,  2009, INRIA Rapport de
  R\'echerche 6930

\bibitem[\protect\citeauthoryear{{Chincarini} \& {Rood}}{{Chincarini} \&
  {Rood}}{1975}]{chincarini1975}
{Chincarini} G.,  {Rood} H.~J.,  1975, \mn@doi [\nat] {10.1038/257294a0}, \href
  {http://adsabs.harvard.edu/abs/1975Natur.257..294C} {257, 294}

\bibitem[\protect\citeauthoryear{{Choi}, {Bond}, {Strauss}, {Coil}, {Davis}  \&
  {Willmer}}{{Choi} et~al.}{2010}]{choi2010}
{Choi} E.,  {Bond} N.~A.,  {Strauss} M.~A.,  {Coil} A.~L.,  {Davis} M.,
  {Willmer} C.~N.~A.,  2010, \mn@doi [\mnras]
  {10.1111/j.1365-2966.2010.16707.x}, \href
  {http://adsabs.harvard.edu/abs/2010MNRAS.406..320C} {406, 320}

\bibitem[\protect\citeauthoryear{{Codis}, {Pichon}, {Devriendt}, {Slyz},
  {Pogosyan}, {Dubois}  \& {Sousbie}}{{Codis} et~al.}{2012}]{codis2012}
{Codis} S.,  {Pichon} C.,  {Devriendt} J.,  {Slyz} A.,  {Pogosyan} D.,
  {Dubois} Y.,   {Sousbie} T.,  2012, \mn@doi [\mnras]
  {10.1111/j.1365-2966.2012.21636.x}, \href
  {http://adsabs.harvard.edu/abs/2012MNRAS.427.3320C} {427, 3320}

\bibitem[\protect\citeauthoryear{{Colberg}}{{Colberg}}{2007}]{colberg2007}
{Colberg} J.,  2007, MNRAS, 375, 337

\bibitem[\protect\citeauthoryear{Colberg, Krughoff  \& Connolly}{Colberg
  et~al.}{2005}]{colberg2005}
Colberg J.~M.,  Krughoff K.~S.,   Connolly A.~J.,  2005, \mnras, 359, 272

\bibitem[\protect\citeauthoryear{{Colberg} et~al.,}{{Colberg}
  et~al.}{2008}]{Colberg2008}
{Colberg} J.~M.,  et~al., 2008, \mn@doi [\mnras]
  {10.1111/j.1365-2966.2008.13307.x}, \href
  {http://adsabs.harvard.edu/abs/2008MNRAS.387..933C} {387, 933}

\bibitem[\protect\citeauthoryear{{Colless} et~al.,}{{Colless}
  et~al.}{2003}]{colless2003}
{Colless} M.,  et~al., 2003, ArXiv Astrophysics e-prints 0306581, \href
  {http://adsabs.harvard.edu/abs/2003astro.ph..6581C} {}

\bibitem[\protect\citeauthoryear{{Colombi}, {Pogosyan}  \&
  {Souradeep}}{{Colombi} et~al.}{2000}]{colombi2000}
{Colombi} S.,  {Pogosyan} D.,   {Souradeep} T.,  2000, \mn@doi [Physical Review
  Letters] {10.1103/PhysRevLett.85.5515}, \href
  {http://adsabs.harvard.edu/abs/2000PhRvL..85.5515C} {85, 5515}

\bibitem[\protect\citeauthoryear{{Courtois}, {Pomar{\`e}de}, {Tully}, {Hoffman}
   \& {Courtois}}{{Courtois} et~al.}{2013}]{courtois2013}
{Courtois} H.~M.,  {Pomar{\`e}de} D.,  {Tully} R.~B.,  {Hoffman} Y.,
  {Courtois} D.,  2013, \mn@doi [\aj] {10.1088/0004-6256/146/3/69}, \href
  {http://adsabs.harvard.edu/abs/2013AJ....146...69C} {146, 69}

\bibitem[\protect\citeauthoryear{{Creasey}, {Scannapieco}, {Nuza}, {Yepes},
  {Gottl{\"o}ber}  \& {Steinmetz}}{{Creasey}
  et~al.}{2015}]{2015ApJ...800L...4C}
{Creasey} P.,  {Scannapieco} C.,  {Nuza} S.~E.,  {Yepes} G.,  {Gottl{\"o}ber}
  S.,   {Steinmetz} M.,  2015, \mn@doi [\apjl] {10.1088/2041-8205/800/1/L4},
  \href {http://adsabs.harvard.edu/abs/2015ApJ...800L...4C} {800, L4}

\bibitem[\protect\citeauthoryear{{Danovich}, {Dekel}, {Hahn}  \&
  {Teyssier}}{{Danovich} et~al.}{2012}]{2012MNRAS.422.1732D}
{Danovich} M.,  {Dekel} A.,  {Hahn} O.,   {Teyssier} R.,  2012, \mn@doi
  [\mnras] {10.1111/j.1365-2966.2012.20751.x}, \href
  {http://adsabs.harvard.edu/abs/2012MNRAS.422.1732D} {422, 1732}

\bibitem[\protect\citeauthoryear{{Danovich}, {Dekel}, {Hahn}, {Ceverino}  \&
  {Primack}}{{Danovich} et~al.}{2015}]{2015MNRAS.449.2087D}
{Danovich} M.,  {Dekel} A.,  {Hahn} O.,  {Ceverino} D.,   {Primack} J.,  2015,
  \mn@doi [\mnras] {10.1093/mnras/stv270}, \href
  {http://adsabs.harvard.edu/abs/2015MNRAS.449.2087D} {449, 2087}

\bibitem[\protect\citeauthoryear{{Davis}, {Efstathiou}, {Frenk}  \&
  {White}}{{Davis} et~al.}{1985}]{1985ApJ...292..371D}
{Davis} M.,  {Efstathiou} G.,  {Frenk} C.~S.,   {White} S.~D.~M.,  1985,
  \mn@doi [\apj] {10.1086/163168}, \href
  {http://adsabs.harvard.edu/abs/1985ApJ...292..371D} {292, 371}

\bibitem[\protect\citeauthoryear{{Dekel} et~al.,}{{Dekel}
  et~al.}{2009a}]{2009Natur.457..451D}
{Dekel} A.,  et~al., 2009a, \mn@doi [\nat] {10.1038/nature07648}, \href
  {http://adsabs.harvard.edu/abs/2009Natur.457..451D} {457, 451}

\bibitem[\protect\citeauthoryear{{Dekel}, {Sari}  \& {Ceverino}}{{Dekel}
  et~al.}{2009b}]{2009ApJ...703..785D}
{Dekel} A.,  {Sari} R.,   {Ceverino} D.,  2009b, \mn@doi [\apj]
  {10.1088/0004-637X/703/1/785}, \href
  {http://adsabs.harvard.edu/abs/2009ApJ...703..785D} {703, 785}

\bibitem[\protect\citeauthoryear{{Doroshkevich}}{{Doroshkevich}}{1970}]{doroshkevich1970}
{Doroshkevich} A.~G.,  1970, \mn@doi [Astrophysics] {10.1007/BF01001625}, \href
  {http://adsabs.harvard.edu/abs/1970Ap......6..320D} {6, 320}

\bibitem[\protect\citeauthoryear{{Doroshkevich}, {Tucker}, {Allam}  \&
  {Way}}{{Doroshkevich} et~al.}{2004}]{2004A&A...418....7D}
{Doroshkevich} A.,  {Tucker} D.~L.,  {Allam} S.,   {Way} M.~J.,  2004, \mn@doi
  [\aap] {10.1051/0004-6361:20031780}, \href
  {http://adsabs.harvard.edu/abs/2004A%26A...418....7D} {418, 7}

\bibitem[\protect\citeauthoryear{{Dressler}}{{Dressler}}{1980}]{dressler1980}
{Dressler} A.,  1980, \mn@doi [\apj] {10.1086/157753}, \href
  {http://adsabs.harvard.edu/abs/1980ApJ...236..351D} {236, 351}

\bibitem[\protect\citeauthoryear{{Eardley} et~al.,}{{Eardley}
  et~al.}{2015}]{2015MNRAS.448.3665E}
{Eardley} E.,  et~al., 2015, \mn@doi [\mnras] {10.1093/mnras/stv237}, \href
  {http://adsabs.harvard.edu/abs/2015MNRAS.448.3665E} {448, 3665}

\bibitem[\protect\citeauthoryear{{Eckert} et~al.,}{{Eckert}
  et~al.}{2015}]{2015Natur.528..105E}
{Eckert} D.,  et~al., 2015, \mn@doi [\nat] {10.1038/nature16058}, \href
  {http://adsabs.harvard.edu/abs/2015Natur.528..105E} {528, 105}

\bibitem[\protect\citeauthoryear{Edelsbrunner \& Harer}{Edelsbrunner \&
  Harer}{2010}]{edelsbrunner2010}
Edelsbrunner H.,  Harer J.,  2010, Computational Topology: An Introduction.
Applied mathematics, American Mathematical Society, \url
  {http://books.google.at/books?id=MDXa6gFRZuIC}

\bibitem[\protect\citeauthoryear{Edelsbrunner, Letscher  \&
  Zomorodian}{Edelsbrunner et~al.}{2002}]{edelsbrunner2002}
Edelsbrunner H.,  Letscher J.,   Zomorodian A.,  2002, \mn@doi [Discrete
  Computat. Geom.] {10.1007/s00454-002-2885-2}, 28, 511

\bibitem[\protect\citeauthoryear{{Einasto} et~al.,}{{Einasto}
  et~al.}{2011}]{einasto2011b}
{Einasto} J.,  et~al., 2011, \aap, 534, A128

\bibitem[\protect\citeauthoryear{{Fairall}}{{Fairall}}{1998}]{fairall1998}
{Fairall} A.~P.,  ed. 1998, {Large-scale structures in the universe}

\bibitem[\protect\citeauthoryear{{Falck} \& {Neyrinck}}{{Falck} \&
  {Neyrinck}}{2015}]{falck2015}
{Falck} B.,  {Neyrinck} M.~C.,  2015, \mn@doi [\mnras] {10.1093/mnras/stv879},
  \href {http://adsabs.harvard.edu/abs/2015MNRAS.450.3239F} {450, 3239}

\bibitem[\protect\citeauthoryear{{Falck}, {Neyrinck}  \& {Szalay}}{{Falck}
  et~al.}{2012}]{falck2012}
{Falck} B.~L.,  {Neyrinck} M.~C.,   {Szalay} A.~S.,  2012, \mn@doi [\apj]
  {10.1088/0004-637X/754/2/126}, \href
  {http://adsabs.harvard.edu/abs/2012ApJ...754..126F} {754, 126}

\bibitem[\protect\citeauthoryear{{Falck}, {Koyama}, {Zhao}  \& {Li}}{{Falck}
  et~al.}{2014}]{2014JCAP...07..058F}
{Falck} B.,  {Koyama} K.,  {Zhao} G.-b.,   {Li} B.,  2014, \mn@doi [\jcap]
  {10.1088/1475-7516/2014/07/058}, \href
  {http://adsabs.harvard.edu/abs/2014JCAP...07..058F} {7, 058}

\bibitem[\protect\citeauthoryear{Florack, Romeny, Koenderink  \&
  Viergever}{Florack et~al.}{1992}]{florack1992}
Florack L. M.~J.,  Romeny B. M. T.~H.,  Koenderink J.~J.,   Viergever M.~A.,
  1992, Image and Vision Computing, 10, 376

\bibitem[\protect\citeauthoryear{{Forero-Romero} \&
  {Gonz{\'a}lez}}{{Forero-Romero} \&
  {Gonz{\'a}lez}}{2015}]{2015ApJ...799...45F}
{Forero-Romero} J.~E.,  {Gonz{\'a}lez} R.,  2015, \mn@doi [\apj]
  {10.1088/0004-637X/799/1/45}, \href
  {http://adsabs.harvard.edu/abs/2015ApJ...799...45F} {799, 45}

\bibitem[\protect\citeauthoryear{{Forero-Romero}, {Hoffman}, {Gottl{\"o}ber},
  {Klypin}  \& {Yepes}}{{Forero-Romero} et~al.}{2009}]{forero2009}
{Forero-Romero} J.~E.,  {Hoffman} Y.,  {Gottl{\"o}ber} S.,  {Klypin} A.,
  {Yepes} G.,  2009, \mn@doi [\mnras] {10.1111/j.1365-2966.2009.14885.x}, \href
  {http://adsabs.harvard.edu/abs/2009MNRAS.396.1815F} {396, 1815}

\bibitem[\protect\citeauthoryear{Forman}{Forman}{1998}]{forman1998}
Forman R.,  1998, \mn@doi [Advances in Mathematics]
  {http://dx.doi.org/10.1006/aima.1997.1650}, 134, 90

\bibitem[\protect\citeauthoryear{{Frenk} et~al.,}{{Frenk}
  et~al.}{1999}]{1999ApJ...525..554F}
{Frenk} C.~S.,  et~al., 1999, \mn@doi [\apj] {10.1086/307908}, \href
  {http://adsabs.harvard.edu/abs/1999ApJ...525..554F} {525, 554}

\bibitem[\protect\citeauthoryear{Geller \& Huchra}{Geller \&
  Huchra}{1989}]{geller1989}
Geller M.~J.,  Huchra J.~P.,  1989, Science, 246, 897

\bibitem[\protect\citeauthoryear{{Genovese}, {Perone-Pacifico}, {Verdinelli}
  \& {Wasserman}}{{Genovese} et~al.}{2010}]{genovese2010}
{Genovese} C.~R.,  {Perone-Pacifico} M.,  {Verdinelli} I.,   {Wasserman} L.,
  2010, preprint, \href {http://adsabs.harvard.edu/abs/2010arXiv1003.5536G} {}
  (\mn@eprint {arXiv} {1003.5536})

\bibitem[\protect\citeauthoryear{{Gillet}, {Ocvirk}, {Aubert}, {Knebe},
  {Libeskind}, {Yepes}, {Gottl{\"o}ber}  \& {Hoffman}}{{Gillet}
  et~al.}{2015}]{2015ApJ...800...34G}
{Gillet} N.,  {Ocvirk} P.,  {Aubert} D.,  {Knebe} A.,  {Libeskind} N.,  {Yepes}
  G.,  {Gottl{\"o}ber} S.,   {Hoffman} Y.,  2015, \mn@doi [\apj]
  {10.1088/0004-637X/800/1/34}, \href
  {http://adsabs.harvard.edu/abs/2015ApJ...800...34G} {800, 34}

\bibitem[\protect\citeauthoryear{Giovanelli \& Haynes}{Giovanelli \&
  Haynes}{1985}]{giovanelli1985}
Giovanelli R.,  Haynes M.~P.,  1985, \aj, 90, 2445

\bibitem[\protect\citeauthoryear{{Goerdt}, {Ceverino}, {Dekel}  \&
  {Teyssier}}{{Goerdt} et~al.}{2015}]{2015MNRAS.454..637G}
{Goerdt} T.,  {Ceverino} D.,  {Dekel} A.,   {Teyssier} R.,  2015, \mn@doi
  [\mnras] {10.1093/mnras/stv2005}, \href
  {http://adsabs.harvard.edu/abs/2015MNRAS.454..637G} {454, 637}

\bibitem[\protect\citeauthoryear{{Gonz{\'a}lez} \& {Padilla}}{{Gonz{\'a}lez} \&
  {Padilla}}{2010}]{gonzalez2010}
{Gonz{\'a}lez} R.~E.,  {Padilla} N.~D.,  2010, \mn@doi [\mnras]
  {10.1111/j.1365-2966.2010.17015.x}, \href
  {http://adsabs.harvard.edu/abs/2010MNRAS.407.1449G} {407, 1449}

\bibitem[\protect\citeauthoryear{{Gonz{\'a}lez} \& {Padilla}}{{Gonz{\'a}lez} \&
  {Padilla}}{2016}]{2016ApJ...829...58G}
{Gonz{\'a}lez} R.~E.,  {Padilla} N.~D.,  2016, \mn@doi [\apj]
  {10.3847/0004-637X/829/1/58}, \href
  {http://adsabs.harvard.edu/abs/2016ApJ...829...58G} {829, 58}

\bibitem[\protect\citeauthoryear{{Gonz{\'a}lez}, {Prieto}, {Padilla}  \&
  {Jimenez}}{{Gonz{\'a}lez} et~al.}{2017}]{2017MNRAS.464.4666G}
{Gonz{\'a}lez} R.~E.,  {Prieto} J.,  {Padilla} N.,   {Jimenez} R.,  2017,
  \mn@doi [\mnras] {10.1093/mnras/stw2715}, \href
  {http://adsabs.harvard.edu/abs/2017MNRAS.464.4666G} {464, 4666}

\bibitem[\protect\citeauthoryear{{Gott}, {Juri{\'c}}, {Schlegel}, {Hoyle},
  {Vogeley}, {Tegmark}, {Bahcall}  \& {Brinkmann}}{{Gott}
  et~al.}{2005}]{2005ApJ...624..463G}
{Gott} III J.~R.,  {Juri{\'c}} M.,  {Schlegel} D.,  {Hoyle} F.,  {Vogeley} M.,
  {Tegmark} M.,  {Bahcall} N.,   {Brinkmann} J.,  2005, \mn@doi [\apj]
  {10.1086/428890}, \href {http://adsabs.harvard.edu/abs/2005ApJ...624..463G}
  {624, 463}

\bibitem[\protect\citeauthoryear{{Graham} \& {Clowes}}{{Graham} \&
  {Clowes}}{1995}]{graham1995}
{Graham} M.,  {Clowes} R.,  1995, MNRAS, 275, 790

\bibitem[\protect\citeauthoryear{Gregory, Thompson  \& Tifft}{Gregory
  et~al.}{1978}]{gregthomp1978}
Gregory S.~A.,  Thompson L.~A.,   Tifft W.~G.,  1978, \baas, 10, 622

\bibitem[\protect\citeauthoryear{{Guo}, {Tempel}  \& {Libeskind}}{{Guo}
  et~al.}{2015}]{2015ApJ...800..112G}
{Guo} Q.,  {Tempel} E.,   {Libeskind} N.~I.,  2015, \mn@doi [\apj]
  {10.1088/0004-637X/800/2/112}, \href
  {http://adsabs.harvard.edu/abs/2015ApJ...800..112G} {800, 112}

\bibitem[\protect\citeauthoryear{{Gurbatov}, {Saichev}  \&
  {Shandarin}}{{Gurbatov} et~al.}{1989}]{gurbatov1989}
{Gurbatov} S.~N.,  {Saichev} A.~I.,   {Shandarin} S.~F.,  1989, \mnras, \href
  {http://adsabs.harvard.edu/abs/1989MNRAS.236..385G} {236, 385}

\bibitem[\protect\citeauthoryear{{Guzzo} et~al.,}{{Guzzo}
  et~al.}{2014}]{vipers2014}
{Guzzo} L.,  et~al., 2014, \mn@doi [\aap] {10.1051/0004-6361/201321489}, \href
  {http://adsabs.harvard.edu/abs/2014A%26A...566A.108G} {566, A108}

\bibitem[\protect\citeauthoryear{Gyulassy}{Gyulassy}{2008}]{gyulassy2008}
Gyulassy A.~G.,  2008, Combinatorial construction of Morse-Smale complexes for
  data analysis and visualization, Ph.D. thesis, UC Davis

\bibitem[\protect\citeauthoryear{{Hahn}}{{Hahn}}{2009}]{hahnphd2009}
{Hahn} O.,  2009, Galaxy Formation in the Cosmic Web, Ph.D. thesis, ETH
  Z\"urich

\bibitem[\protect\citeauthoryear{{Hahn}, {Porciani}, {Carollo}  \&
  {Dekel}}{{Hahn} et~al.}{2007a}]{hahn2007a}
{Hahn} O.,  {Porciani} C.,  {Carollo} C.~M.,   {Dekel} A.,  2007a, \mn@doi
  [\mnras] {10.1111/j.1365-2966.2006.11318.x}, \href
  {http://adsabs.harvard.edu/abs/2007MNRAS.375..489H} {375, 489}

\bibitem[\protect\citeauthoryear{{Hahn}, {Carollo}, {Porciani}  \&
  {Dekel}}{{Hahn} et~al.}{2007b}]{hahn2007b}
{Hahn} O.,  {Carollo} C.~M.,  {Porciani} C.,   {Dekel} A.,  2007b, \mn@doi
  [\mnras] {10.1111/j.1365-2966.2007.12249.x}, \href
  {http://adsabs.harvard.edu/abs/2007MNRAS.381...41H} {381, 41}

\bibitem[\protect\citeauthoryear{{Heinrich}, {Stoica}  \& {Tran}}{{Heinrich}
  et~al.}{2012}]{Heinrich:12}
{Heinrich} P.,  {Stoica} R.~S.,   {Tran} V.~C.,  2012, Spatial Statistics, 2,
  47

\bibitem[\protect\citeauthoryear{{He{\ss}}, {Kitaura}  \&
  {Gottl{\"o}ber}}{{He{\ss}} et~al.}{2013}]{hess2013}
{He{\ss}} S.,  {Kitaura} F.-S.,   {Gottl{\"o}ber} S.,  2013, \mn@doi [MNRAS]
  {10.1093/mnras/stt1428}, \href
  {http://adsabs.harvard.edu/abs/2013MNRAS.435.2065H} {435, 2065}

\bibitem[\protect\citeauthoryear{Hidding}{Hidding}{2017}]{hidding2017}
Hidding J.,  2017, The Phase-Space Geometry of the Cosmic Web, Ph.D. thesis,
  University of Groningen

\bibitem[\protect\citeauthoryear{{Hidding}, {van de Weygaert}, {Vegter},
  {Jones}  \& {Teillaud}}{{Hidding} et~al.}{2012}]{hidding2012}
{Hidding} J.,  {van de Weygaert} R.,  {Vegter} G.,  {Jones} B.~J.~T.,
  {Teillaud} M.,  2012, preprint, \href
  {http://adsabs.harvard.edu/abs/2012arXiv1205.1669H} {} (\mn@eprint {arXiv}
  {1205.1669})

\bibitem[\protect\citeauthoryear{{Hidding}, {Shandarin}  \& {van de
  Weygaert}}{{Hidding} et~al.}{2014}]{hidding2014}
{Hidding} J.,  {Shandarin} S.~F.,   {van de Weygaert} R.,  2014, \mn@doi
  [\mnras] {10.1093/mnras/stt2142}, \href
  {http://adsabs.harvard.edu/abs/2014MNRAS.437.3442H} {437, 3442}

\bibitem[\protect\citeauthoryear{{Hirv}, {Pelt}, {Saar}, {Tago}, {Tamm},
  {Tempel}  \& {Einasto}}{{Hirv} et~al.}{2017}]{hirv2017}
{Hirv} A.,  {Pelt} J.,  {Saar} E.,  {Tago} E.,  {Tamm} A.,  {Tempel} E.,
  {Einasto} M.,  2017, \aap, 599, A31

\bibitem[\protect\citeauthoryear{{Hoffman}, {Metuki}, {Yepes}, {Gottl{\"o}ber},
  {Forero-Romero}, {Libeskind}  \& {Knebe}}{{Hoffman}
  et~al.}{2012}]{2012MNRAS.425.2049H}
{Hoffman} Y.,  {Metuki} O.,  {Yepes} G.,  {Gottl{\"o}ber} S.,  {Forero-Romero}
  J.~E.,  {Libeskind} N.~I.,   {Knebe} A.,  2012, \mn@doi [\mnras]
  {10.1111/j.1365-2966.2012.21553.x}, \href
  {http://adsabs.harvard.edu/abs/2012MNRAS.425.2049H} {425, 2049}

\bibitem[\protect\citeauthoryear{{Hoffman}, {Pomar{\`e}de}, {Tully}  \&
  {Courtois}}{{Hoffman} et~al.}{2017}]{hoffman2017}
{Hoffman} Y.,  {Pomar{\`e}de} D.,  {Tully} R.~B.,   {Courtois} H.~M.,  2017,
  \mn@doi [Nature Astronomy] {10.1038/s41550-016-0036}, \href
  {http://adsabs.harvard.edu/abs/2017NatAs...1E..36H} {1, 0036}

\bibitem[\protect\citeauthoryear{{Hoyle}}{{Hoyle}}{1951}]{hoyle1951}
{Hoyle} F.,  1951, in Problems of Cosmical Aerodynamics. p.~195

\bibitem[\protect\citeauthoryear{{Huchra} et~al.,}{{Huchra}
  et~al.}{2012}]{huchra2012}
{Huchra} J.~P.,  et~al., 2012, \mn@doi [ApJS] {10.1088/0067-0049/199/2/26},
  \href {http://adsabs.harvard.edu/abs/2012ApJS..199...26H} {199, 26}

\bibitem[\protect\citeauthoryear{{Ibata} et~al.,}{{Ibata}
  et~al.}{2013}]{ibata2013}
{Ibata} R.~A.,  et~al., 2013, \mn@doi [\nat] {10.1038/nature11717}, \href
  {http://adsabs.harvard.edu/abs/2013Natur.493...62I} {493, 62}

\bibitem[\protect\citeauthoryear{Icke}{Icke}{1973}]{icke1973}
Icke V.,  1973, \aap, 27, 1

\bibitem[\protect\citeauthoryear{{Icke}}{{Icke}}{1984}]{icke1984}
{Icke} V.,  1984, \mn@doi [\mnras] {10.1093/mnras/206.1.1P}, \href
  {http://adsabs.harvard.edu/abs/1984MNRAS.206P...1I} {206, 1P}

\bibitem[\protect\citeauthoryear{{J{\~o}eveer}, {Einasto}  \&
  {Tago}}{{J{\~o}eveer} et~al.}{1978}]{joeveer1978b}
{J{\~o}eveer} M.,  {Einasto} J.,   {Tago} E.,  1978, \mn@doi [\mnras]
  {10.1093/mnras/185.2.357}, 185, 357

\bibitem[\protect\citeauthoryear{{Jones}, {van de Weygaert}  \&
  {Arag{\'o}n-Calvo}}{{Jones} et~al.}{2010}]{jones2010}
{Jones} B.~J.~T.,  {van de Weygaert} R.,   {Arag{\'o}n-Calvo} M.~A.,  2010,
  \mn@doi [\mnras] {10.1111/j.1365-2966.2010.17202.x}, \href
  {http://adsabs.harvard.edu/abs/2010MNRAS.408..897J} {408, 897}

\bibitem[\protect\citeauthoryear{Kirshner, Oemler, Schechter  \&
  Shectman}{Kirshner et~al.}{1981}]{kirshner1981}
Kirshner R.~P.,  Oemler A.~J.,  Schechter P.~L.,   Shectman S.~A.,  1981,
  \apjl, 248, L57

\bibitem[\protect\citeauthoryear{{Kirshner}, {Oemler}, {Schechter}  \&
  {Shectman}}{{Kirshner} et~al.}{1987}]{kirshner1987}
{Kirshner} R.~P.,  {Oemler} Jr. A.,  {Schechter} P.~L.,   {Shectman} S.~A.,
  1987, \mn@doi [\apj] {10.1086/165080}, \href
  {http://adsabs.harvard.edu/abs/1987ApJ...314..493K} {314, 493}

\bibitem[\protect\citeauthoryear{{Kitaura}}{{Kitaura}}{2013}]{kitaura2013}
{Kitaura} F.-S.,  2013, \mn@doi [MNRAS] {10.1093/mnrasl/sls029}, \href
  {http://adsabs.harvard.edu/abs/2013MNRAS.429L..84K} {429, L84}

\bibitem[\protect\citeauthoryear{{Kitaura} \& {Angulo}}{{Kitaura} \&
  {Angulo}}{2012}]{2012MNRAS.425.2443K}
{Kitaura} F.-S.,  {Angulo} R.~E.,  2012, \mn@doi [\mnras]
  {10.1111/j.1365-2966.2012.21614.x}, \href
  {http://adsabs.harvard.edu/abs/2012MNRAS.425.2443K} {425, 2443}

\bibitem[\protect\citeauthoryear{{Knebe}, {Gill}, {Gibson}, {Lewis}, {Ibata}
  \& {Dopita}}{{Knebe} et~al.}{2004}]{2004ApJ...603....7K}
{Knebe} A.,  {Gill} S.~P.~D.,  {Gibson} B.~K.,  {Lewis} G.~F.,  {Ibata} R.~A.,
   {Dopita} M.~A.,  2004, \mn@doi [\apj] {10.1086/381306}, \href
  {http://adsabs.harvard.edu/abs/2004ApJ...603....7K} {603, 7}

\bibitem[\protect\citeauthoryear{{Knebe} et~al.,}{{Knebe}
  et~al.}{2011}]{Knebe2011}
{Knebe} A.,  et~al., 2011, \mn@doi [\mnras] {10.1111/j.1365-2966.2011.18858.x},
  \href {http://adsabs.harvard.edu/abs/2011MNRAS.415.2293K} {415, 2293}

\bibitem[\protect\citeauthoryear{{Knebe} et~al.,}{{Knebe}
  et~al.}{2013}]{2013MNRAS.435.1618K}
{Knebe} A.,  et~al., 2013, \mn@doi [\mnras] {10.1093/mnras/stt1403}, \href
  {http://adsabs.harvard.edu/abs/2013MNRAS.435.1618K} {435, 1618}

\bibitem[\protect\citeauthoryear{{Kofman}, {Pogosian}  \& {Shandarin}}{{Kofman}
  et~al.}{1990}]{kofman1990}
{Kofman} L.,  {Pogosian} D.,   {Shandarin} S.,  1990, \mn@doi [\mnras]
  {10.1093/mnras/242.2.200}, \href
  {http://adsabs.harvard.edu/abs/1990MNRAS.242..200K} {242, 200}

\bibitem[\protect\citeauthoryear{{Kofman}, {Pogosyan}, {Shandarin}  \&
  {Melott}}{{Kofman} et~al.}{1992}]{kofman1992}
{Kofman} L.,  {Pogosyan} D.,  {Shandarin} S.~F.,   {Melott} A.~L.,  1992,
  \mn@doi [\apj] {10.1086/171517}, \href
  {http://adsabs.harvard.edu/abs/1992ApJ...393..437K} {393, 437}

\bibitem[\protect\citeauthoryear{{Kraan-Korteweg}, {Cluver}, {Bilicki},
  {Jarrett}, {Colless}, {Elagali}, {B{\"o}hringer}  \& {Chon}}{{Kraan-Korteweg}
  et~al.}{2017}]{2017MNRAS.466L..29K}
{Kraan-Korteweg} R.~C.,  {Cluver} M.~E.,  {Bilicki} M.,  {Jarrett} T.~H.,
  {Colless} M.,  {Elagali} A.,  {B{\"o}hringer} H.,   {Chon} G.,  2017, \mn@doi
  [\mnras] {10.1093/mnrasl/slw229}, \href
  {http://adsabs.harvard.edu/abs/2017MNRAS.466L..29K} {466, L29}

\bibitem[\protect\citeauthoryear{{Kuutma}, {Tamm}  \& {Tempel}}{{Kuutma}
  et~al.}{2017}]{kuutma2017}
{Kuutma} T.,  {Tamm} A.,   {Tempel} E.,  2017, \mn@doi [\aap]
  {10.1051/0004-6361/201730526}, \href
  {http://adsabs.harvard.edu/abs/2017A%26A...600L...6K} {600, L6}

\bibitem[\protect\citeauthoryear{{Lahav}, {Santiago}, {Webster}, {Strauss},
  {Davis}, {Dressler}  \& {Huchra}}{{Lahav} et~al.}{2000}]{2000MNRAS.312..166L}
{Lahav} O.,  {Santiago} B.~X.,  {Webster} A.~M.,  {Strauss} M.~A.,  {Davis} M.,
   {Dressler} A.,   {Huchra} J.~P.,  2000, \mn@doi [\mnras]
  {10.1046/j.1365-8711.2000.03145.x}, \href
  {http://adsabs.harvard.edu/abs/2000MNRAS.312..166L} {312, 166}

\bibitem[\protect\citeauthoryear{{Lavaux} \& {Wandelt}}{{Lavaux} \&
  {Wandelt}}{2010}]{lavaux2010}
{Lavaux} G.,  {Wandelt} B.~D.,  2010, \mn@doi [\mnras]
  {10.1111/j.1365-2966.2010.16197.x}, \href
  {http://adsabs.harvard.edu/abs/2010MNRAS.403.1392L} {403, 1392}

\bibitem[\protect\citeauthoryear{{Lavaux} \& {Wandelt}}{{Lavaux} \&
  {Wandelt}}{2012}]{lavaux2012}
{Lavaux} G.,  {Wandelt} B.~D.,  2012, \mn@doi [\apj]
  {10.1088/0004-637X/754/2/109}, \href
  {http://adsabs.harvard.edu/abs/2012ApJ...754..109L} {754, 109}

\bibitem[\protect\citeauthoryear{{Leclercq}, {Jasche}, {Sutter}, {Hamaus}  \&
  {Wandelt}}{{Leclercq} et~al.}{2015a}]{leclercq2015b}
{Leclercq} F.,  {Jasche} J.,  {Sutter} P.~M.,  {Hamaus} N.,   {Wandelt} B.,
  2015a, \mn@doi [\jcap] {10.1088/1475-7516/2015/03/047}, \href
  {http://adsabs.harvard.edu/abs/2015JCAP...03..047L} {3, 047}

\bibitem[\protect\citeauthoryear{{Leclercq}, {Jasche}  \& {Wandelt}}{{Leclercq}
  et~al.}{2015b}]{leclercq2015}
{Leclercq} F.,  {Jasche} J.,   {Wandelt} B.,  2015b, \mn@doi [\jcap]
  {10.1088/1475-7516/2015/06/015}, \href
  {http://adsabs.harvard.edu/abs/2015JCAP...06..015L} {6, 015}

\bibitem[\protect\citeauthoryear{{Leclercq}, {Jasche}  \& {Wandelt}}{{Leclercq}
  et~al.}{2015c}]{leclercq2015c}
{Leclercq} F.,  {Jasche} J.,   {Wandelt} B.,  2015c, \mn@doi [\aap]
  {10.1051/0004-6361/201526006}, \href
  {http://adsabs.harvard.edu/abs/2015A%26A...576L..17L} {576, L17}

\bibitem[\protect\citeauthoryear{{Leclercq}, {Jasche}, {Lavaux}  \&
  {Wandelt}}{{Leclercq} et~al.}{2016}]{2016arXiv160100093L}
{Leclercq} F.,  {Jasche} J.,  {Lavaux} G.,   {Wandelt} B.,  2016, preprint,
  \href {http://adsabs.harvard.edu/abs/2016arXiv160100093L} {} (\mn@eprint
  {arXiv} {1601.00093})

\bibitem[\protect\citeauthoryear{{Lee} \& {Park}}{{Lee} \&
  {Park}}{2009}]{leepark2009}
{Lee} J.,  {Park} D.,  2009, \mn@doi [\apjl] {10.1088/0004-637X/696/1/L10},
  \href {http://adsabs.harvard.edu/abs/2009ApJ...696L..10L} {696, L10}

\bibitem[\protect\citeauthoryear{Lee \& Pen}{Lee \& Pen}{2000}]{leepen2000}
Lee J.,  Pen U.-L.,  2000, ApJ, 532, L5

\bibitem[\protect\citeauthoryear{Li, Sone  \& Doi}{Li et~al.}{2003}]{li2003}
Li Q.,  Sone S.,   Doi K.,  2003, \mn@doi [Medical Physics]
  {10.1118/1.1581411}, 30, 2040

\bibitem[\protect\citeauthoryear{{Libeskind}, {Hoffman}, {Knebe}, {Steinmetz},
  {Gottl{\"o}ber}, {Metuki}  \& {Yepes}}{{Libeskind}
  et~al.}{2012}]{2012MNRAS.421L.137L}
{Libeskind} N.~I.,  {Hoffman} Y.,  {Knebe} A.,  {Steinmetz} M.,
  {Gottl{\"o}ber} S.,  {Metuki} O.,   {Yepes} G.,  2012, \mn@doi [\mnras]
  {10.1111/j.1745-3933.2012.01222.x}, \href
  {http://adsabs.harvard.edu/abs/2012MNRAS.421L.137L} {421, L137}

\bibitem[\protect\citeauthoryear{{Libeskind}, {Hoffman}, {Forero-Romero},
  {Gottl{\"o}ber}, {Knebe}, {Steinmetz}  \& {Klypin}}{{Libeskind}
  et~al.}{2013a}]{2013MNRAS.428.2489L}
{Libeskind} N.~I.,  {Hoffman} Y.,  {Forero-Romero} J.,  {Gottl{\"o}ber} S.,
  {Knebe} A.,  {Steinmetz} M.,   {Klypin} A.,  2013a, \mn@doi [\mnras]
  {10.1093/mnras/sts216}, \href
  {http://adsabs.harvard.edu/abs/2013MNRAS.428.2489L} {428, 2489}

\bibitem[\protect\citeauthoryear{{Libeskind}, {Hoffman}, {Steinmetz},
  {Gottl{\"o}ber}, {Knebe}  \& {Hess}}{{Libeskind}
  et~al.}{2013b}]{2013ApJ...766L..15L}
{Libeskind} N.~I.,  {Hoffman} Y.,  {Steinmetz} M.,  {Gottl{\"o}ber} S.,
  {Knebe} A.,   {Hess} S.,  2013b, \mn@doi [\apjl]
  {10.1088/2041-8205/766/2/L15}, \href
  {http://adsabs.harvard.edu/abs/2013ApJ...766L..15L} {766, L15}

\bibitem[\protect\citeauthoryear{{Libeskind}, {Hoffman}  \&
  {Gottl{\"o}ber}}{{Libeskind} et~al.}{2014a}]{2014MNRAS.441.1974L}
{Libeskind} N.~I.,  {Hoffman} Y.,   {Gottl{\"o}ber} S.,  2014a, \mn@doi
  [\mnras] {10.1093/mnras/stu629}, \href
  {http://adsabs.harvard.edu/abs/2014MNRAS.441.1974L} {441, 1974}

\bibitem[\protect\citeauthoryear{{Libeskind}, {Knebe}, {Hoffman}  \&
  {Gottl{\"o}ber}}{{Libeskind} et~al.}{2014b}]{2014MNRAS.443.1274L}
{Libeskind} N.~I.,  {Knebe} A.,  {Hoffman} Y.,   {Gottl{\"o}ber} S.,  2014b,
  \mn@doi [\mnras] {10.1093/mnras/stu1216}, \href
  {http://adsabs.harvard.edu/abs/2014MNRAS.443.1274L} {443, 1274}

\bibitem[\protect\citeauthoryear{{Libeskind}, {Hoffman}, {Tully}, {Courtois},
  {Pomar{\`e}de}, {Gottl{\"o}ber}  \& {Steinmetz}}{{Libeskind}
  et~al.}{2015a}]{2015MNRAS.452.1052L}
{Libeskind} N.~I.,  {Hoffman} Y.,  {Tully} R.~B.,  {Courtois} H.~M.,
  {Pomar{\`e}de} D.,  {Gottl{\"o}ber} S.,   {Steinmetz} M.,  2015a, \mn@doi
  [\mnras] {10.1093/mnras/stv1302}, \href
  {http://adsabs.harvard.edu/abs/2015MNRAS.452.1052L} {452, 1052}

\bibitem[\protect\citeauthoryear{{Libeskind}, {Tempel}, {Hoffman}, {Tully}  \&
  {Courtois}}{{Libeskind} et~al.}{2015b}]{2015MNRAS.453L.108L}
{Libeskind} N.~I.,  {Tempel} E.,  {Hoffman} Y.,  {Tully} R.~B.,   {Courtois}
  H.,  2015b, \mn@doi [\mnras] {10.1093/mnrasl/slv099}, \href
  {http://adsabs.harvard.edu/abs/2015MNRAS.453L.108L} {453, L108}

\bibitem[\protect\citeauthoryear{{Lietzen} et~al.,}{{Lietzen}
  et~al.}{2016}]{2016A&A...588L...4L}
{Lietzen} H.,  et~al., 2016, \mn@doi [\aap] {10.1051/0004-6361/201628261},
  \href {http://adsabs.harvard.edu/abs/2016A%26A...588L...4L} {588, L4}

\bibitem[\protect\citeauthoryear{Lindeberg}{Lindeberg}{1998}]{lindeberg1998}
Lindeberg T.,  1998, Int. J. Comput. Vision, 30, 79

\bibitem[\protect\citeauthoryear{{Lynden-Bell}, {Faber}, {Burstein}, {Davies},
  {Dressler}, {Terlevich}  \& {Wegner}}{{Lynden-Bell}
  et~al.}{1988}]{1988ApJ...326...19L}
{Lynden-Bell} D.,  {Faber} S.~M.,  {Burstein} D.,  {Davies} R.~L.,  {Dressler}
  A.,  {Terlevich} R.~J.,   {Wegner} G.,  1988, \mn@doi [\apj]
  {10.1086/166066}, \href {http://adsabs.harvard.edu/abs/1988ApJ...326...19L}
  {326, 19}

\bibitem[\protect\citeauthoryear{{Mart{\'{\i}}nez}, {Muriel}  \&
  {Coenda}}{{Mart{\'{\i}}nez} et~al.}{2016}]{2016MNRAS.455..127M}
{Mart{\'{\i}}nez} H.~J.,  {Muriel} H.,   {Coenda} V.,  2016, \mn@doi [\mnras]
  {10.1093/mnras/stv2295}, \href
  {http://adsabs.harvard.edu/abs/2016MNRAS.455..127M} {455, 127}

\bibitem[\protect\citeauthoryear{{Metuki}, {Libeskind}, {Hoffman}, {Crain}  \&
  {Theuns}}{{Metuki} et~al.}{2015}]{2015MNRAS.446.1458M}
{Metuki} O.,  {Libeskind} N.~I.,  {Hoffman} Y.,  {Crain} R.~A.,   {Theuns} T.,
  2015, \mn@doi [\mnras] {10.1093/mnras/stu2166}, \href
  {http://adsabs.harvard.edu/abs/2015MNRAS.446.1458M} {446, 1458}

\bibitem[\protect\citeauthoryear{{Metuki}, {Libeskind}  \& {Hoffman}}{{Metuki}
  et~al.}{2016}]{2016MNRAS.460..297M}
{Metuki} O.,  {Libeskind} N.~I.,   {Hoffman} Y.,  2016, \mn@doi [\mnras]
  {10.1093/mnras/stw979}, \href
  {http://adsabs.harvard.edu/abs/2016MNRAS.460..297M} {460, 297}

\bibitem[\protect\citeauthoryear{{Milnor}}{{Milnor}}{1963}]{milnor1963}
{Milnor} J.,  1963, Journal of Mathematical Physics

\bibitem[\protect\citeauthoryear{Morse}{Morse}{1934}]{morse1934}
Morse M.,  1934, American Mathematical Society Colloquium Publication, 18, 1

\bibitem[\protect\citeauthoryear{Murphy, Eke  \& Frenk}{Murphy
  et~al.}{2011}]{Murphy2011}
Murphy D. N.~A.,  Eke V.~R.,   Frenk C.~S.,  2011, Monthly Notices of the Royal
  Astronomical Society, 413, 2288

\bibitem[\protect\citeauthoryear{{Nevalainen} et~al.,}{{Nevalainen}
  et~al.}{2015}]{2015A&A...583A.142N}
{Nevalainen} J.,  et~al., 2015, \mn@doi [\aap] {10.1051/0004-6361/201526443},
  \href {http://adsabs.harvard.edu/abs/2015A%26A...583A.142N} {583, A142}

\bibitem[\protect\citeauthoryear{{Neyrinck}}{{Neyrinck}}{2008}]{neyrinck2008}
{Neyrinck} M.~C.,  2008, \mn@doi [\mnras] {10.1111/j.1365-2966.2008.13180.x},
  \href {http://adsabs.harvard.edu/abs/2008MNRAS.386.2101N} {386, 2101}

\bibitem[\protect\citeauthoryear{{Neyrinck}}{{Neyrinck}}{2012}]{neyrinck2012}
{Neyrinck} M.~C.,  2012, \mn@doi [\mnras] {10.1111/j.1365-2966.2012.21956.x},
  \href {http://adsabs.harvard.edu/abs/2012MNRAS.427..494N} {427, 494}

\bibitem[\protect\citeauthoryear{Novikov, Colombi  \& Dor{\'e}}{Novikov
  et~al.}{2006}]{novikov2006}
Novikov D.,  Colombi S.,   Dor{\'e} O.,  2006, \mnras, 366, 1201

\bibitem[\protect\citeauthoryear{{Nuza}, {Kitaura}, {He{\ss}}, {Libeskind}  \&
  {M{\"u}ller}}{{Nuza} et~al.}{2014}]{2014MNRAS.445..988N}
{Nuza} S.~E.,  {Kitaura} F.-S.,  {He{\ss}} S.,  {Libeskind} N.~I.,
  {M{\"u}ller} V.,  2014, \mn@doi [\mnras] {10.1093/mnras/stu1746}, \href
  {http://adsabs.harvard.edu/abs/2014MNRAS.445..988N} {445, 988}

\bibitem[\protect\citeauthoryear{Okabe, Boots, Sugihara  \& Chiu}{Okabe
  et~al.}{2000}]{okabe2000}
Okabe A.,  Boots B.,  Sugihara K.,   Chiu S.~N.,  2000, Spatial tessellations:
  Concepts and applications of {V}oronoi diagrams, 2nd edn.
Probability and Statistics, Wiley, NYC

\bibitem[\protect\citeauthoryear{{Pahwa} et~al.,}{{Pahwa}
  et~al.}{2016}]{2016MNRAS.457..695P}
{Pahwa} I.,  et~al., 2016, \mn@doi [\mnras] {10.1093/mnras/stv2930}, \href
  {http://adsabs.harvard.edu/abs/2016MNRAS.457..695P} {457, 695}

\bibitem[\protect\citeauthoryear{{Pan}, {Vogeley}, {Hoyle}, {Choi}  \&
  {Park}}{{Pan} et~al.}{2012}]{pan2012}
{Pan} D.~C.,  {Vogeley} M.~S.,  {Hoyle} F.,  {Choi} Y.-Y.,   {Park} C.,  2012,
  \mn@doi [\mnras] {10.1111/j.1365-2966.2011.20197.x}, \href
  {http://adsabs.harvard.edu/abs/2012MNRAS.421..926P} {421, 926}

\bibitem[\protect\citeauthoryear{{Park} \& {Lee}}{{Park} \&
  {Lee}}{2007}]{parklee2007}
{Park} D.,  {Lee} J.,  2007, \mn@doi [Physical Review Letters]
  {10.1103/PhysRevLett.98.081301}, \href
  {http://adsabs.harvard.edu/abs/2007PhRvL..98h1301P} {98, 081301}

\bibitem[\protect\citeauthoryear{{Peebles}}{{Peebles}}{1969}]{peebles1969}
{Peebles} P.~J.~E.,  1969, \mn@doi [\apj] {10.1086/149876}, \href
  {http://adsabs.harvard.edu/abs/1969ApJ...155..393P} {155, 393}

\bibitem[\protect\citeauthoryear{Peebles}{Peebles}{1980}]{peebles1980}
Peebles P.~J.~E.,  1980, The Large Scale Structure of the Universe, Princeton
  University Press

\bibitem[\protect\citeauthoryear{{Pichon}, {Codis}, {Pogosyan}, {Dubois},
  {Desjacques}  \& {Devriendt}}{{Pichon} et~al.}{2016}]{pichon2016}
{Pichon} C.,  {Codis} S.,  {Pogosyan} D.,  {Dubois} Y.,  {Desjacques} V.,
  {Devriendt} J.,  2016, in {van de Weygaert} R.,  {Shandarin} S.,  {Saar} E.,
   {Einasto} J.,  eds,  IAU Symposium Vol. 308, The Zeldovich Universe: Genesis
  and Growth of the Cosmic Web. pp 421--432 (\mn@eprint {arXiv} {1409.2608}),
  \mn@doi{10.1017/S1743921316010309}

\bibitem[\protect\citeauthoryear{{Pisani}, {Sutter}, {Hamaus}, {Alizadeh},
  {Biswas}, {Wandelt}  \& {Hirata}}{{Pisani} et~al.}{2015}]{pisani2015}
{Pisani} A.,  {Sutter} P.~M.,  {Hamaus} N.,  {Alizadeh} E.,  {Biswas} R.,
  {Wandelt} B.~D.,   {Hirata} C.~M.,  2015, \mn@doi [\prd]
  {10.1103/PhysRevD.92.083531}, \href
  {http://adsabs.harvard.edu/abs/2015PhRvD..92h3531P} {92, 083531}

\bibitem[\protect\citeauthoryear{{Planck Collaboration} et~al.,}{{Planck
  Collaboration} et~al.}{2014}]{2014A&A...571A..16P}
{Planck Collaboration} et~al., 2014, \mn@doi [\aap]
  {10.1051/0004-6361/201321591}, \href
  {http://adsabs.harvard.edu/abs/2014A%26A...571A..16P} {571, A16}

\bibitem[\protect\citeauthoryear{{Platen}, {van de Weygaert}  \&
  {Jones}}{{Platen} et~al.}{2007}]{platen2007}
{Platen} E.,  {van de Weygaert} R.,   {Jones} B.~J.~T.,  2007, \mn@doi [\mnras]
  {10.1111/j.1365-2966.2007.12125.x}, \href
  {http://adsabs.harvard.edu/abs/2007MNRAS.380..551P} {380, 551}

\bibitem[\protect\citeauthoryear{{Platen}, {van de Weygaert}  \&
  {Jones}}{{Platen} et~al.}{2008}]{platen2008}
{Platen} E.,  {van de Weygaert} R.,   {Jones} B.~J.~T.,  2008, \mn@doi [\mnras]
  {10.1111/j.1365-2966.2008.13019.x}, \href
  {http://adsabs.harvard.edu/abs/2008MNRAS.387..128P} {387, 128}

\bibitem[\protect\citeauthoryear{{Pomar{\`e}de}, {Tully}, {Hoffman}  \&
  {Courtois}}{{Pomar{\`e}de} et~al.}{2015}]{2015ApJ...812...17P}
{Pomar{\`e}de} D.,  {Tully} R.~B.,  {Hoffman} Y.,   {Courtois} H.~M.,  2015,
  \mn@doi [\apj] {10.1088/0004-637X/812/1/17}, \href
  {http://adsabs.harvard.edu/abs/2015ApJ...812...17P} {812, 17}

\bibitem[\protect\citeauthoryear{{Poudel}, {Hein{\"a}m{\"a}ki}, {Tempel},
  {Einasto}, {Lietzen}  \& {Nurmi}}{{Poudel}
  et~al.}{2017}]{2017A&A...597A..86P}
{Poudel} A.,  {Hein{\"a}m{\"a}ki} P.,  {Tempel} E.,  {Einasto} M.,  {Lietzen}
  H.,   {Nurmi} P.,  2017, \mn@doi [\aap] {10.1051/0004-6361/201629639}, \href
  {http://adsabs.harvard.edu/abs/2017A%26A...597A..86P} {597, A86}

\bibitem[\protect\citeauthoryear{{Proust} et~al.,}{{Proust}
  et~al.}{2006}]{2006A&A...447..133P}
{Proust} D.,  et~al., 2006, \mn@doi [\aap] {10.1051/0004-6361:20052838}, \href
  {http://adsabs.harvard.edu/abs/2006A%26A...447..133P} {447, 133}

\bibitem[\protect\citeauthoryear{Ramachandra \& Shandarin}{Ramachandra \&
  Shandarin}{2015}]{Ramachandara_Shandarin:15}
Ramachandra N.~S.,  Shandarin S.~F.,  2015, \mn@doi [Monthly Notices of the
  Royal Astronomical Society] {10.1093/mnras/stv1389}, 452, 1643

\bibitem[\protect\citeauthoryear{Ramachandra \& Shandarin}{Ramachandra \&
  Shandarin}{2017}]{Ramachandra2017}
Ramachandra N.~S.,  Shandarin S.~F.,  2017, \mn@doi [Monthly Notices of the
  Royal Astronomical Society] {10.1093/mnras/stx183}, 467, 1748

\bibitem[\protect\citeauthoryear{{Romano-D{\'{\i}}az} \& {van de
  Weygaert}}{{Romano-D{\'{\i}}az} \& {van de Weygaert}}{2007}]{romanodiaz2007}
{Romano-D{\'{\i}}az} E.,  {van de Weygaert} R.,  2007, \mn@doi [\mnras]
  {10.1111/j.1365-2966.2007.12190.x}, \href
  {http://adsabs.harvard.edu/abs/2007MNRAS.382....2R} {382, 2}

\bibitem[\protect\citeauthoryear{{Sahni} \& {Coles}}{{Sahni} \&
  {Coles}}{1995}]{1995PhR...262....1S}
{Sahni} V.,  {Coles} P.,  1995, \mn@doi [\physrep]
  {10.1016/0370-1573(95)00014-8}, \href
  {http://adsabs.harvard.edu/abs/1995PhR...262....1S} {262, 1}

\bibitem[\protect\citeauthoryear{{Sahni}, {Sathyaprakah}  \&
  {Shandarin}}{{Sahni} et~al.}{1994}]{sahni1994}
{Sahni} V.,  {Sathyaprakah} B.~S.,   {Shandarin} S.~F.,  1994, \mn@doi [\apj]
  {10.1086/174464}, \href {http://adsabs.harvard.edu/abs/1994ApJ...431...20S}
  {431, 20}

\bibitem[\protect\citeauthoryear{Sato, Nakajima, Shiraga, Atsumi, Yoshida,
  Koller, Gerig  \& Kikinis}{Sato et~al.}{1998}]{sato1998}
Sato Y.,  Nakajima S.,  Shiraga N.,  Atsumi H.,  Yoshida S.,  Koller T.,  Gerig
  G.,   Kikinis R.,  1998, \mn@doi [Medical Image Analysis] {DOI:
  10.1016/S1361-8415(98)80009-1}, 2, 143

\bibitem[\protect\citeauthoryear{{Scannapieco} et~al.,}{{Scannapieco}
  et~al.}{2012}]{2012MNRAS.423.1726S}
{Scannapieco} C.,  et~al., 2012, \mn@doi [\mnras]
  {10.1111/j.1365-2966.2012.20993.x}, \href
  {http://adsabs.harvard.edu/abs/2012MNRAS.423.1726S} {423, 1726}

\bibitem[\protect\citeauthoryear{{Schaap} \& {van de Weygaert}}{{Schaap} \&
  {van de Weygaert}}{2000}]{schaapwey2000}
{Schaap} W.~E.,  {van de Weygaert} R.,  2000, \aap, \href
  {http://adsabs.harvard.edu/abs/2000A%26A...363L..29S} {363, L29}

\bibitem[\protect\citeauthoryear{{Schaye} et~al.,}{{Schaye}
  et~al.}{2015}]{eagle2015}
{Schaye} J.,  et~al., 2015, \mn@doi [\mnras] {10.1093/mnras/stu2058}, \href
  {http://adsabs.harvard.edu/abs/2015MNRAS.446..521S} {446, 521}

\bibitem[\protect\citeauthoryear{{Shandarin}}{{Shandarin}}{2011}]{shandarin2011}
{Shandarin} S.~F.,  2011, \mn@doi [\jcap] {10.1088/1475-7516/2011/05/015},
  \href {http://adsabs.harvard.edu/abs/2011JCAP...05..015S} {5, 15}

\bibitem[\protect\citeauthoryear{Shandarin \& Sunyaev}{Shandarin \&
  Sunyaev}{2009}]{shandsuny2009}
Shandarin S.~F.,  Sunyaev R.,  2009, \aap, 500, 19

\bibitem[\protect\citeauthoryear{Shandarin \& Zel'dovich}{Shandarin \&
  Zel'dovich}{1989}]{shandzeld1989}
Shandarin S.,  Zel'dovich Y.,  1989, Rev. Mod. Phys., 61, 185

\bibitem[\protect\citeauthoryear{Shandarin, Sheth  \& Sahni}{Shandarin
  et~al.}{2004}]{shandarin2004}
Shandarin S.~F.,  Sheth J.~V.,   Sahni V.,  2004, \mnras, 353, 162

\bibitem[\protect\citeauthoryear{{Shandarin}, {Habib}  \&
  {Heitmann}}{{Shandarin} et~al.}{2012}]{shandarin2012}
{Shandarin} S.,  {Habib} S.,   {Heitmann} K.,  2012, \mn@doi [\prd]
  {10.1103/PhysRevD.85.083005}, \href
  {http://adsabs.harvard.edu/abs/2012PhRvD..85h3005S} {85, 083005}

\bibitem[\protect\citeauthoryear{{Shapley}}{{Shapley}}{1930}]{1930BHarO.874....9S}
{Shapley} H.,  1930, Harvard College Observatory Bulletin, \href
  {http://adsabs.harvard.edu/abs/1930BHarO.874....9S} {874, 9}

\bibitem[\protect\citeauthoryear{Shectman, Landy, Oemler, Tucker, Lin, Kirshner
   \& Schechter}{Shectman et~al.}{1996}]{shectman1996}
Shectman S.~A.,  Landy S.~D.,  Oemler A.,  Tucker D.~L.,  Lin H.,  Kirshner
  R.~P.,   Schechter P.~L.,  1996, \apj, 470, 172

\bibitem[\protect\citeauthoryear{{Shen}, {Abel}, {Mo}  \& {Sheth}}{{Shen}
  et~al.}{2006}]{shen2006}
{Shen} J.,  {Abel} T.,  {Mo} H.~J.,   {Sheth} R.~K.,  2006, \mn@doi [\apj]
  {10.1086/504513}, \href {http://adsabs.harvard.edu/abs/2006ApJ...645..783S}
  {645, 783}

\bibitem[\protect\citeauthoryear{{Sheth} \& {Tormen}}{{Sheth} \&
  {Tormen}}{1999}]{1999MNRAS.308..119S}
{Sheth} R.~K.,  {Tormen} G.,  1999, \mn@doi [\mnras]
  {10.1046/j.1365-8711.1999.02692.x}, \href
  {http://adsabs.harvard.edu/abs/1999MNRAS.308..119S} {308, 119}

\bibitem[\protect\citeauthoryear{Sheth \& van~de Weygaert}{Sheth \& van~de
  Weygaert}{2004}]{shethwey2004}
Sheth R.,  van~de Weygaert R.,  2004, \mnras, 350, 517

\bibitem[\protect\citeauthoryear{Shivashankar, Pranav, Natarajan, van~de
  Weygaert, Bos  \& Rieder}{Shivashankar et~al.}{2016}]{shivashankar2016}
Shivashankar N.,  Pranav P.,  Natarajan V.,  van~de Weygaert R.,  Bos E.~P.,
  Rieder S.,  2016, IEEE Transactions on Visualization and Computer Graphics,
  22, 1745

\bibitem[\protect\citeauthoryear{{Sorce} et~al.,}{{Sorce}
  et~al.}{2016}]{sorce2016}
{Sorce} J.~G.,  et~al., 2016, \mnras, 455, 2078

\bibitem[\protect\citeauthoryear{{Sousbie}}{{Sousbie}}{2011}]{sousbie2011}
{Sousbie} T.,  2011, \mn@doi [\mnras] {10.1111/j.1365-2966.2011.18394.x}, \href
  {http://adsabs.harvard.edu/abs/2011MNRAS.414..350S} {414, 350}

\bibitem[\protect\citeauthoryear{{Sousbie}, {Pichon}, {Colombi}, {Novikov}  \&
  {Pogosyan}}{{Sousbie} et~al.}{2008a}]{sousbie2008}
{Sousbie} T.,  {Pichon} C.,  {Colombi} S.,  {Novikov} D.,   {Pogosyan} D.,
  2008a, \mn@doi [\mnras] {10.1111/j.1365-2966.2007.12685.x}, \href
  {http://adsabs.harvard.edu/abs/2008MNRAS.383.1655S} {383, 1655}

\bibitem[\protect\citeauthoryear{{Sousbie}, {Pichon}, {Courtois}, {Colombi}  \&
  {Novikov}}{{Sousbie} et~al.}{2008b}]{sousbie2008b}
{Sousbie} T.,  {Pichon} C.,  {Courtois} H.,  {Colombi} S.,   {Novikov} D.,
  2008b, \mn@doi [\apjl] {10.1086/523669}, \href
  {http://adsabs.harvard.edu/abs/2008ApJ...672L...1S} {672, L1}

\bibitem[\protect\citeauthoryear{{Sousbie}, {Colombi}  \& {Pichon}}{{Sousbie}
  et~al.}{2009}]{sousbie2009}
{Sousbie} T.,  {Colombi} S.,   {Pichon} C.,  2009, \mn@doi [\mnras]
  {10.1111/j.1365-2966.2008.14244.x}, \href
  {http://adsabs.harvard.edu/abs/2009MNRAS.393..457S} {393, 457}

\bibitem[\protect\citeauthoryear{{Sousbie}, {Pichon}  \& {Kawahara}}{{Sousbie}
  et~al.}{2011}]{sousbie2011b}
{Sousbie} T.,  {Pichon} C.,   {Kawahara} H.,  2011, \mn@doi [\mnras]
  {10.1111/j.1365-2966.2011.18395.x}, \href
  {http://adsabs.harvard.edu/abs/2011MNRAS.414..384S} {414, 384}

\bibitem[\protect\citeauthoryear{{Springel}}{{Springel}}{2005}]{2005MNRAS.364.1105S}
{Springel} V.,  2005, \mn@doi [\mnras] {10.1111/j.1365-2966.2005.09655.x},
  \href {http://adsabs.harvard.edu/abs/2005MNRAS.364.1105S} {364, 1105}

\bibitem[\protect\citeauthoryear{{Springel} et~al.,}{{Springel}
  et~al.}{2005}]{springmillen2005}
{Springel} V.,  et~al., 2005, \mn@doi [\nat] {10.1038/nature03597}, \href
  {http://adsabs.harvard.edu/abs/2005Natur.435..629S} {435, 629}

\bibitem[\protect\citeauthoryear{{Stoica}, {Gregori}  \& {Mateu}}{{Stoica}
  et~al.}{2005}]{stoica2005}
{Stoica} R.~S.,  {Gregori} P.,   {Mateu} J.,  2005, Stochastic Processes and
  their Applications, 115, 1860

\bibitem[\protect\citeauthoryear{{Stoica}, {Mart{\'{\i}}nez}  \&
  {Saar}}{{Stoica} et~al.}{2007}]{2007JRSSC..56....1S}
{Stoica} R.~S.,  {Mart{\'{\i}}nez} V.~J.,   {Saar} E.,  2007, \mn@doi [Journal
  of the Royal Statistical Society: Series C]
  {10.1111/j.1467-9876.2007.00587.}, \href
  {http://adsabs.harvard.edu/abs/2007JRSSC..56....1S} {56, 459}

\bibitem[\protect\citeauthoryear{{Stoica}, {Mart{\'{\i}}nez}  \&
  {Saar}}{{Stoica} et~al.}{2010}]{2010A&A...510A..38S}
{Stoica} R.~S.,  {Mart{\'{\i}}nez} V.~J.,   {Saar} E.,  2010, \mn@doi [\aap]
  {10.1051/0004-6361/200912823}, \href
  {http://adsabs.harvard.edu/abs/2010A%26A...510A..38S} {510, A38}

\bibitem[\protect\citeauthoryear{{Sutter}, {Lavaux}, {Wandelt}  \&
  {Weinberg}}{{Sutter} et~al.}{2012}]{sutter2012}
{Sutter} P.~M.,  {Lavaux} G.,  {Wandelt} B.~D.,   {Weinberg} D.~H.,  2012,
  \mn@doi [\apj] {10.1088/0004-637X/761/1/44}, \href
  {http://adsabs.harvard.edu/abs/2012ApJ...761...44S} {761, 44}

\bibitem[\protect\citeauthoryear{{Sutter}, {Carlesi}, {Wandelt}  \&
  {Knebe}}{{Sutter} et~al.}{2015}]{2015MNRAS.446L...1S}
{Sutter} P.~M.,  {Carlesi} E.,  {Wandelt} B.~D.,   {Knebe} A.,  2015, \mn@doi
  [\mnras] {10.1093/mnrasl/slu155}, \href
  {http://adsabs.harvard.edu/abs/2015MNRAS.446L...1S} {446, L1}

\bibitem[\protect\citeauthoryear{{Tegmark} et~al.,}{{Tegmark}
  et~al.}{2004}]{tegmark2004}
{Tegmark} M.,  et~al., 2004, \mn@doi [\apj] {10.1086/382125}, \href
  {http://adsabs.harvard.edu/abs/2004ApJ...606..702T} {606, 702}

\bibitem[\protect\citeauthoryear{{Tempel} \& {Libeskind}}{{Tempel} \&
  {Libeskind}}{2013}]{tempel2013}
{Tempel} E.,  {Libeskind} N.~I.,  2013, \mn@doi [\apjl]
  {10.1088/2041-8205/775/2/L42}, \href
  {http://adsabs.harvard.edu/abs/2013ApJ...775L..42T} {775, L42}

\bibitem[\protect\citeauthoryear{{Tempel} \& {Tamm}}{{Tempel} \&
  {Tamm}}{2015}]{2015A&A...576L...5T}
{Tempel} E.,  {Tamm} A.,  2015, \mn@doi [\aap] {10.1051/0004-6361/201525827},
  \href {http://adsabs.harvard.edu/abs/2015A%26A...576L...5T} {576, L5}

\bibitem[\protect\citeauthoryear{{Tempel}, {Stoica}  \& {Saar}}{{Tempel}
  et~al.}{2013}]{tempel2012}
{Tempel} E.,  {Stoica} R.~S.,   {Saar} E.,  2013, \mnras, 428, 1827

\bibitem[\protect\citeauthoryear{{Tempel}, {Stoica}, {Mart{\'{\i}}nez},
  {Liivam{\"a}gi}, {Castellan}  \& {Saar}}{{Tempel} et~al.}{2014}]{tempel2014}
{Tempel} E.,  {Stoica} R.~S.,  {Mart{\'{\i}}nez} V.~J.,  {Liivam{\"a}gi} L.~J.,
   {Castellan} G.,   {Saar} E.,  2014, \mn@doi [\mnras]
  {10.1093/mnras/stt2454}, \href
  {http://adsabs.harvard.edu/abs/2014MNRAS.438.3465T} {438, 3465}

\bibitem[\protect\citeauthoryear{{Tempel}, {Guo}, {Kipper}  \&
  {Libeskind}}{{Tempel} et~al.}{2015}]{2015MNRAS.450.2727T}
{Tempel} E.,  {Guo} Q.,  {Kipper} R.,   {Libeskind} N.~I.,  2015, \mn@doi
  [\mnras] {10.1093/mnras/stv919}, \href
  {http://adsabs.harvard.edu/abs/2015MNRAS.450.2727T} {450, 2727}

\bibitem[\protect\citeauthoryear{{Tempel}, {Stoica}, {Kipper}  \&
  {Saar}}{{Tempel} et~al.}{2016}]{2016AC....16...17T}
{Tempel} E.,  {Stoica} R.~S.,  {Kipper} R.,   {Saar} E.,  2016, \mn@doi
  [Astronomy and Computing] {10.1016/j.ascom.2016.03.004}, \href
  {http://adsabs.harvard.edu/abs/2016A%26C....16...17T} {16, 17}

\bibitem[\protect\citeauthoryear{{Trowland}}{{Trowland}}{2013}]{trowlandphd2013}
{Trowland} H.~E.,  2013, Spinning galaxies within the large scale structure of
  the Universe, Ph.D. thesis, University of Sydney

\bibitem[\protect\citeauthoryear{{Trowland}, {Lewis}  \&
  {Bland-Hawthorn}}{{Trowland} et~al.}{2013}]{trowland2013}
{Trowland} H.~E.,  {Lewis} G.~F.,   {Bland-Hawthorn} J.,  2013, \mn@doi [\apj]
  {10.1088/0004-637X/762/2/72}, \href
  {http://adsabs.harvard.edu/abs/2013ApJ...762...72T} {762, 72}

\bibitem[\protect\citeauthoryear{{Tully} \& {Fisher}}{{Tully} \&
  {Fisher}}{1987}]{tully_atlas1987}
{Tully} R.~B.,  {Fisher} J.~R.,  1987, {Atlas of Nearby Galaxies}

\bibitem[\protect\citeauthoryear{{Tully}, {Shaya}, {Karachentsev}, {Courtois},
  {Kocevski}, {Rizzi}  \& {Peel}}{{Tully} et~al.}{2008}]{tully2008}
{Tully} R.~B.,  {Shaya} E.~J.,  {Karachentsev} I.~D.,  {Courtois} H.~M.,
  {Kocevski} D.~D.,  {Rizzi} L.,   {Peel} A.,  2008, \mn@doi [\apj]
  {10.1086/527428}, \href {http://adsabs.harvard.edu/abs/2008ApJ...676..184T}
  {676, 184}

\bibitem[\protect\citeauthoryear{{Tully}, {Courtois}, {Hoffman}  \&
  {Pomar{\`e}de}}{{Tully} et~al.}{2014}]{tully2014}
{Tully} R.~B.,  {Courtois} H.,  {Hoffman} Y.,   {Pomar{\`e}de} D.,  2014,
  \mn@doi [\nat] {10.1038/nature13674}, \href
  {http://adsabs.harvard.edu/abs/2014Natur.513...71T} {513, 71}

\bibitem[\protect\citeauthoryear{{Vogelsberger} et~al.,}{{Vogelsberger}
  et~al.}{2014}]{illustris2014}
{Vogelsberger} M.,  et~al., 2014, \mn@doi [\mnras] {10.1093/mnras/stu1536},
  \href {http://adsabs.harvard.edu/abs/2014MNRAS.444.1518V} {444, 1518}

\bibitem[\protect\citeauthoryear{{Wojtak}, {Knebe}, {Watson}, {Iliev},
  {He{\ss}}, {Rapetti}, {Yepes}  \& {Gottl{\"o}ber}}{{Wojtak}
  et~al.}{2014}]{2014MNRAS.438.1805W}
{Wojtak} R.,  {Knebe} A.,  {Watson} W.~A.,  {Iliev} I.~T.,  {He{\ss}} S.,
  {Rapetti} D.,  {Yepes} G.,   {Gottl{\"o}ber} S.,  2014, \mn@doi [\mnras]
  {10.1093/mnras/stt2321}, \href
  {http://adsabs.harvard.edu/abs/2014MNRAS.438.1805W} {438, 1805}

\bibitem[\protect\citeauthoryear{{Wu}, {Batuski}  \& {Khalil}}{{Wu}
  et~al.}{2009}]{wu2009}
{Wu} Y.,  {Batuski} D.~J.,   {Khalil} A.,  2009, \mn@doi [\apj]
  {10.1088/0004-637X/707/2/1160}, \href
  {http://adsabs.harvard.edu/abs/2009ApJ...707.1160W} {707, 1160}

\bibitem[\protect\citeauthoryear{{Zel'dovich}}{{Zel'dovich}}{1970}]{zeldovich1970}
{Zel'dovich} Y.~B.,  1970, \aap, \href
  {http://adsabs.harvard.edu/abs/1970A%26A.....5...84Z} {5, 84}

\bibitem[\protect\citeauthoryear{{Zel'dovich}}{{Zel'dovich}}{1974}]{1974FizSz..24..304Z}
{Zel'dovich} Y.~B.,  1974, Fizika Sz., \href
  {http://adsabs.harvard.edu/abs/1974FizSz..24..304Z} {24, 304}

\bibitem[\protect\citeauthoryear{{Zeldovich}, {Einasto}  \&
  {Shandarin}}{{Zeldovich} et~al.}{1982}]{zeldovich1982}
{Zeldovich} I.~B.,  {Einasto} J.,   {Shandarin} S.~F.,  1982, \mn@doi [\nat]
  {10.1038/300407a0}, \href {http://adsabs.harvard.edu/abs/1982Natur.300..407Z}
  {300, 407}

\bibitem[\protect\citeauthoryear{{de Lapparent}, V.  \& {Huchra}}{{de
  Lapparent} et~al.}{1986}]{lapparent1986}
{de Lapparent} V. {Geller} M.,   {Huchra} J.,  1986, \apj, 302, L1

\bibitem[\protect\citeauthoryear{{de Vaucouleurs}}{{de
  Vaucouleurs}}{1953}]{1953AJ.....58...30D}
{de Vaucouleurs} G.,  1953, \mn@doi [\aj] {10.1086/106805}, \href
  {http://adsabs.harvard.edu/abs/1953AJ.....58...30D} {58, 30}

\bibitem[\protect\citeauthoryear{{van Haarlem} \& {van de Weygaert}}{{van
  Haarlem} \& {van de Weygaert}}{1993}]{haarlemwey1993}
{van Haarlem} M.,  {van de Weygaert} R.,  1993, \apj, 418, 544

\bibitem[\protect\citeauthoryear{{van de Weygaert}}{{van de
  Weygaert}}{1994}]{weygaert1994}
{van de Weygaert} R.,  1994, \aap, \href
  {http://adsabs.harvard.edu/abs/1994A\%26A...283..361V} {283, 361}

\bibitem[\protect\citeauthoryear{{van de Weygaert}}{{van de
  Weygaert}}{2016}]{weyiau308}
{van de Weygaert} R.,  2016, in {van de Weygaert} R.,  {Shandarin} S.,  {Saar}
  E.,   {Einasto} J.,  eds,  IAU Symposium Vol. 308, The Zeldovich Universe:
  Genesis and Growth of the Cosmic Web. pp 493--523 (\mn@eprint {arXiv}
  {1611.01222}), \mn@doi{10.1017/S1743921316010504}

\bibitem[\protect\citeauthoryear{van~de Weygaert \& Bertschinger}{van~de
  Weygaert \& Bertschinger}{1996}]{weyedb1996}
van~de Weygaert R.,  Bertschinger E.,  1996, \mnras, 281, 84

\bibitem[\protect\citeauthoryear{{van de Weygaert} \& {Bond}}{{van de Weygaert}
  \& {Bond}}{2008}]{weybond2008}
{van de Weygaert} R.,  {Bond} J.~R.,  2008, in {Plionis} M.,  {L{\'o}pez-Cruz}
  O.,   {Hughes} D.,  eds,  Lecture Notes in Physics, Berlin Springer Verlag
  Vol. 740, A Pan-Chromatic View of Clusters of Galaxies and the Large-Scale
  Structure. p.~335

\bibitem[\protect\citeauthoryear{{van de Weygaert} \& {Schaap}}{{van de
  Weygaert} \& {Schaap}}{2009}]{weyschaap2009}
{van de Weygaert} R.,  {Schaap} W.,  2009, in {Mart{\'{\i}}nez} V.~J.,  {Saar}
  E.,  {Mart{\'{\i}}nez-Gonz{\'a}lez} E.,   {Pons-Border{\'{\i}}a} M.-J.,  eds,
   Lecture Notes in Physics, Berlin Springer Verlag Vol. 665, Data Analysis in
  Cosmology. pp 291--413

\bibitem[\protect\citeauthoryear{{van de Weygaert}, {Shandarin}, {Saar}  \&
  {Einasto}}{{van de Weygaert} et~al.}{2016}]{iau308}
{van de Weygaert} R.,  {Shandarin} S.,  {Saar} E.,   {Einasto} J.,  eds, 2016,
  {The Zeldovich Universe: Genesis and Growth of the Cosmic Web}  IAU Symposium
  Vol. 308, \mn@doi{10.1017/S174392131601098X.
}

\makeatother
\end{thebibliography}

\section*{Affiliations}
\noindent
{\it
$^1$Leibniz-Institute f\"ur Astrophysik Potsdam (AIP), An der Sternwarte 16, D-14482 Potsdam, Germany\\
$^2$Kapteyn Astronomical Institute, University of Groningen, P.O. Box 800, 9700 AV Groningen, The Netherlands\\
$^3$Institute for Computational Cosmology, Durham Univerity, South Road, DH1 3LE, United Kingdom\\
$^4$Institute of Theoretical Astrophysics, University of Oslo, PO Box 1029 Blindern, N-0315, Oslo, Norway \\
$^5$Tartu Observatory, Observatooriumi 1, 61602 T\~oravere, Estonia \\
$^{6}$Kavli Institute for Particle Astrophysics and Cosmology, Stanford University, Menlo Park, CA 94025, USA\\
$^{7}$Department of Physics, Stanford University, Stanford, CA 94305, USA\\
$^{8}$NASA Ames Research Center N232, Moffett Field, Mountain View, CA 94035, U.S.A.\\
$^{9}$Instituto Astronomico de Ensenada,UNAM, Mexico\\
$^{10}$Departamento de F\'isica, Universidad de los Andes, Cra. 1 No. 18A-10, Edificio Ip, Bogot\'a, Colombia\\
$^{11}$Instituto de Astrof{\'i}sica, Pontificia Universidad Catolica de
Chile, Santiago, Chile\\
$^{12}$Centro de Astro-Ingenier{\'i}a, Pontificia Universidad Catolica
de Chile, Santiago, Chile\\
$^{13}$Observatoire de la Cote d'Azur, Laboratoire Lagrange, Boulevard de l'Observatoire, CS, 34229, 06304 NICE, France\\
$^{14}$Institute of Cosmology and Gravitation, University of Portsmouth, Portsmouth PO1 3FX, UK\\
$^{15}$Janusz Gil Institute of Astronomy, University of Zielona G\'ora, ul. Szafrana 2, 65-516 Zielona G\'ora, Poland\\
$^{16}$Racah Institute of Physics, Hebrew University of Jerusalem, Givat Ram, Jerusalem, 91904 Israel\\
$^{17}$Instituto de Astrof\`{i}sica de Canarias (IAC), C/V\'{i}a L\'{a}ctea, s/n, E-38200, La Laguna, Tenerife, Spain\\
$^{18}$Departamento Astrof\'{i}sica, Universidad de La Laguna (ULL), E-38206 La Laguna, Tenerife, Spain\\
$^{19}$Departamento de F\'isica Te\'{o}rica, M\'{o}dulo 15, Facultad de Ciencias, Universidad Aut\'{o}noma de Madrid, 28049 Madrid, Spain\\
$^{20}$Astro-UAM, UAM, Unidad Asociada CSIC\\
$^{21}$Scuola Normale Superiore, Piazza dei Cavalieri 7, I-56126 Pisa, Italy\\
$^{22}$Universidad de Buenos Aires, Facultad de Ciencias Exactas y Naturales, Buenos Aires, Argentina\\
$^{23}$CONICET-Universidad de Buenos Aires, Instituto de Astronom\'{\i}a y F\'{\i}sica del Espacio (IAFE), Buenos Aires, Argentina\\
$^{24}$Department of Physics and Astronomy, University of Kansas, Lawrence, Kansas 66045, USA\\
$^{25}$ICRAR, M468, University of Western Australia, Crawley, WA 6009, Australia \\
$^{26}$Universit\'e de Lorraine, Institut Elie Cartan de Lorraine, 54506 Vandoeuvre-l\'es-Nancy Cedex, France\\
$^{27}$Institut de M\'ecanique C\'eleste et Calcul des Eph\'em\'erides (IMCCE), Observatoire de Paris, 75014 Paris, France\\
$^{28}$Institut d'Astrophysique de Paris, CNRS UMR 7095 and UPMC, 98bis, bd Arago, F-75014 Paris, France
}

\end{document}